\documentclass[11pt,epsf,epsfig,aps,floatfix,preprint,nofootinbib]{revtex4-1}
\usepackage{graphics}
\usepackage{graphicx}
\usepackage{setspace}
\usepackage{multirow}
\usepackage[english]{babel}
\usepackage{fullpage}
\usepackage{amssymb}
\usepackage[final]{pdfpages}
\usepackage{amsmath}
\usepackage{verbatim}
\usepackage{pstricks}
\usepackage{slashed}

\newcommand{\nc}{\newcommand}
\nc{\beq}{\begin{equation}}
\nc{\eeq}{\end{equation}}
\nc{\barray}{\begin{eqnarray}}
\nc{\earray}{\end{eqnarray}}
\nc{\barrayn}{\begin{eqnarray*}}
\nc{\earrayn}{\end{eqnarray*}}
\nc{\bcenter}{\begin{center}}
\nc{\ecenter}{\end{center}}
\nc{\ket}[1]{| #1 \rangle}
\nc{\bra}[1]{\langle #1 |}
\nc{\0}{\ket{0}}
\nc{\mc}{\mathcal}
\nc{\er}[1]{(\ref{eq:#1})}
\nc{\onehalf}{\frac{1}{2}}
\nc{\partialbar}{\bar{\partial}}
\nc{\psit}{\widetilde{\psi}}
\nc{\Tr}{\mbox{Tr}}
\nc{\ev}{\;\mathrm{eV}}
\nc{\mev}{\;\mathrm{MeV}}
\nc{\gev}{\;\mathrm{GeV}}

\def\tN{\widetilde{N}}
\def\chii0{\chi_i^0}
\def\chij0{\chi_j^0}

\newcommand{\bi}{\begin{itemize}}
\newcommand{\ei}{\end{itemize}}

\newcommand{\gsim}{\lower.7ex\hbox{$\;\stackrel{\textstyle>}{\sim}\;$}}
\newcommand{\lsim}{\lower.7ex\hbox{$\;\stackrel{\textstyle<}{\sim}\;$}}

\begin{document}

\setlength{\baselineskip}{0.22in}

\begin{flushright}MCTP-11-34\\
\end{flushright}

\vspace{0.2cm}

\title{The Dark Matter Inverse Problem: \\ Extracting Particle Physics from Scattering Events}

\author{
Samuel D. McDermott, Hai-Bo Yu, Kathryn M. Zurek
}

\vspace*{0.2cm}

\affiliation{
Michigan Center for Theoretical Physics, Department of Physics, University of Michigan, Ann Arbor, MI 48109
}

\date{\today}

\begin{abstract}
\noindent
The primary observable in dark matter direct detection is the spectrum of scattering events.
We simulate multiple positive direct detection signals (on germanium, xenon, and argon targets) to explore the extent to which the underlying particle physics, manifested in the momentum dependence of the operator mediating the scattering, can be extracted.  Taking into account realization (Poisson) noise, a single target nucleus with 300 events has limited power to discriminate operators with momentum dependence differing by $q^{\pm2}$ for a wide range of dark matter masses from 10 GeV to 1 TeV. With the inclusion of multiple targets (or a factor of several more events on a single target), the discrimination of operators with different momentum dependence becomes very strong at the 95\% C.L. for dark matter candidates of mass 50 GeV and above. On the other hand, operator discrimination remains poor for 10 GeV candidates until  multiple experiments each collect 1000 or more events. 
\end{abstract}

\maketitle

\tableofcontents

\section{Introduction}

Dark matter (DM) dominates the dynamics of matter on large scales in our universe, yet the theory which describes its underlying interactions, both with itself and with the Standard Model (SM) sector, remains hidden.  
Viable, testable models that are capable of both describing the DM and satisfying the observed constraints abound in the literature.  The most popular DM candidate is the weakly interacting massive particle (WIMP).  A well-studied example of a WIMP is the lightest supersymmetric particle in the minimal supersymmetric standard model (see {\it e.g.} \cite{Jungman:1995df} for a review).
This candidate satisfies the observational constraints that it is long-lived, cold, and weakly interacting.  Its mass is typically predicted to be between 10 GeV and 1 TeV, and its interaction with ordinary particles is mediated by an operator that gives rise to momentum-independent interactions.

Direct detection experiments, such as XENON100 \cite{Aprile:2010um} and CDMS-II \cite{Ahmed:2009zw}, have achieved a high level of sensitivity to a WIMP in this mass range.   In general, the focus has been on ruling out DM candidates with the most standard characteristics -- momentum independent scattering in the mass range well above 10 GeV.  However, some recent experimental hints  have led to more detailed investigations of other possibilities.    In particular, the DAMA-LIBRA \cite{Bernabei:2010mq}, CoGeNT \cite{Aalseth:2010vx}, and CRESST-II \cite{Angloher:2011uu} experiments have excesses which may be consistent with a DM mass of order 10 GeV, leading to a renewed theoretical interest in DM candidates with mass in this range.  However, these results are in tension with the null results of XENON10 \cite{S2,XENON10}, XENON100 \cite{Aprile:2010um}, and CDMS \cite{lowthresh}.  This has revitalized interest in DM candidates that have non-standard features that might allow one to evade the constraints from the null experiments.  In particular, operators that mediate interactions that give rise to momentum-dependent scattering rates, first considered in \cite{Pospelov:2000bq,kurylov,Bagnasco}, have received renewed interest \cite{Chang:2009yt,Fitzpatrick:2010br,An:2010kc,Banks,Chang:2010en,darkdark}.

Independent of these results, direct detection experiments with increasing sensitivity are continuing to probe the WIMP DM hypothesis. The XENON collaboration will soon begin construction on a XENON1T phase \cite{xenon}; a 3.6 ton liquid argon detector operated by the DEAP/CLEAN collaboration, DEAP-3600 \cite{Kos:2010zz}, is under construction; and 100 kg and 1 ton cryogenic germanium detectors are being planned \cite{cdmsnew} by the Super-CDMS/GEODM collaborations.  If there is a positive signal from one of these experiments, the signal must be confirmed by multiple experiments with different targets.  Furthermore, one would like to determine experimentally whether the scattering is in fact mediated by a standard spin-independent operator or whether the underlying particle physics is more complex, with non-standard types of momentum dependence in the scattering. Additionally, recent theoretical work has focused on the feasibility of extracting different types of information from scattering events \cite{other,Pato:2010zk}.

In this paper we explore the dark matter inverse problem; that is, the capability of direct detection experiments to extract the underlying particle physics mediating the scattering of a DM particle.  In practice this question boils down to how well the momentum dependence in an operator can be mimicked by other operators for a given target and DM mass.  We simulate detection events using different types of interactions with varying momentum and velocity dependence, then we fit these events by a wide variety of interactions. In principle, one can distinguish interaction types by just checking the recoil spectra for an individual target.  For example, if DM scatters with the target nucleus through an interaction with a positive momentum dependence, one would expect events to drop in low recoil energy bins, while the event rate increases exponentially towards low recoil energy for a standard momentum independent interaction.  However, if only one target is available, the capability for distinguishing operators through the spectra in this way is limited by the experimental noise and detector threshold.  With multiple nuclear targets one may also compare the overlap of the DM preferred regions on different target nuclei to determine the correct operator.  We find that, for similar cross-sections, analysis of high mass DM particles leaves distinctive qualitative imprints that allow one to extract the momentum dependence in the scattering. When two operators cannot be distinguished between each other, it is because they have a very similar momentum dependence. On the other hand, operator discrimination for a 10 GeV candidate is poor.  We conclude that a significantly lower threshold, or much improved statistics, than is available for the next generation of experiments will be necessary in order to extract the particle physics mediating the scattering for a light DM candidate.

We begin the paper by briefly reviewing direct detection basics.
In Section III we discuss our methodology, including detector effects, analysis methods, and conventions.
In Section IV we go over our results.
Finally, we summarize our conclusions in Section V.

\section{Preliminaries}

We compare the spectra for standard spin- and momentum-independent scattering against operators with momentum dependence. We call operators with $n$ additional powers of momentum transfer $q^n$-type scattering, where $n = \pm 2,\pm4$.  We also consider scattering via anapole and dipole operators, which feature mixed momentum dependence (constant and $q^2$ dependence for the former operator, and $q^2$ and $q^4$ dependence for the latter).  In this section we review the rate relations for the various types of operators and define our conventions for reporting results. 

For generic scattering the observed differential rate of observation of DM particles may be written
\beq
\label{rate}
\frac{dR}{dE_R}(E_R)=\frac{\rho_0}{m_{\rm DM}m_N} \int d^3\vec{v} ~v~ f(v_0,\vec{v}_e;v_{min},v_{esc}) \epsilon_{\rm eff}(E_R) \frac{d \sigma_{{\rm DM}- N}}{dE_R}(v,A,E_R).
\eeq
Here, $\rho_0$ is the local DM density; $m_{\rm DM}$ is the DM mass; $m_N$ is the detector nucleus mass; the velocity distribution is given by $f(v_0,\vec{v}_e;v_{min},v_{esc})$; $\epsilon_{\rm eff}(E_R)$ is the detector efficiency, which may depend on the energy of recoil; and the differential cross-section $d \sigma_{{\rm DM}- N}/dE_R$ describes the interaction.
The total number of events is given by integrating the differential rate as
\beq
\label{spec}
N_i(E_R^{min},E_R^{max})=\int_{E_R^{min,i}}^{E_R^{max,i}}dE_R \frac{d R}{dE_R} \epsilon_{\rm exp},
\eeq
where $\epsilon_{\rm exp}$ is the exposure, generally given in terms of kilograms of exposed material multiplied by the days of duration of the experiment.

All of the particle physics information present in Eq.~\eqref{rate} is contained in the final term, $d \sigma_{{\rm DM}- N}/dE_R$, while the astrophysical considerations are reflected in $\rho_0$ and $f(v_0,\vec{v}_e;v_{min},v_{esc})$.
The remaining effects, such as $\epsilon_{\rm eff}(E_R)$, $\epsilon_{\rm exp}$, and $m_N$, are detector-specific.
The objective of this paper is to extract information about the operator content of $d \sigma_{{\rm DM}- N}/dE_R$ based solely on observables.
We fix the astrophysical parameters by the best current observations and assume the detector effects are well understood.
In particular, we take $\rho_0=0.4$ GeV/cm$^3$ and assume a Maxwell-Boltzmann velocity distribution with a smooth exponential cutoff at the escape velocity:
\beq
f(v_0,\vec{v}_e;v_{min},v_{esc})=\left\{ \begin{array}{l l}
\frac{1}{N} \left( e^{-(\vec{v}_e+\vec{v}_{\rm DM}) \cdot (\vec{v}_e+\vec{v}_{\rm DM})/v_0^2} - e^{-v_{esc}^2/v_0^2} \right) & v<v_{esc} \\
0 & v>v_{esc}
\end{array} ,
\right.
\label{veldis}
\eeq
where we take $v_{esc}=544$ km/s, $v_0=220$ km/s, and $v_e=232$ km/s.
The normalization $N$ for this distribution can be found in the appendix of \cite{Fitzpatrick:2010br}.
We find that our results are insensitive to changes in these velocity parameters.
Note here that we are only interested in extracting the correct particle physics interaction. This differs from recent work where the goal was to marginalize over astrophysical uncertainties while extracting the basic physical parameters \cite{Pato:2010zk}. Astrophysical uncertainty will enlarge the preferred region in the relevant parameter space for a single experiment, but our results involve a comparison of multiple experiments for which the astrophysics are the same. We have checked that our results are unchanged as long as we use the same inputs for the different experiments and do not marginalize over the unknown quantities.

The differential scattering cross-section is related to the DM-nucleus scattering cross-section $\sigma_N$ via
\beq
\label{scatt}
\frac{d \sigma_{{\rm DM}- N}}{dE_R}=\frac{m_N \sigma_N}{2\mu_N^2 v^2},
\eeq
where $\mu_N$ is the reduced mass of the DM-nucleus pair and $v$ is their relative velocity.
For standard spin- and momentum-independent scattering, which we denote ``std", the results are quoted in terms of the cross-section for scattering off a nucleon, $\sigma_n^{std}$:
\beq
\label{si}
\sigma_N^{std}=\sigma_n^{std}\frac{\mu_N^2}{\mu_n^2}\frac{\left[f_pZ+f_n(A-Z)\right]^2}{f_p^2}F_1^2(A,E_R),
\eeq
where $\mu_n$ is the reduced mass of the DM-nucleon pair.
We take $f_p=f_n=1$ here, and assume a standard Helm form factor, $F_1(A,E_R) =3 j_1(q r_0)/(q r_0) e^{-(qs)^2 {\rm fm}^2 /2}$, with $s = 0.9$ and  $r_0 = \left((1.23 A^{1/3}-0.6)^2+7/3 (0.52\pi)^2-5 s^2\right)^{1/2} \mbox{ fm}$  \cite{Fricke:1995zz}.  

For spin-independent, momentum-dependent scattering, the results will be reported according to the convention \cite{Chang:2009yt}:
\beq
\label{qn}
\sigma_N^{q^n}=\left(\frac{q}{q_0}\right)^n\sigma_N^{std}.
\eeq
In this work, we take $q_0=50{\rm~MeV}$ for all values of $n$\footnote{Particle physics realizations of momentum-dependent scattering cross sections are discussed in~\cite{Chang:2009yt}. Note that \cite{Chang:2009yt} assumes that the mediator mass is heavier than the momentum transfer so that $n=2,4$ are positive. If the mediator mass is smaller than the momentum transfer, we can have $n=-2,-4$.}. 
We allow $\sigma_n$ to float for both the standard and $q^n$ type operators to satisfy normalization conditions described below.
In general, this will lead to $\sigma_n \sim \mathcal{O}(10^{-45} )\mbox{ cm}^2$.
Note that the momentum-transfer dependence of the $q^n$-type operators serves to introduce extra powers of the recoil energy, not extra powers of the DM velocity. 

Lastly, we define our conventions for operators with mixed momentum dependence, the anapole ($\bar{\chi}\gamma^\mu\gamma_5\chi A_\mu$) and dipole ($\bar{\chi}\sigma^{\mu\nu}\chi F_{\mu\nu}$) operators.  The scattering cross-sections for these operators are \cite{Fitzpatrick:2010br}
\begin{align}
\nonumber
\sigma_N^{an}&=\frac{\mu_N^2}{4\pi M^4} \left\{4 v^2 Z^2 F_1^2(A,E_R)-q^2 \left[ \frac{1}{\mu_N^2} Z^2 F_1^2(A,E_R) -2A^2 \frac{J+1}{3J} \frac{b_N^2}{m_N^2 b_n^2} F_2^2(A,E_R) \right] \right\},
\\
\label{dip}
\sigma_N^{dip}&=\frac{4\mu_N^2}{\pi M^4\Lambda^2} \Bigg\{4 q^2 v^2 Z^2 F_1^2(A,E_R) \nonumber \\ 
&\qquad -q^4 \left[ \left(\frac{2}{m_{\rm DM}m_N}+\frac{1}{m_N^2}\right) Z^2 F_1^2(A,E_R) -2A^2 \frac{J+1}{3J} \frac{b_N^2}{m_N^2 b_n^2}F_2^2(A,E_R)  \right] \Bigg\} .
\end{align}
Here, $M$ is the mass of a mediator particle that couples an off-shell photon to the DM particle, while $\Lambda$ is a confinement or compositeness scale that describes the magnetic dipole moment physics; we take it to be 100 MeV in all of our analysis.
The mediator mass $M$ is \emph{a priori} unknown. In general, it floats around a fiducial value of a few hundred GeV.
$J$ is the spin of the target nucleus, $b_N$ is the magnetic moment of the target nucleus, and $b_n=e/2m_p$ is the Bohr magneton.
The form factor $F_2(A,E_R)$ for the spin coupling term is important for spin-odd nuclei and is taken from \cite{Bednyakov:2006ux}. Since the isoscalar and isovector couplings are model-dependent, we choose to take $a_p = a_n$ in this work. We find that the results are qualitatively insensitive to these exact values.

The normalized interaction strength that we display on the plots for both anapole and dipole moment operators is
\beq
\widehat{\sigma}_n=\frac{\mu_n^2
}{4\pi M^4} \simeq 1.05 \times 10^{-39} \left( \frac{\mu_n}{m_n} \right)^2 \left(\frac{M}{400~{\rm GeV}} \right)^{-4} {\rm~cm^2}.
\eeq
The fiducial value of $M$ and thus of $\widehat{\sigma}_n$ differs depending on the type of interaction and the mass of the DM, but these values are chosen to obey the same normalization conditions alluded to above.
Our normalization conventions are described in greater detail in the next section.

\section{Methodology}

We simulate mock experimental (``input'') data, and compare the goodness-of-fit of theoretical (``fit") spectra against the input data.  Ideal experimental data is first generated via a model, then smeared according to both Poisson statistics and the finite energy resolution of the experiment.  We then fit theoretical models (convoluted to take into account detector resolution) to the input data using a log-likelihood ratio test.  We describe our methodology in this section.

\subsection{Detector and Statistical Effects}

In order to fully simulate a direct detection experiment, one must fold detector effects into the rate given in Eq.~\eqref{rate}.
A convolution integral (accounting for the finite energy resolution of the detector) and Poisson noise (expected in a counting type experiment) are the most important of these effects.
For low mass DM in particular, these considerations are equally important in drawing conclusions.
We describe here how we take these effects into account.
\\~\\ \noindent \textbf{Convolution.}
In order to compare a theoretical model against an input data set, the rates must be smeared to take into account detector effects and noise. Both the input data and the theoretical spectra are smeared according to the energy resolution, because a realistic detector does not have perfect energy resolution.
The observed rate is well modeled by convolving the expected rate with an approximately Gaussian distribution whose width is a function of detector element and true recoil energy:
\beq
\label{conv}
\frac{d\widetilde{R}}{dE_R}=\int dE' \frac{dR}{dE_R}(E') \frac{1}{\sqrt{2\pi}~\sigma_{det}(A,E')} \exp \left[ -\frac{(E_R-E')^2}{2\sigma_{det}^2(A,E')} \right].
\eeq
For all isotopes present in a given detector we define the detector energy resolutions to be
\begin{align}
\sigma_{det}({\rm Ge},E)&=\sqrt{(0.3)^2+ (0.06)^2 E/{\rm keV}} ~{\rm keV,}\\
\sigma_{det}({\rm Xe},E)&=0.6 \sqrt{E/{\rm keV}}~ {\rm keV,}\\
\sigma_{det}({\rm Ar},E)&=0.7 \sqrt{E/{\rm keV}}~ {\rm keV}.
\end{align}
In all cases we take a flat efficiency.
The isotopic abundances are crucial for extracting information in the anapole and dipole cases. We list the spins, magnetic moments, and abundances of the important spin-odd nuclei for each element in Table \ref{phys}, and take the abundances for the spin-even nuclei to be set by naturally occurring levels.

\begin{table}
\centering
\begin{tabular}{| l | r | r | r |  l | r |}
\hline
Element & $A$ & $~Z$ & $J$ & $b_N/b_n$ & Abundance \\ \hline
Germanium & 73 & 32 & 9/$2^+$ &$ -0.879$ & $7.73\%$ \\ \hline
Xenon & 129 & 54 & $1/2^+$ & $-0.778$ & $26.4\%$ \\ \hline
Xenon & 131 & 54 & $3/2^+$ & $+0.692$ &$ 21.2\%$ \\ \hline
\end{tabular}
\caption{Physical properties of stable spin-odd nuclei present in detectors. The dominant argon nucleus has $A=40$, and there are no spin-odd isotopes with appreciable abundance.}\label{phys}
\end{table}

Integrating the convoluted rate in Eq.~\eqref{conv} over the recoil energy range in each bin $i$ gives rise to binned observation numbers
\beq
\label{number}
\tN_i(E_R^{min},E_R^{max})=\int_{E_R^{min,i}}^{E_R^{max,i}}dE_R \frac{d\widetilde{R}}{dE_R} \epsilon_{exp}.
\eeq
These binned observation numbers produce the observable quantities that make up both our input data and the data used in our fits.
We will refer to the vector of data points generated by Eq.~\eqref{number} as the ``convoluted spectrum."
\\~\\ \noindent \textbf{Poisson Noise.}
We expect that our input data are one of an ensemble of many other data that could be produced by the expected model.
Since the input data that come from any given experiment are produced by counting, we expect that the input data will deviate from the convoluted spectra given in Eq.~\eqref{number} at a level determined by Poisson statistics.
To simulate the observed events, we thus introduce stochastic Poisson noise distributed about the spectrum given in Eq.~\eqref{number}.

We have checked that, as expected, the average CLC generated by many noisy data sets converges on a CLC described by data generated by inputs from Eq.~\eqref{number} with no noise.
\\~\\ \noindent \textbf{Log-Likelihood Ratio Test.}
We use a log-likelihood ratio to provide an estimate of the fit parameters that best reproduce a set of input data.
For input data $n_i$ in the $i$-th energy bin (which has been smeared in energy as described above and in the number of events according to Poisson statistics) and theoretical parameters $\theta$ that produce a convoluted spectrum with values $\nu_i$, the log-likelihood ratio is
\beq
\label{llhood}
\widetilde{L}=2\sum_{i=1}^N\left[\nu_i({\bf \theta})-n_i+n_i\ln \frac{n_i}{\nu_i({\bf \theta})} \right] + {\rm ~const.},
\eeq
where if $n_i=0$ the last term in the brackets is set to zero.\footnote{The value of $\nu_i$ should never be exactly zero because even at high energy the convolution integral will receive some contribution from events in the low energy tail of the convolution.}  The parameters of interest are the DM-nucleon cross section and DM mass. The parameters that minimize Eq.~\eqref{llhood} provide the point of best fit, and contours of constant $\widetilde{L}$ values give confidence level regions. To define the 95\% confidence level regions we take the region inside of which $\widetilde{L} \leq \widetilde{L}_{\rm min}+\Delta \widetilde{L}_{95}^{\rm 2~d.o.f.}$, where for instance $\Delta \widetilde{L}_{95}^{\rm 2~d.o.f.}=5.99$ \cite{PDG}.

\subsection{Defining the Spectra}

We generate a vector of input data in the $i$ energy bins, $n_i$, for all three target nuclei assuming interactions derived from standard, anapole, dipole, $q^2$, and $q^{-4}$ operators.  Recall that the input data are smeared in energy as well as with Poisson noise.
These spectra are obtained for DM masses of 10, 50, and 250 GeV.
This is then compared to the vector of theoretical data in the $i$ energy bins, $\nu_i$, which are the convoluted spectra for standard, anapole, dipole, $q^{\pm2}$, and $q^{\pm4}$ operators.

For xenon and germanium experiments, we take the energy range of the experiment to be 5-60 keV, binned in a 1 keV interval from 5 to 13 keV, 2 keV from 13 to 25 keV, and 5 keV from 25 to 60 keV.
For the argon experiment, a higher threshold is assumed, taking the energy range of the experiment to be 20-60 keV, binned in 1 keV from 20 to 28 keV, 2 keV from 28 to 40 keV, and 5 keV from 40 to 60 keV.
These bins are moderately coarser than current and upcoming experimental capabilities, but we find that the results are insensitive to bin size.
We have checked that more than doubling the number of allowed bins and using bins as narrow as $0.4{\rm~keV}$ does not affect our results.
In all cases, properly accounting for the effect of the noise and the energy resolution are more important than binning choices for reaching robust conclusions.

Our conventions for the total rate may be fixed in two ways.
First, we take the exposure of all targets to be 2 ton-years and fix the total number of events on germanium to be 300 for 50 and 250 GeV DM candidates and 100 for a 10 GeV DM candidate.
We allow $\sigma_n$ and $\widehat{\sigma}_n$ to vary so as to achieve the desired number of events on germanium.
For a fixed exposure more events are observed on a xenon target due to the higher atomic number, while fewer are observed on an argon target.
To check the robustness of our results and their sensitivity to statistics, we also calculate results when an equal number of events are obtained on all targets (300 events for 50 and 250 GeV DM test masses, and 100 events for 10 GeV DM test mass). This will allow us to disentangle the effect of statistics from the effects of particle physics. We find that while the constraining power of individual experiments changes, we are able to obtain good discrimination in both cases.

Due to threshold effects, fewer events are expected to result for 10 GeV DM candidates for a given scattering cross-section.
As a result of these fewer events, noise is an important limiting factor in our ability to definitively extract the correct operator mediating the interaction.
For light DM, kinematics is also very limiting.
For these reasons, our conclusions for light DM are less sensitive to the total number of events: even with as many as 1000 events on a single target (which for a germanium target corresponds to 20 ton-years of exposure for a cross-section $\sigma_n \sim 10^{-44}{\rm~cm^2}$) we do not obtain good discrimination for a 10 GeV candidate.

\section{Results}

There are two possible measures for operator discrimination that will be obvious from our results: the absolute goodness-of-fit per degree of freedom, $\widetilde{L}_{\rm min}$/d.o.f., and the overlap of the 95\% preferred regions from fits to spectra generated by different target nuclei.
A mismatch of the input data and theoretical model should give high log-likelihood values, and comparing the $\widetilde{L}_{\rm min}$/d.o.f. values for different trial operators can provide a clue to nature of the interaction that generated the data.
Similarly, different operators will give rise to different preferred regions as observed on multiple targets.
Overlapping confidence level curves (CLCs) can indicate that the momentum dependence scales with target nucleus as expected for the hypothesized interaction and disjoint CLCs can indicate the converse.
As detailed below, considering the value of $\widetilde{L}_{\rm min}$ for the combined data from all targets gives a test that is sensitive to both of these measures.

Two subsets of the results are shown in Figs.~(\ref{SI}) and (\ref{dipole}) for input data corresponding to standard and dipole mediated scattering. More complete results are available in the appendix and online \cite{dminv}. The subsets shown in Figs.~\eqref{SI} and \eqref{dipole} are chosen to represent illustrative contrasts to the input data; we describe our particular choices in more detail below.
Each figure corresponds to one set of input data (labeled by $n_i$) generated according to the methodology described in the previous section. The individual panels of each figure correspond to fits of a theoretical model (labeled by $\nu_i$) to the input data.
We divide the plots in Figs.~(\ref{SI}) and (\ref{dipole}) so that the left panels show our results for the 10 and 50 GeV DM candidates for scattering on all three targets: xenon, germanium and argon.\footnote{We omit the argon curves in the 10 GeV case because argon events generated for the 10 GeV analysis were essentially compatible with zero. Our argon curves are therefore closer to exclusion curves than preferred regions.
In all cases the germanium and xenon CLCs fit safely underneath the argon ``exclusion curve."}
The right panels show the same information for 250 GeV candidates.
In the appendix, all CLCs are shown on the same plot.
The colors on the curves correspond to the minimum log-likelihood per degree of freedom\footnote{For high mass, ${\rm d.o.f.(Ge,Xe)}=19$, while ${\rm d.o.f.(Ar)}=16$, corresponding to the number of $E_R$ bins minus the two fit parameters, $m_{DM}$ and $\sigma_n$. For low mass candidates we have many fewer degrees of freedom because fewer bins are filled: ${\rm d.o.f.(Ge)}=6$ and ${\rm d.o.f.(Xe)}=1-5$, depending on the operator.} for the fit of each operator to the data, with darker colors representing the better fits.
We have checked that different realizations of noise do not change our ability to extract information about the particle physics.

In this section, we will start by going through a simplified analysis of the scattering kinematics that explains the shapes of the CLCs in Figs.~\eqref{SI} and \eqref{dipole}.
Next, we will explain how the tests discussed here allow one to extract the correct particle physics that underlies direct detection events.
We will do this both when all exposures are the same and when the number of observed events is the same on all targets. This allows us to determine the effects of statistics on our results.
Finally, we will compare models using the p-value test. This test will be a quantitative measure of both of the qualitative tests described above.

\subsection{Contour Shapes from Scattering Kinematics}

We discuss the shapes of the CLCs that we observe in Figs.~\eqref{SI} and \eqref{dipole}. The kinematics of elastic scattering between DM particles and detector target nucleus provides a straightforward handle on the shape of these preferred regions.
For the purpose of illustration, we take standard scattering and simplify the differential event rate given in Eq.~(\ref{rate}) with the limit $v_{esc} \to \infty$, ignoring solar motion, terrestrial motion, detector efficiency, and detector resolution.
This recovers the result \cite{Jungman:1995df}
\begin{equation}
\label{rate2}
\frac{dR}{dE_R}(E_R)\simeq\frac{\rho_0\sigma_N}{\sqrt{\pi}m_{\rm DM}\mu^2_Nv_0}\exp\left(-\frac{E_Rm_N}{2\mu^2_Nv^2_0}\right).
\end{equation}
The balance between the exponential term and the linear term on the right-hand side of Eq.~(\ref{rate2}) gives most of the important kinematic information. This balance is the key to understanding the basic content of our plots.  To better understand the implications of scattering kinematics on the operator discrimination, we consider two extreme limits first.

If the DM mass is much less the target nucleus mass then $\mu_N\sim m_{\rm DM}$ and we can approximate the differential rate by 
\begin{equation}
\frac{dR}{dE_R}(E_R)\sim\frac{\rho_0\sigma_N}{\sqrt{\pi}m^3_{\rm DM}v_0}\exp\left(-\frac{E_Rm_N}{2m^2_{\rm DM}v^2_0}\right).
\label{lowmassrate}
\end{equation}
We can immediately see that this rate is very sensitive to the DM mass.  If we increase the DM mass the scattering rate will increase sharply because of the exponential term. Our numerical results clearly reflect this sensitive dependence.
From fits of the 10 GeV case shown in Figs.~\eqref{SI} and \eqref{dipole} we can see that all the 10 GeV contours are very narrow.  By contrast, the event rate is linearly proportional to the scattering cross section and the shift has a rather mild dependence on the target mass. Hence, detectors have poor cross-section discrimination capabilities for the low mass case.  

At the other end of the mass range, if the DM is much heavier than the detector nucleus mass then the differential event rate is given to first order by 
\begin{equation}
\label{hmd}
\frac{dR}{dE_R}(E_R)\sim\frac{\rho_0\sigma_N}{\sqrt{\pi}m_{\rm DM}m^2_{N}v_0}\exp\left(-\frac{E_R}{2m_{N}v^2_0}\right).
\end{equation}
The exponential term is independent of the DM mass in this limit, and it becomes an overall factor for any given large DM mass. The differential event rate now just scales as $dR/dE_R\propto\sigma_n/m_{\rm DM}$. In this limit the target nucleus is not able to determine the DM mass because one always can rescale $\sigma_n$ to compensate the change of DM mass while keeping $\sigma_n/m_{\rm DM}$ unchanged.  This so-called high-mass degeneracy \cite{Pato:2010zk} is observed in Figs.~\eqref{SI} and \eqref{dipole}.  The allowed mass range can be unbounded for all three targets when high-mass degeneracy is important.

These features explain certain coarse aspects of the CLC plots. Now we examine the physics that leads to the separation of the CLCs in the parameter space, and we investigate what this can tell us about the particle physics underlying the data. We do this analysis first with equal exposures on all targets. We will also present results with the same number of events  recorded by all targets.

\subsection{Operator Discrimination with Equal Exposures on All Targets}

\begin{figure}[b]
\begin{center}
\includegraphics[height=2.3in,width=2.5in]{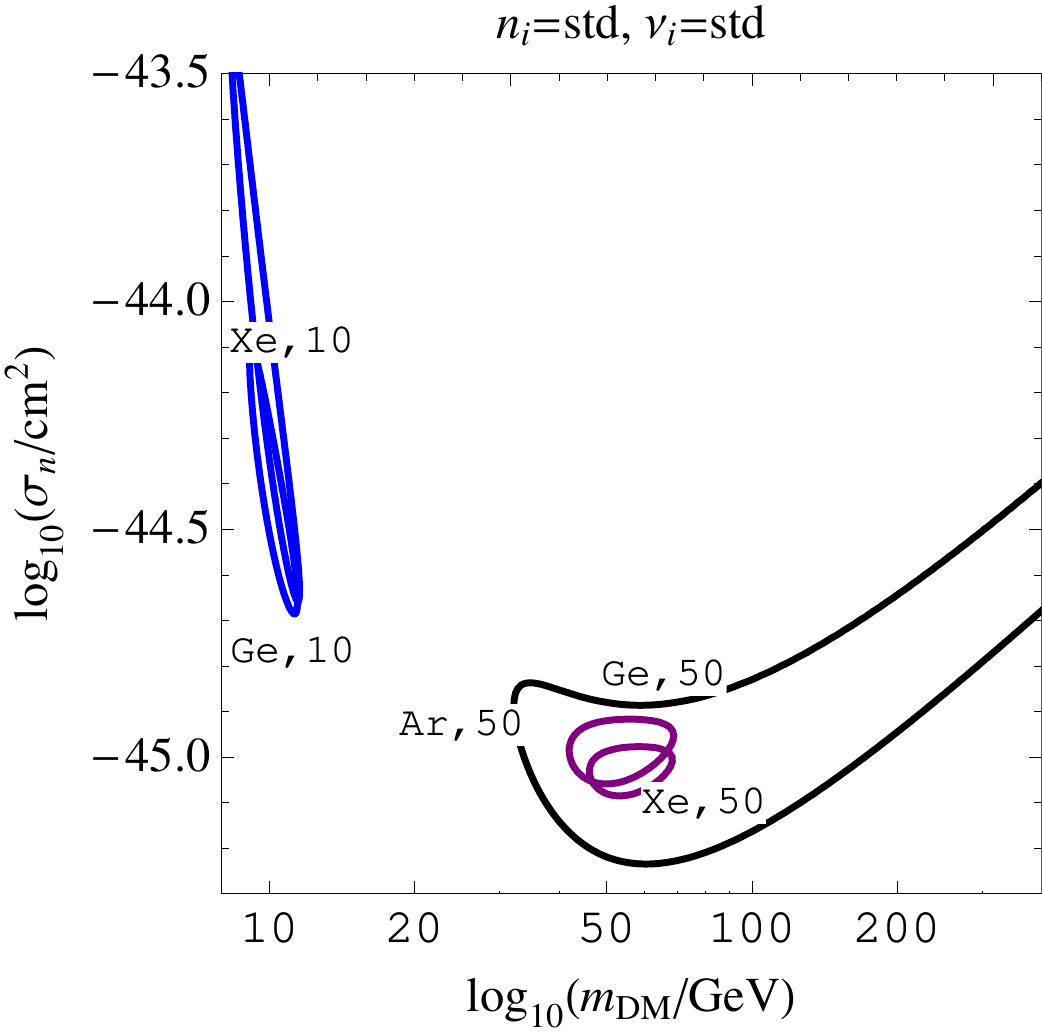}~~~~~~~~
\includegraphics[height=2.3in,width=2.5in]{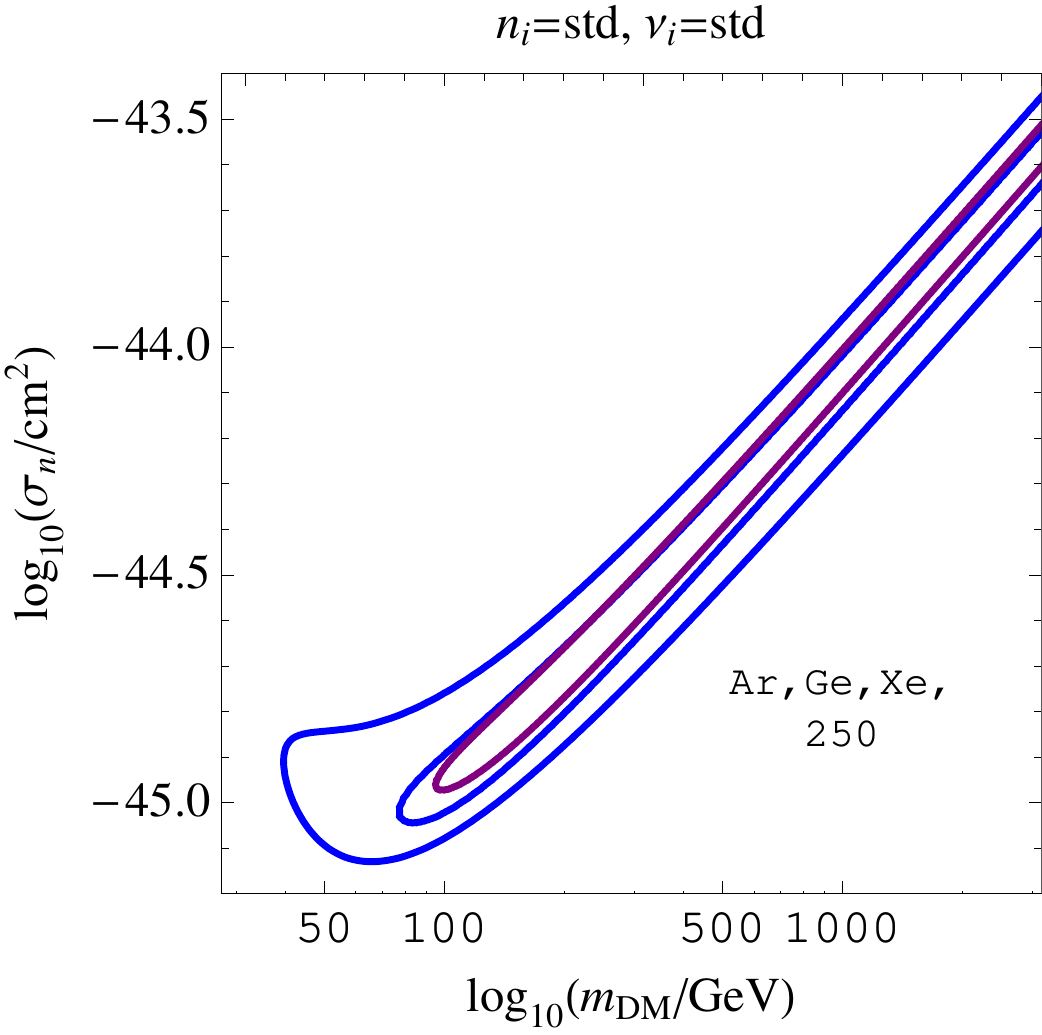}
\includegraphics[height=2.3in,width=2.5in]{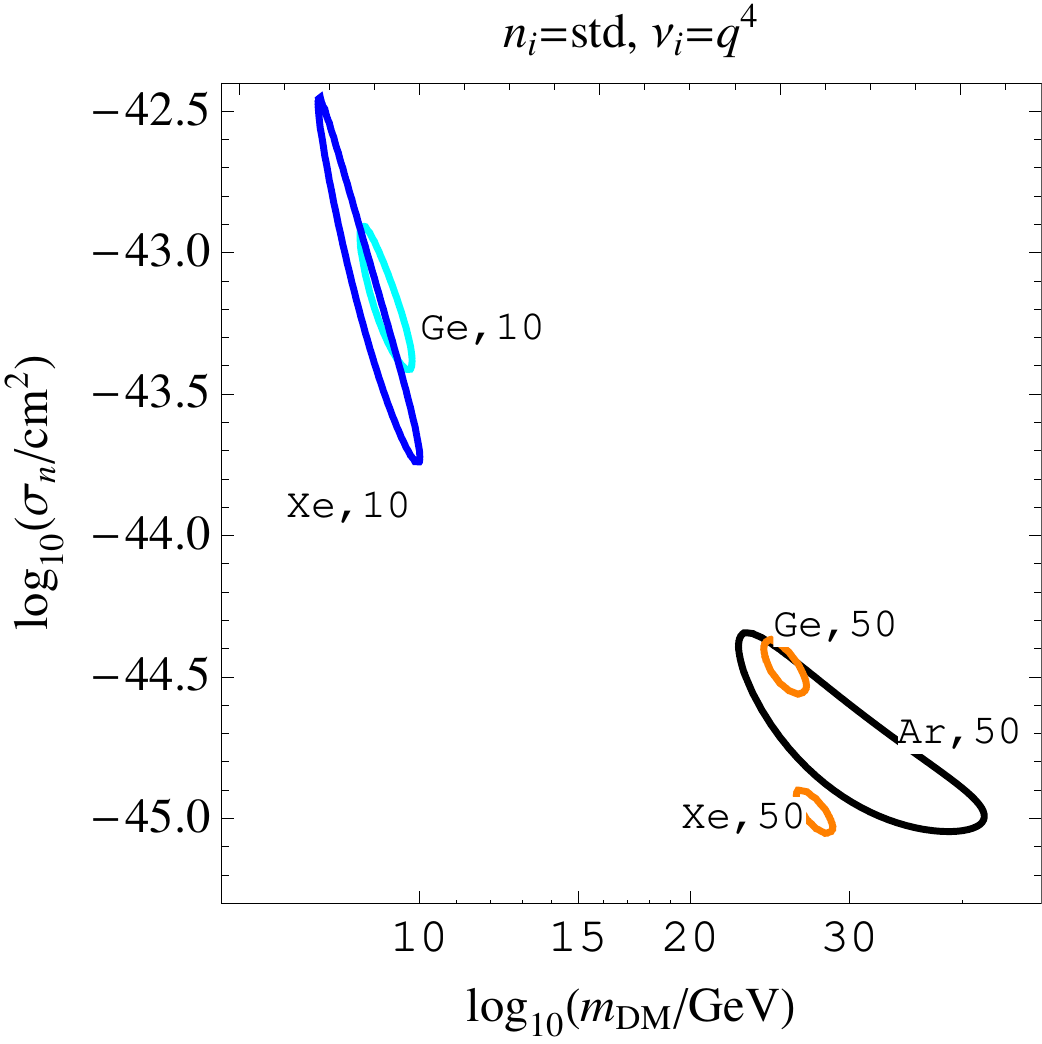}~~~~~~~~
\includegraphics[height=2.3in,width=2.5in]{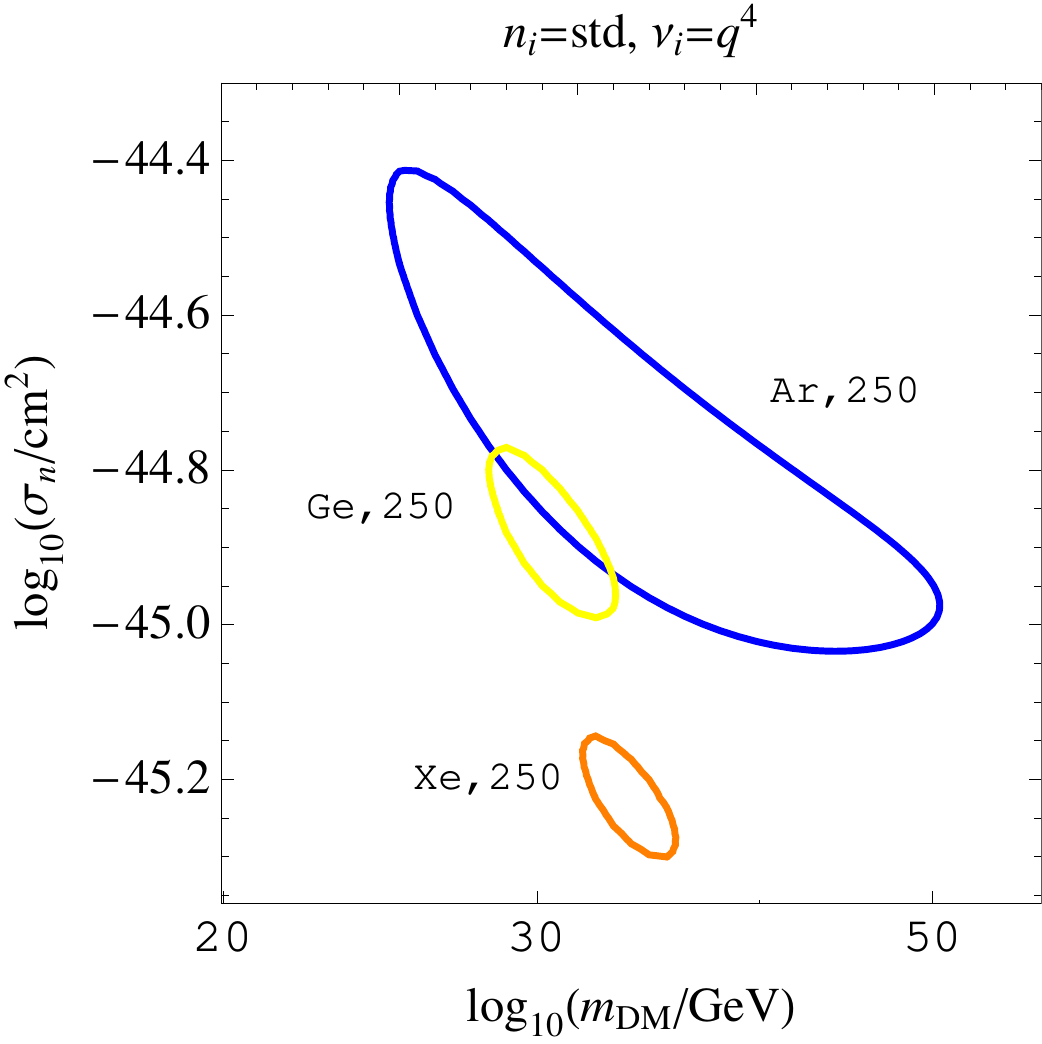}
\includegraphics[height=2.3in,width=2.5in]{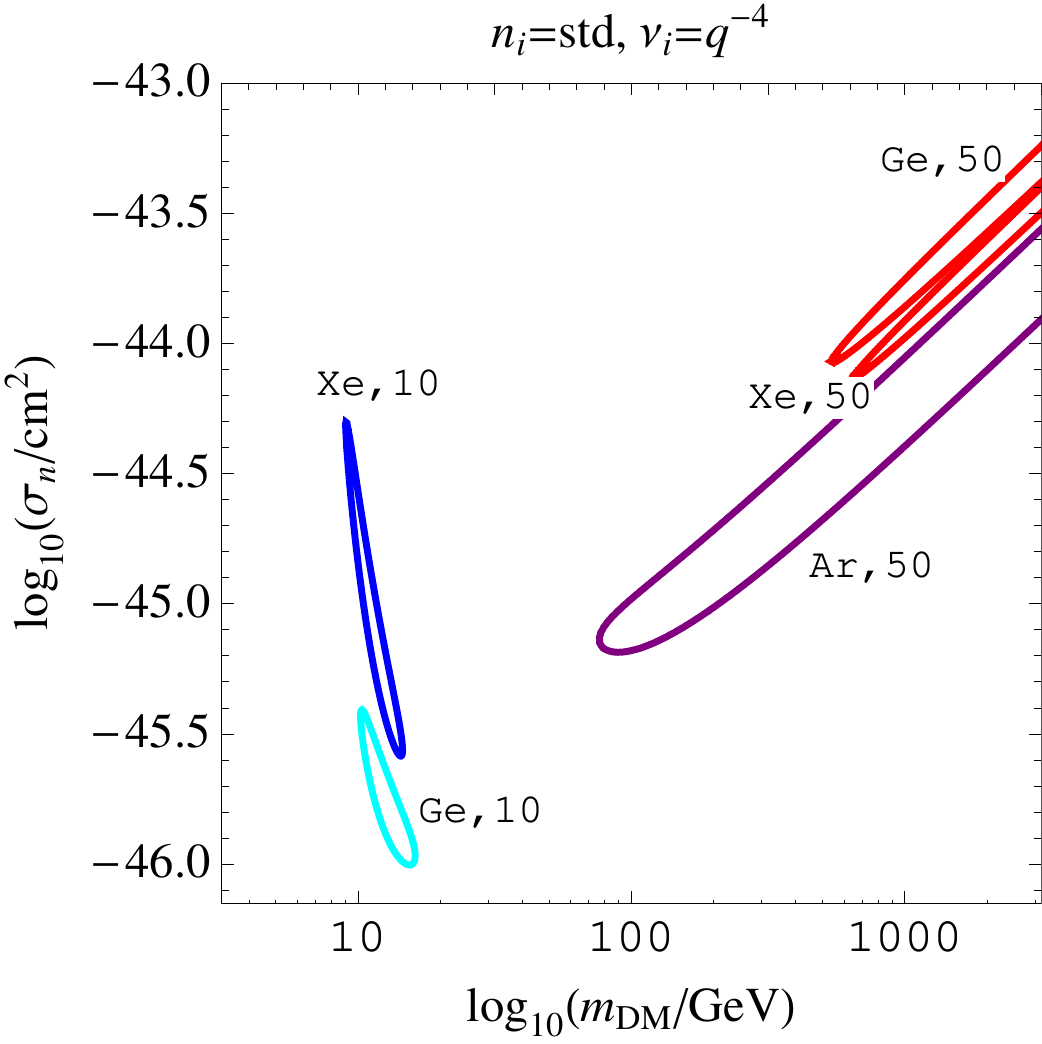}~~~~~~~~
\includegraphics[height=2.3in,width=2.5in]{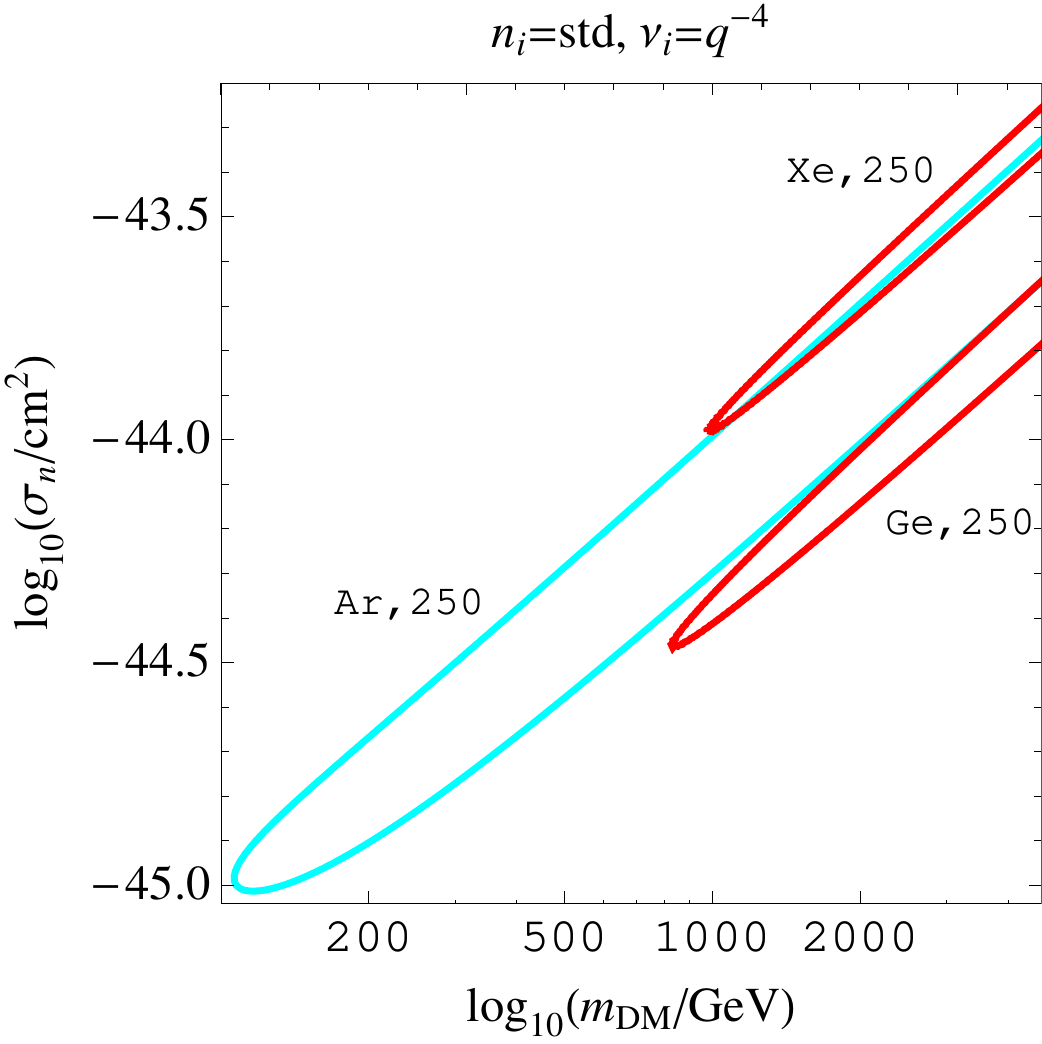}
\qquad \includegraphics[height=.5in,width=3.8in]{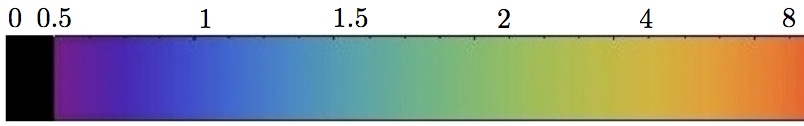}
\caption{95\% CLCs for a 10, 50, and 250 GeV particle interacting through an $n_i=$ standard operator with equal exposures, such that there are 300 events on a germanium target and more (fewer) on a xenon (argon) target. Comparisons are made to $\nu_i=$ standard, $q^4$, and $q^{-4}$ operators. The colors represent the value of $\widetilde{L}_{\rm min}/$d.o.f. The standard interaction can be distinguished from the $q^{\pm4}$ operators via both CLC overlap and the values of $\widetilde{L}_{\rm min}/$d.o.f. As can be seen in Fig.~(10) in the appendix, $\widetilde{L}_{\rm min}/$d.o.f. is less powerful for distinguishing $n_i=$ standard from $\nu_i= q^{\pm2}$, though overlap remains robust.}
\label{SI}
\end{center}
\end{figure}

\begin{figure}[b]
\begin{center}
\includegraphics[height=2.3in,width=2.5in]{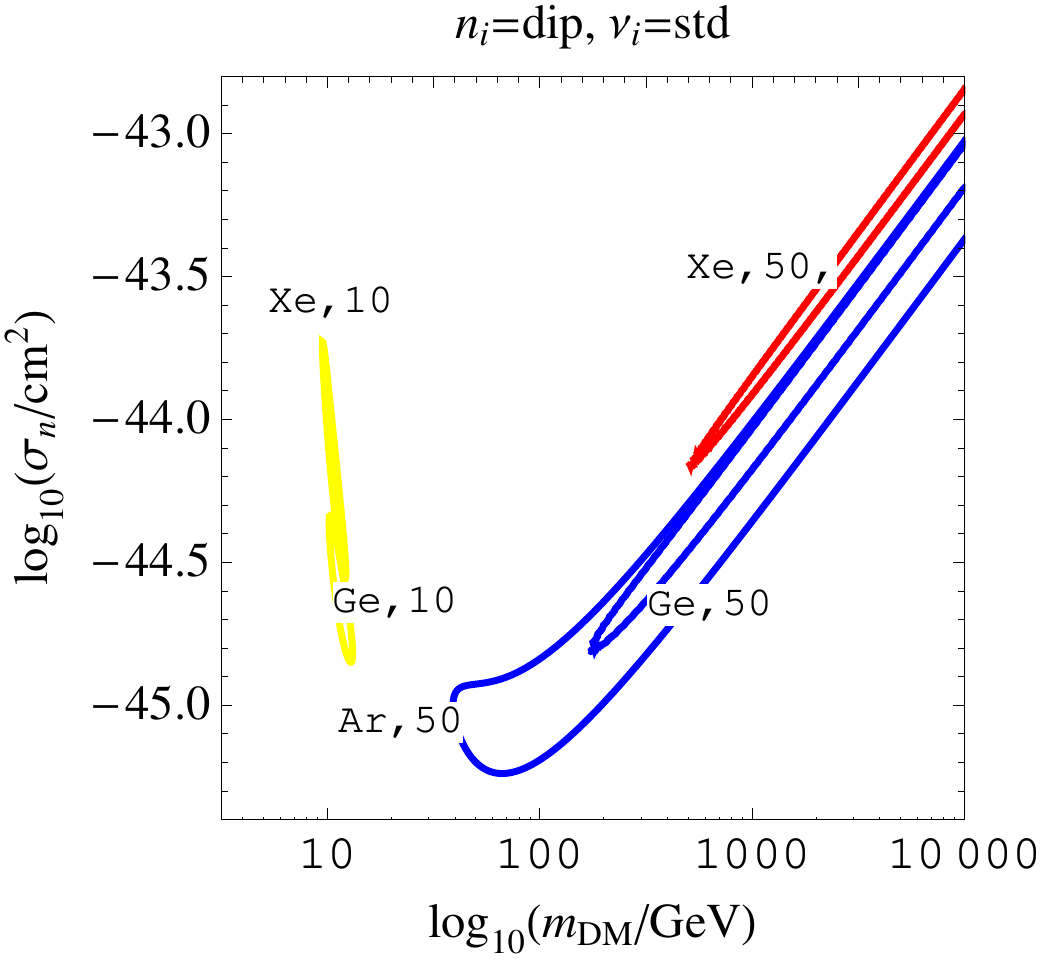}~~~~~~~~
\includegraphics[height=2.3in,width=2.5in]{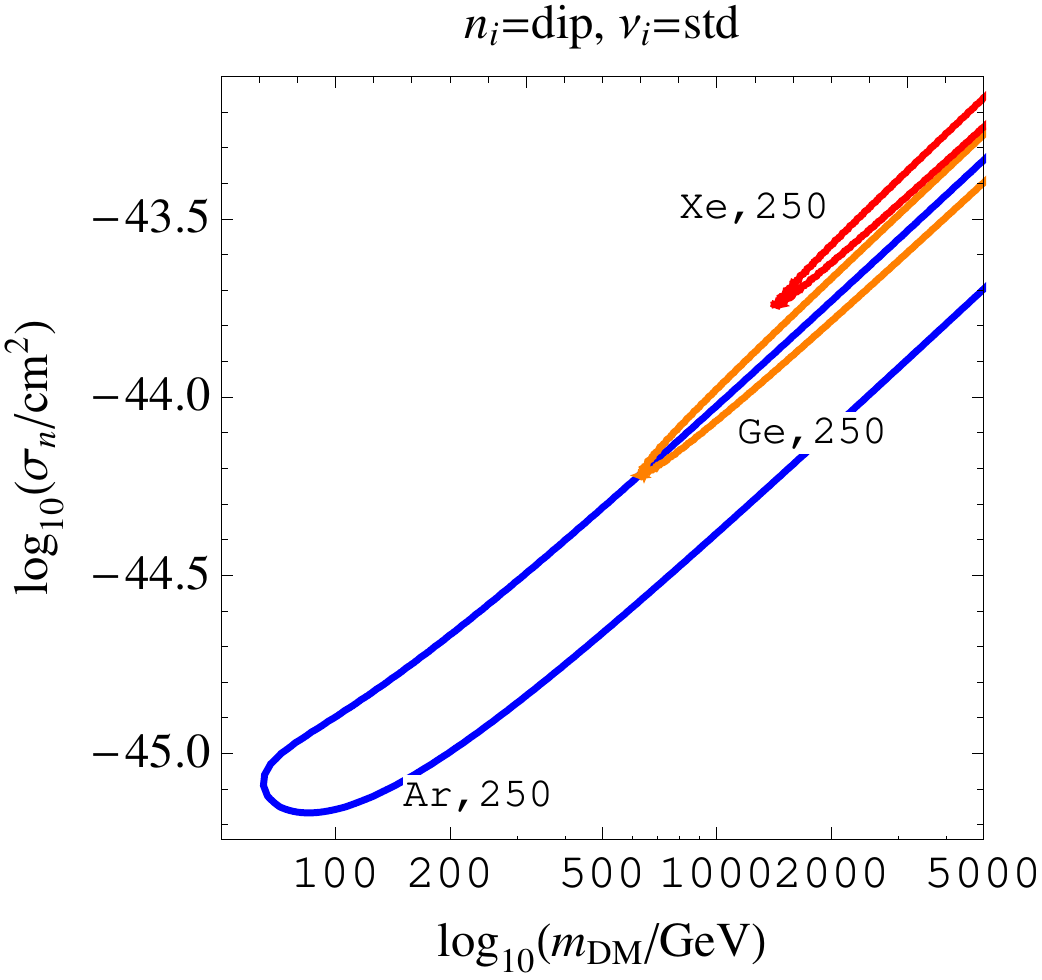}
\includegraphics[height=2.3in,width=2.5in]{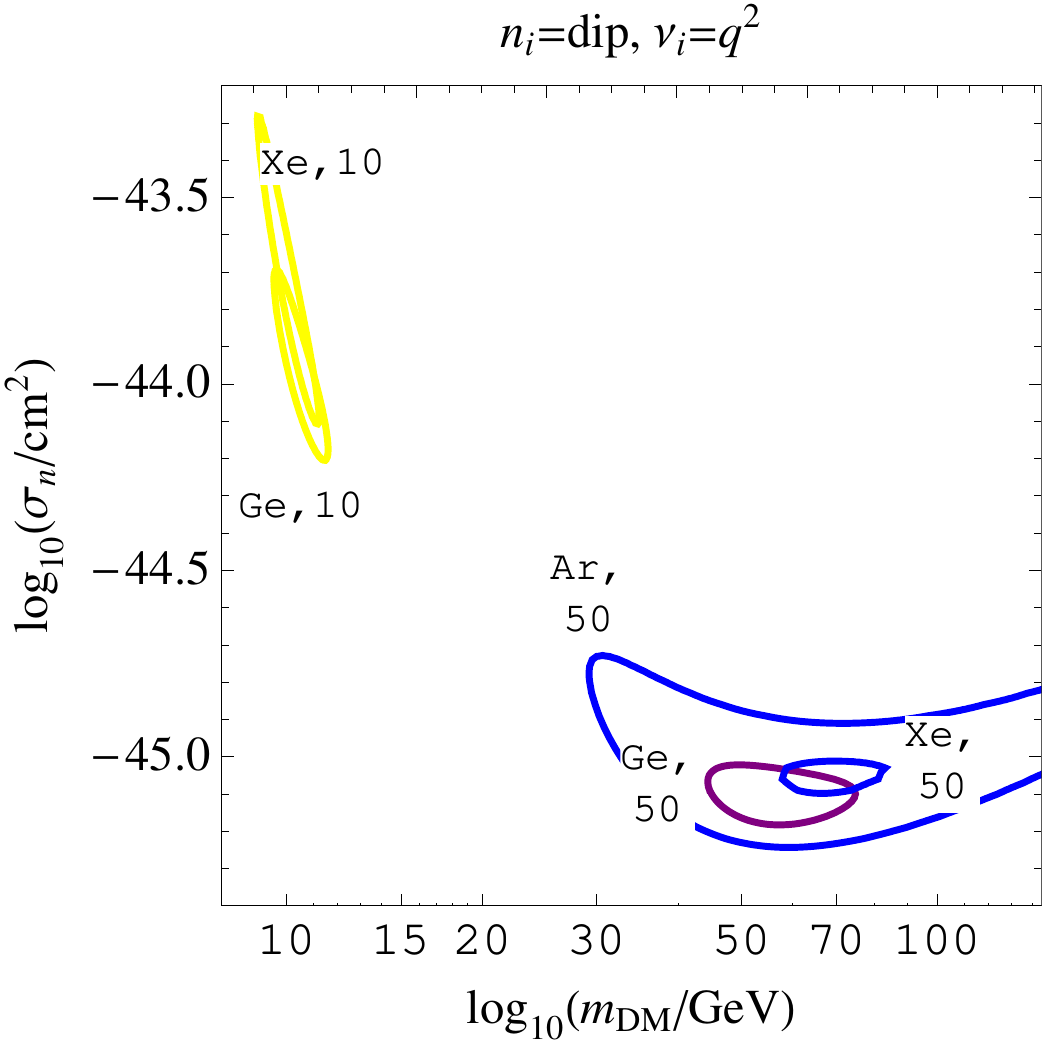}~~~~~~~~
\includegraphics[height=2.3in,width=2.5in]{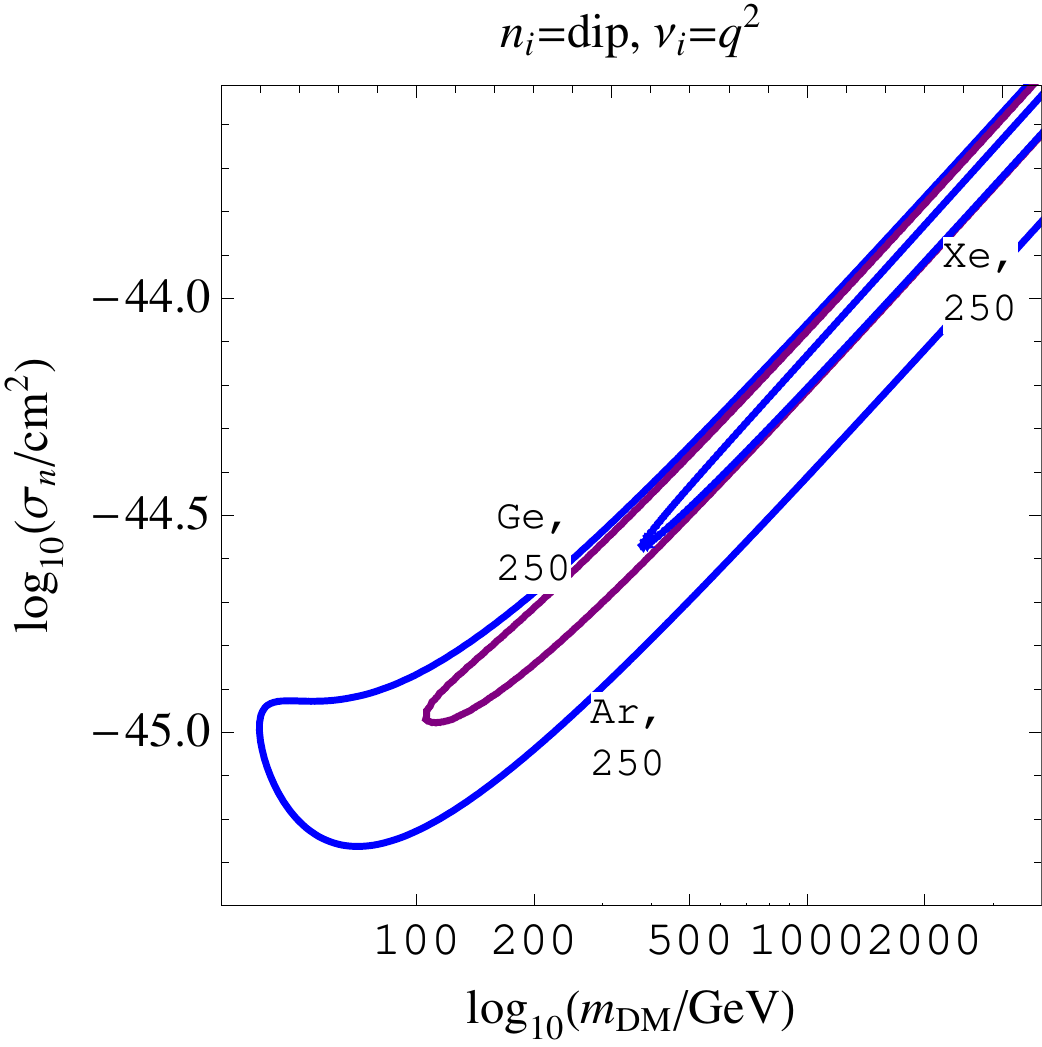}
\includegraphics[height=2.3in,width=2.5in]{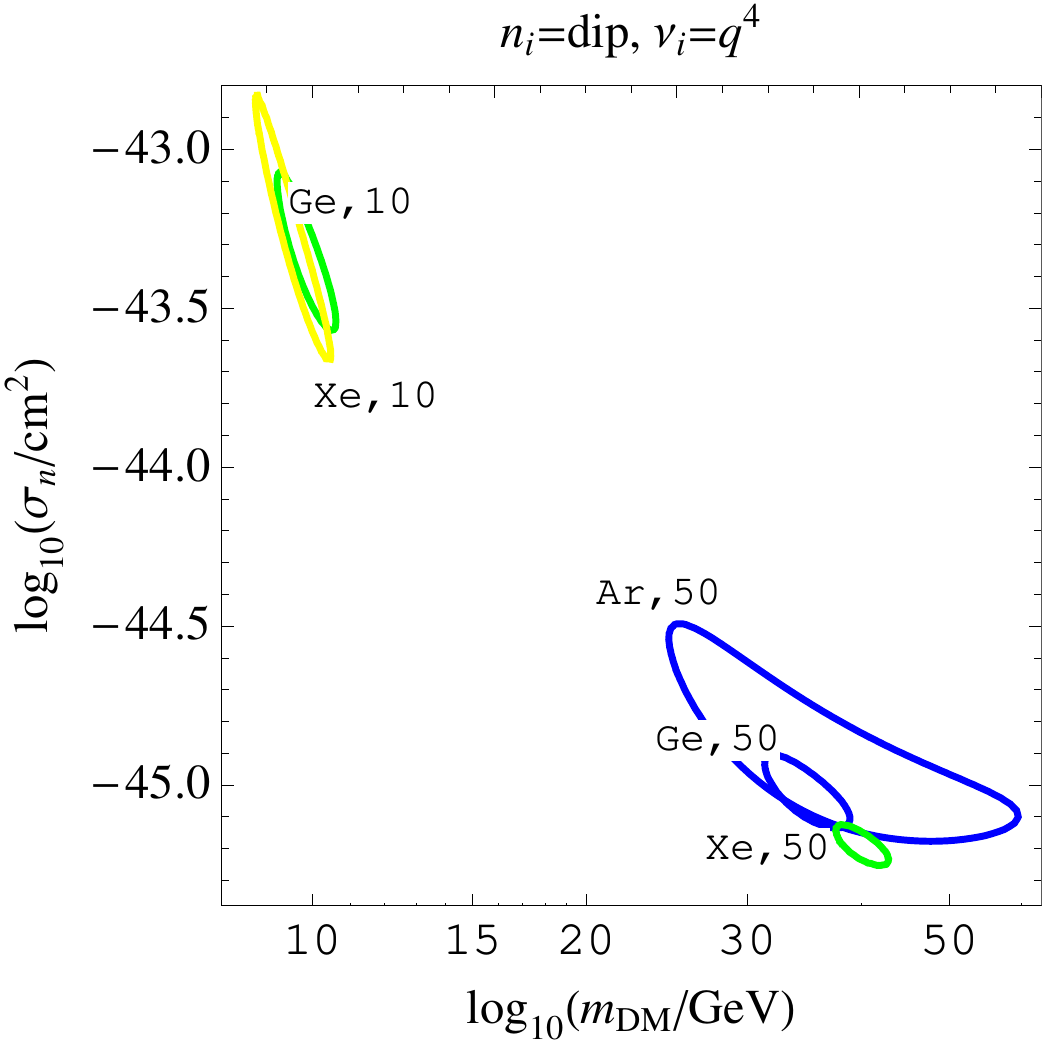}~~~~~~~~
\includegraphics[height=2.3in,width=2.5in]{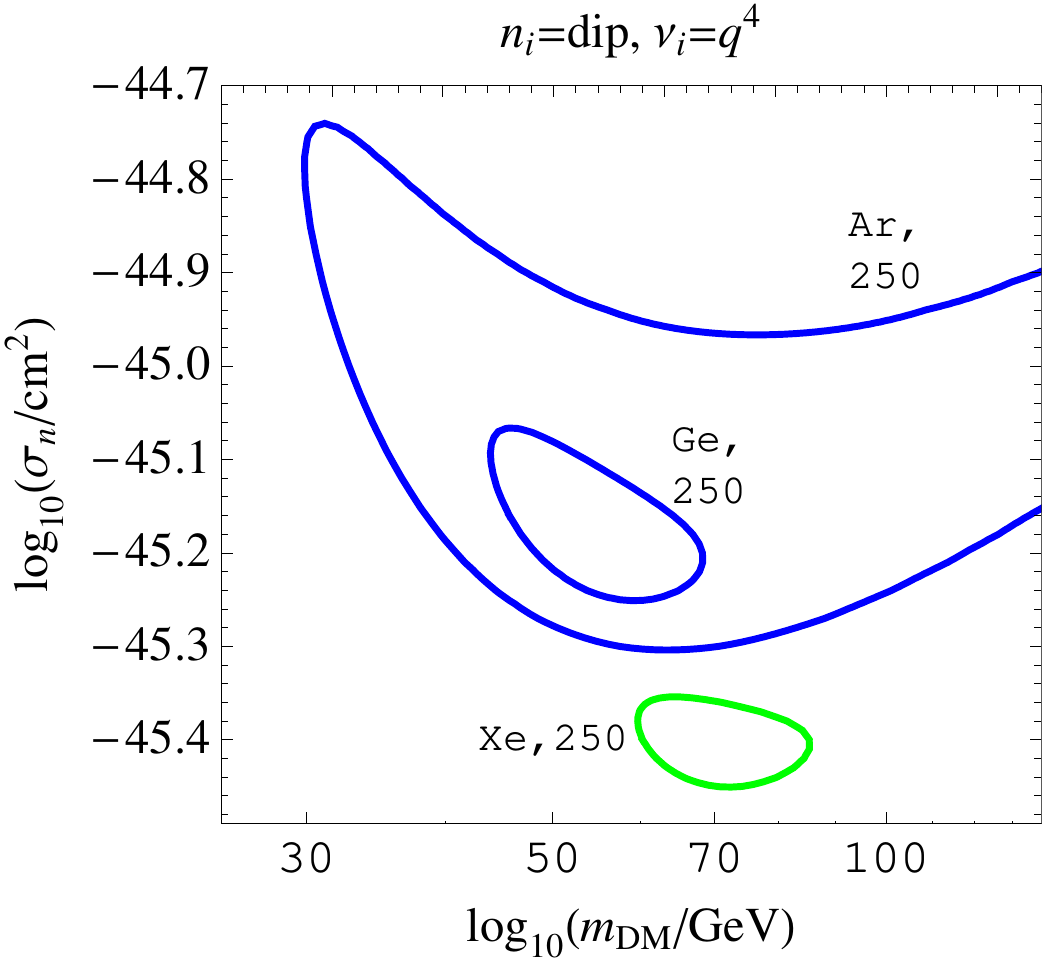}
\qquad \includegraphics[height=.5in,width=3.8in]{bar2.jpg}
\caption{95\% CLCs for a 10, 50, and 250 GeV particle interacting through an $n_i=$ dipole operator with equal exposures. Comparisons are made to $\nu_i=$ standard, $q^2$, and $q^4$ operators. The colors represent the value of $\widetilde{L}_{\rm min}/$ d.o.f. The dipole moment operator is dominated by a term proportional to $q^2$, and is indistinguishable from this $q^2$ operator. The dipole interaction can be distinguished from the $q^{4}{\rm~and~standard}$ operators via CLC overlap, which is largely powered by the statistics of the xenon target. As can be seen in Fig.~(12) in the appendix, overlap remains powerful for distinguishing $n_i=$ dipole from other operators.}
\label{dipole}
\end{center}
\end{figure}

For operator discrimination with equal exposures, the most important feature of Figs.~\eqref{SI} and \eqref{dipole} is that, when the input data and theoretical model match ($n_i=\nu_i$), the CLCs overlap regardless of candidate mass. When the theoretical model is mismatched with the input data, the 50 GeV and 250 GeV CLCs separate, though much less separation is apparent in the 10 GeV case. Thus for low mass candidates the overlap test does not strongly discriminate between models, while for high mass candidates the overlap test allows strong discrimination. Much of this power is derived from the strong statistics gained from scattering off of a xenon target. By contrast, we see in Fig.~\eqref{dipole}, where the momentum dependence of the operators differs by $q^2$, that, for scattering off a single target nucleus, the log-likelihood per degree-of-freedom test can be limited by noise and does not provide a very good discriminant between theoretical models -- the fit can be just as good on a single target with the incorrect theoretical model as it is with the correct theoretical model, even for high mass candidates.  This is especially true for argon, which has the fewest number of events and smallest energy range. Xenon, however, has sufficient statistics to be able to extract the correct operator mediating the scattering on its own. The good discrimination of xenon is what makes the overlap between the xenon, germanium, and argon CLCs a sensitive test.

We now investigate how CLC separation occurs. In Fig.~\eqref{SI} we compare input data corresponding to standard momentum independent scattering to scattering from $q^{\pm4}$ interactions. We specifically choose $q^{\pm4}$ because these operators have the most exaggerated kinematic effects. These comparisons will lead to the largest separations in parameter space and show the widest disparity in the values of $\widetilde{L}_{\rm min}/$d.o.f.
The differential event rate of the mismatched operators is given by
\begin{equation}
\label{ratew}
\frac{dR^{q^{\pm4}}}{dE_R}(E_R)\simeq\frac{A^2 \rho_0\sigma^{std}_n}{\sqrt{\pi}m_{\rm DM}\mu^2_Nv_0}\left(\frac{\sqrt{2E_Rm_N}}{q_0} \right)^{\pm4} \exp\left(-\frac{E_R m_N}{2\mu^2_{N}v^2_0}\right).
\end{equation}
The most noticeable effect of the additional momentum dependence, as shown in the second and third rows of panels of Fig.~({\ref{SI}}), is that on all targets the incorrect operators prefer different DM mass ranges than the value chosen by the correct operator: $q^4$ scattering chooses low mass and $q^{-4}$ scattering prefers high mass.
\begin{figure}[t]
\begin{center}
\includegraphics[height=3.1in,width=3.1in]{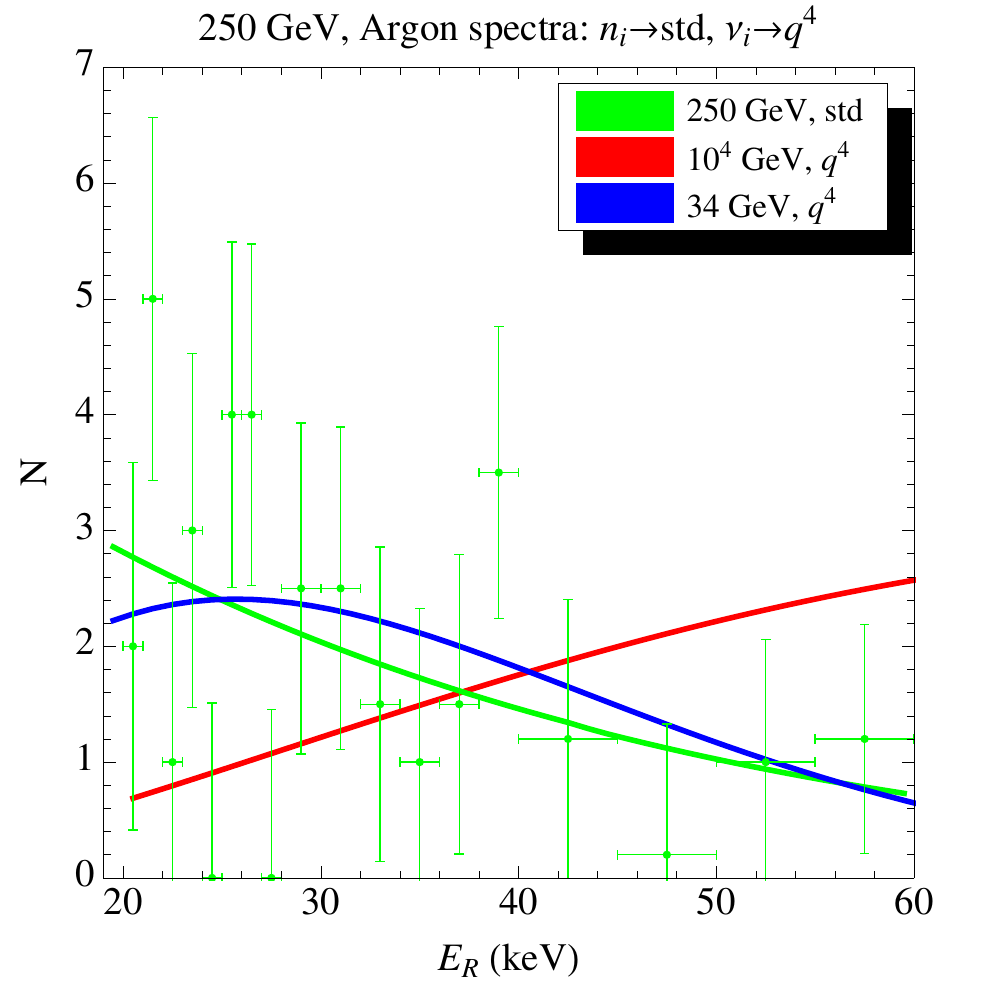}
\includegraphics[height=3.1in,width=3.1in]{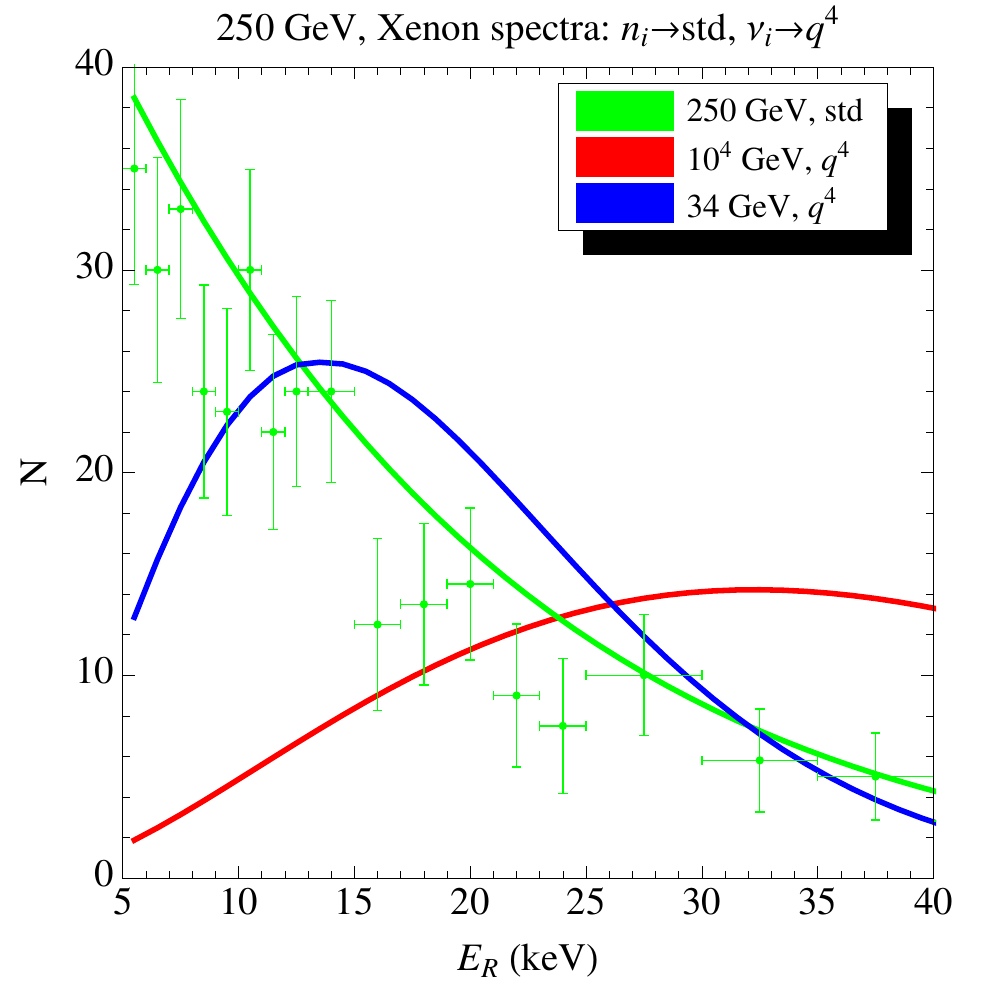}
\caption{Effect of momentum dependence and DM mass on fit spectra. The data are from an $n_i=$ standard interaction and are depicted with Poisson error bars.  The best fit spectrum for $\nu_i=$ standard is shown as a green solid line. We also show two $\nu_i=q^{4}$ spectra. The $q^4$ spectrum in blue is at the best fit point and the $q^4$ spectrum in red is taken at high mass to illustrate the improvement in goodness of fit from reducing $m_{\rm DM}$. We can also see the effect of xenon's higher event rate and lower energy threshold on its ability to determine the correct operator.}
\label{spectraq4} 
\end{center}
\end{figure}

\begin{figure}[t]
\begin{center}
\includegraphics[height=3.1in,width=3.1in]{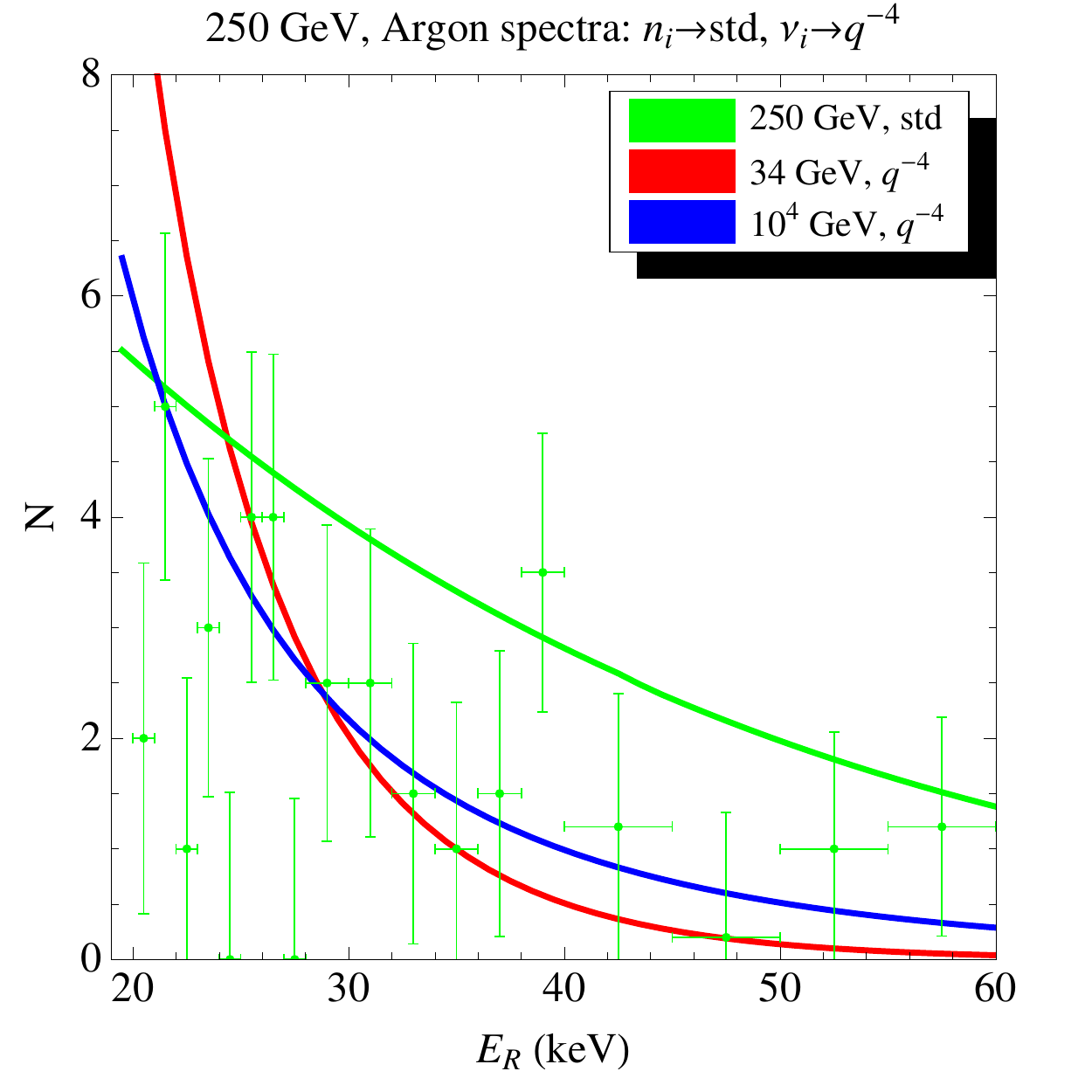}
\includegraphics[height=3.1in,width=3.1in]{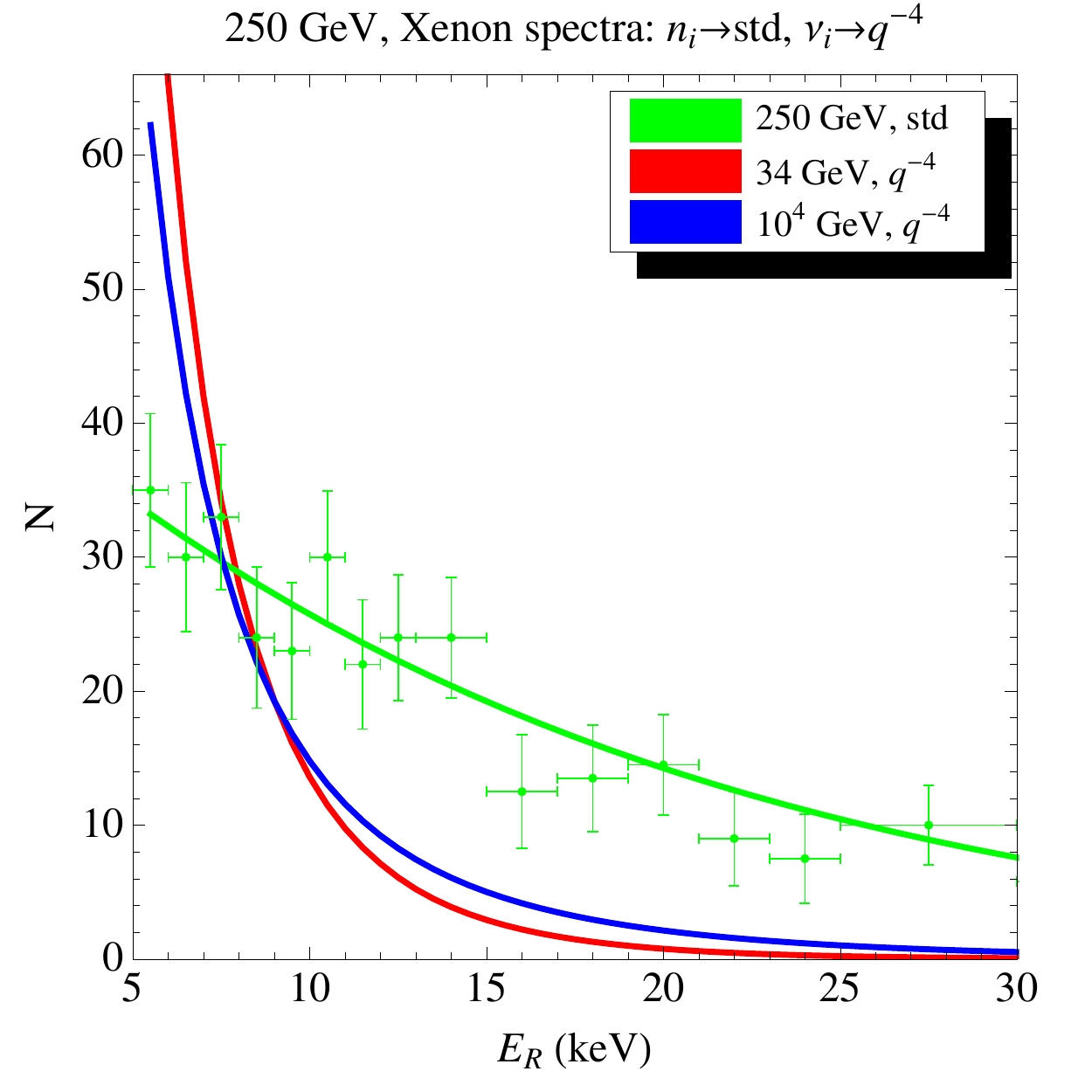}
\caption{Effect of momentum dependence and DM mass on fit spectra. The data are from an $n_i=$ standard interaction and are depicted with Poisson error bars.  The best fit spectrum for $\nu_i=$ standard is shown as a green solid line. We also show two $\nu_i=q^{-4}$ spectra. The $q^{-4}$ spectrum in blue is at the best fit point and the $q^{-4}$ spectrum in red is taken at low mass to illustrate the improvement in goodness of fit from increasing $m_{\rm DM}$. We can also see the effect of xenon's higher event rate and lower energy threshold on its ability to determine the correct operator.}
\label{xenonspectra3} 
\end{center}
\end{figure}

This behavior is driven by the shape of the event distributions. As seen in Fig.~(\ref{spectraq4}), the $q^{+n}$ operator with a lower mass candidate can mimic the scattering of a higher mass candidate with a standard spectrum. This occurs because as one decreases $m_{\rm DM}$ the event number increases at low energy and decreases at high energy. Likewise, we can consider a $q^{-4}$ operator, shown in the bottom row of Fig.~(\ref{SI}). In this case the event rate is suppressed at high recoil but increased at low recoil. In Fig.~\eqref{xenonspectra3} we see that larger DM masses are necessary to suppress the divergent low energy tail of the $q^{-4}$ operator.

\begin{figure}[t]
\begin{center}
\includegraphics[height=2.1in,width=2.1in]{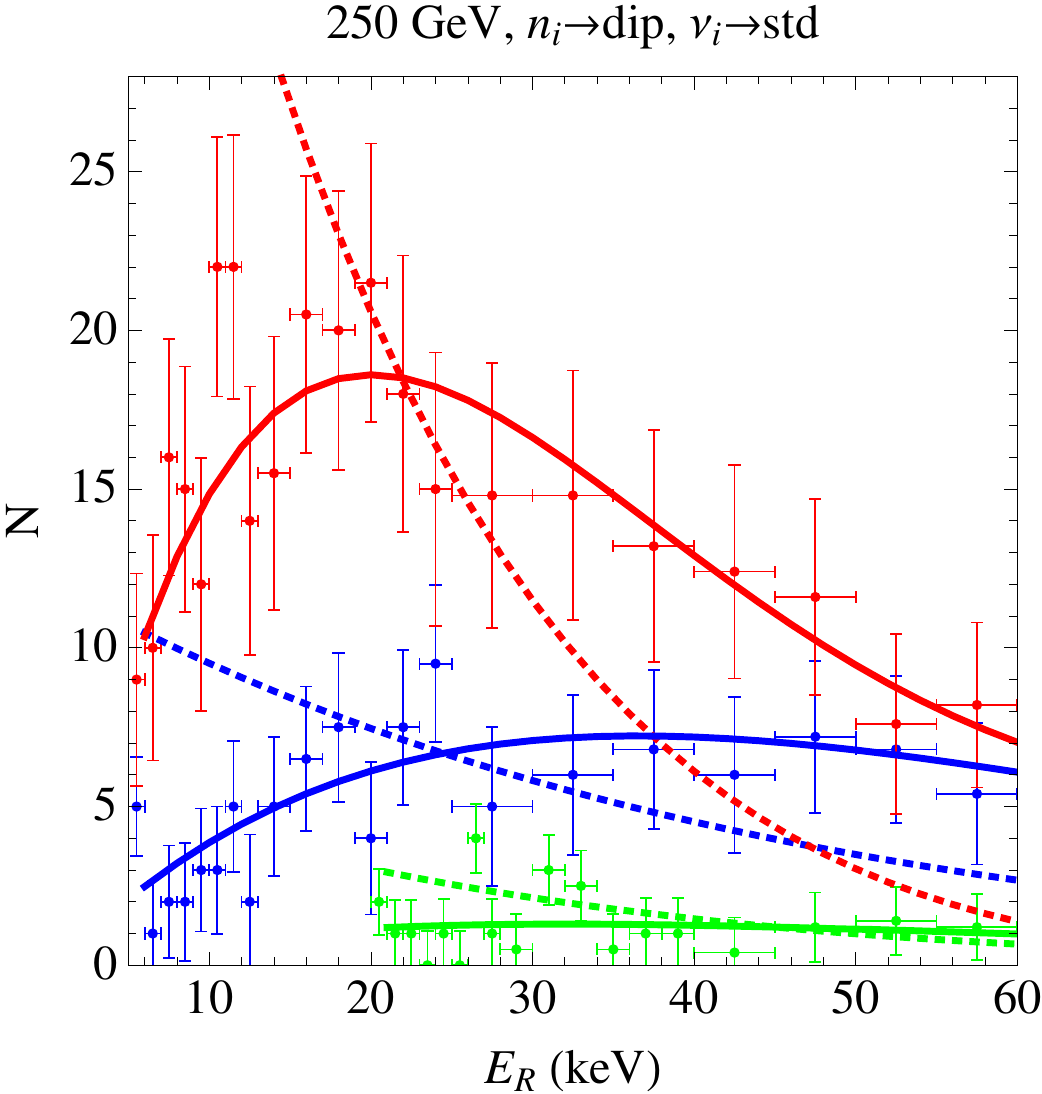}
\includegraphics[height=2.1in,width=2.1in]{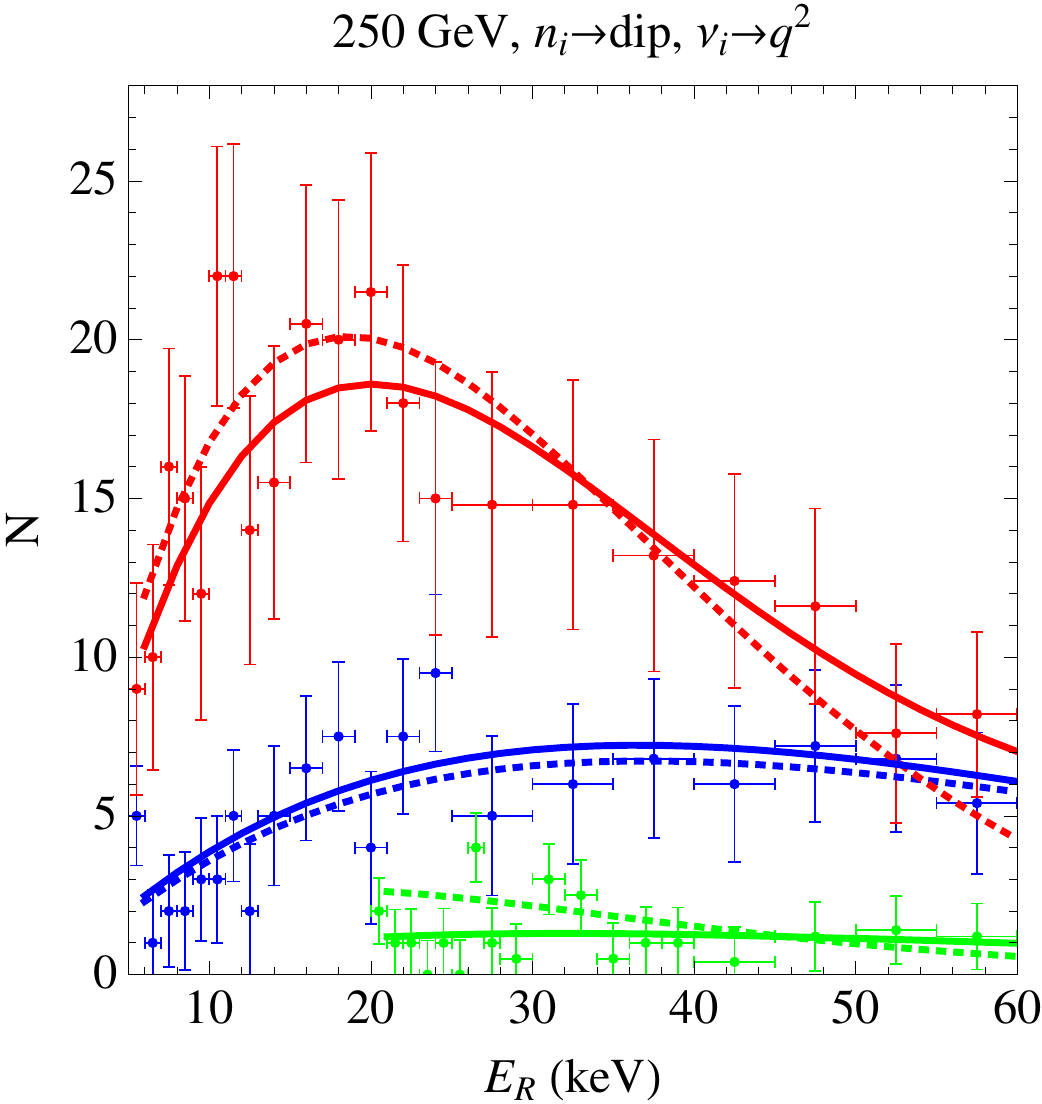}
\includegraphics[height=2.1in,width=2.1in]{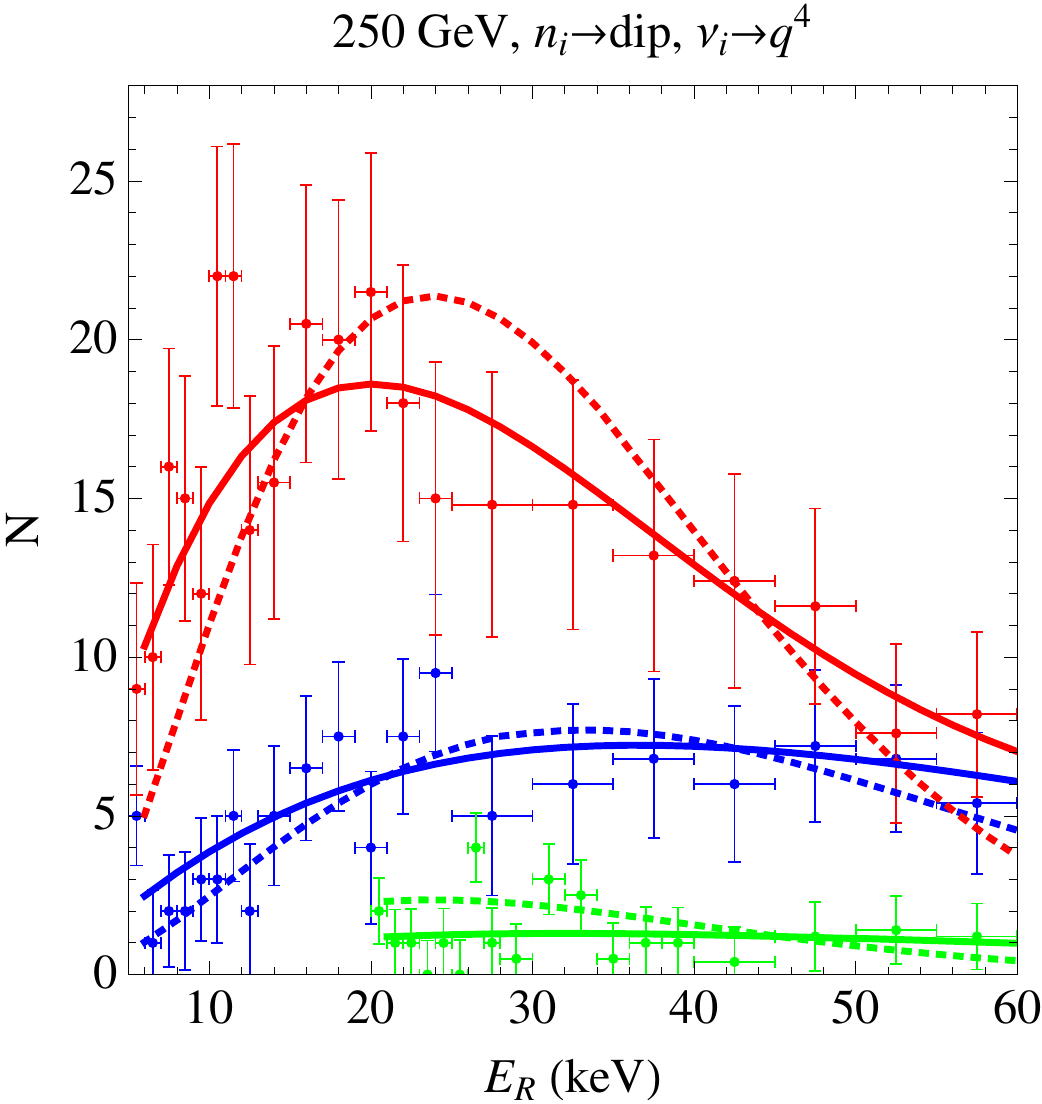}
\caption{Fits of $\nu_i={\rm std,~} q^2, {\rm~and~q^4}$ operators to $n_i=$ dipole data. The data are shown as points with Poisson error bars. The dipole best fit spectrum is shown as a solid line and the other operator best fit spectra are shown as dotted lines. Xenon is red, germanium is blue, and argon is green.}
\label{dipspectra} 
\end{center}
\end{figure}

Changing the momentum dependence by only two powers, {\it e.g.} comparing $n_i=$ std with $\nu_i=q^{\pm2}$, offers less contrast. In this case, fits for individual elements can in general be acceptable (with $\widetilde{L}_{\rm min}/$d.o.f. $\lesssim 1$), and the overlap test seems to be an important discriminant. This can be seen for the $n_i=$ dipole case displayed in Fig.~\eqref{dipole}, where we compare the dipole input data to standard, $q^2$, and $q^4$ interactions.
We see that discrimination is marginal for an individual target (since $\widetilde{L}_{\rm min}/$d.o.f can be good), but the overlap appears to provide conclusive discrimination against the standard and $q^4$-type interactions.
By contrast, $q^2$ and dipole interactions look nearly identical. 
This is because the $q^2$ term in Eq.~\eqref{dip} dominates the $q^4$ term for the elements and DM masses examined here, as seen in Fig.~\eqref{dipspectra}. As expected, the standard spectra are discrepant with the dipole spectra due to a divergent low mass tail, the $q^4$ operator suffers from being overly suppressed at low energy, and the $q^2$ operator provides very good fits. For the same reason, we see in the appendix that there is a degeneracy between the standard and anapole operators. Although the anapole operator has a non-standard velocity-dependence, its spectrum is very similar to momentum independent scattering. Since the velocity dependence does not affect the spectrum and the $q^2$ piece is subdominant, the anapole and standard operators may be said to have the same momentum dependence. Likewise, the dipole and $q^2$ operators share their own momentum dependence.

Based on the collected plots in the appendix, we can extend these conclusions. For example, on argon, which has a high energy threshold and small event number, several incorrect operators can give a decent global fit to the data and a low value of $\widetilde{L}_{\rm min}$/d.o.f.  Even on germanium the wrong operator can mimic the true operator if the momentum dependence of the operator differs by a power of $q^2$ or less. This is not true for xenon, which has a good $\widetilde{L}_{\rm min}$/d.o.f. only for the correct operator. Thus the $\widetilde{L}_{\rm min}$/d.o.f. test works for xenon but not for argon. Additionally, it appears evident from the figures collected in the appendix that the overlap test, combining data from all targets, also works very well for high mass candidates. Nonetheless, the overlap test described here is in some sense another manifestation of the $\widetilde{L}_{\rm min}$ test. This comes about because of the relatively large number of events on the xenon target, which increase its value of $\widetilde{L}_{\rm min}$ (since, for large event numbers, $\widetilde{L}_{\rm min}$ scales with the number of events) and shrink its CLC relative to the other CLCs. The small xenon CLCs are in turn more precise in selecting a preferred region in parameter space, and thus are the single most important factor in the overlap test. In this way, the overlap and $\widetilde{L}_{\rm min}$ tests are linked. This can even be seen in the 10 GeV case, for example, by comparing the CLCs in the appendix from the $n_i=q^2{\rm~and~standard}$ data sets. In the former case the xenon cross section is enhanced relative to germanium by the ratio of the xenon and germanium atomic ratios $A_{\rm Xe}/A_{\rm Ge}$, which increases the xenon sensitivity, shrinks its CLCs, and allows modest CLC separation. In the $n_i=$ standard case there is no such enhancement and no CLC separation is evident. Now we turn to investigating the impact of statistical effects on our conclusions by considering the case where all targets observe the same number of events.

\subsection{Operator Discrimination with Equal Event Numbers on All Targets}

\begin{figure}[h]
\begin{center}
\includegraphics[height=2.3in,width=2.5in]{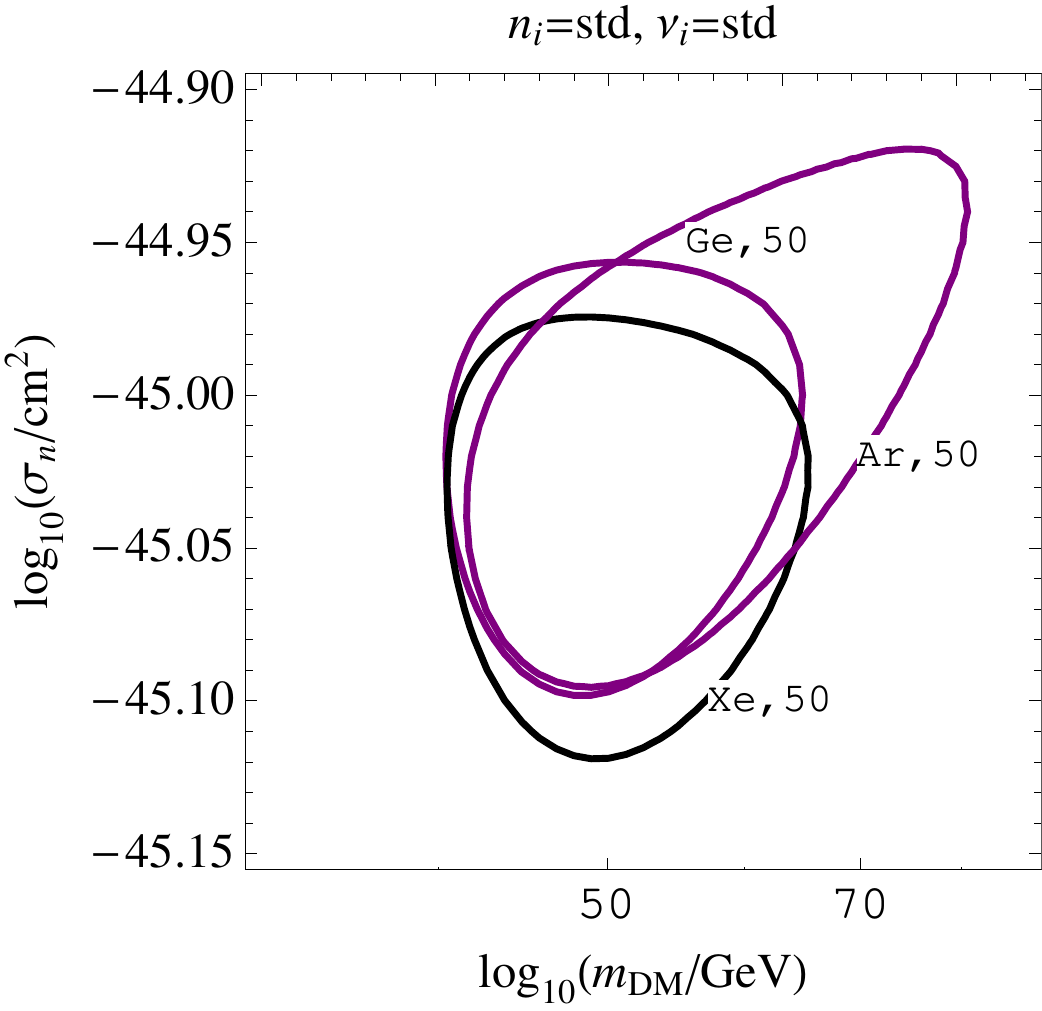}~~~~~~~~
\includegraphics[height=2.3in,width=2.5in]{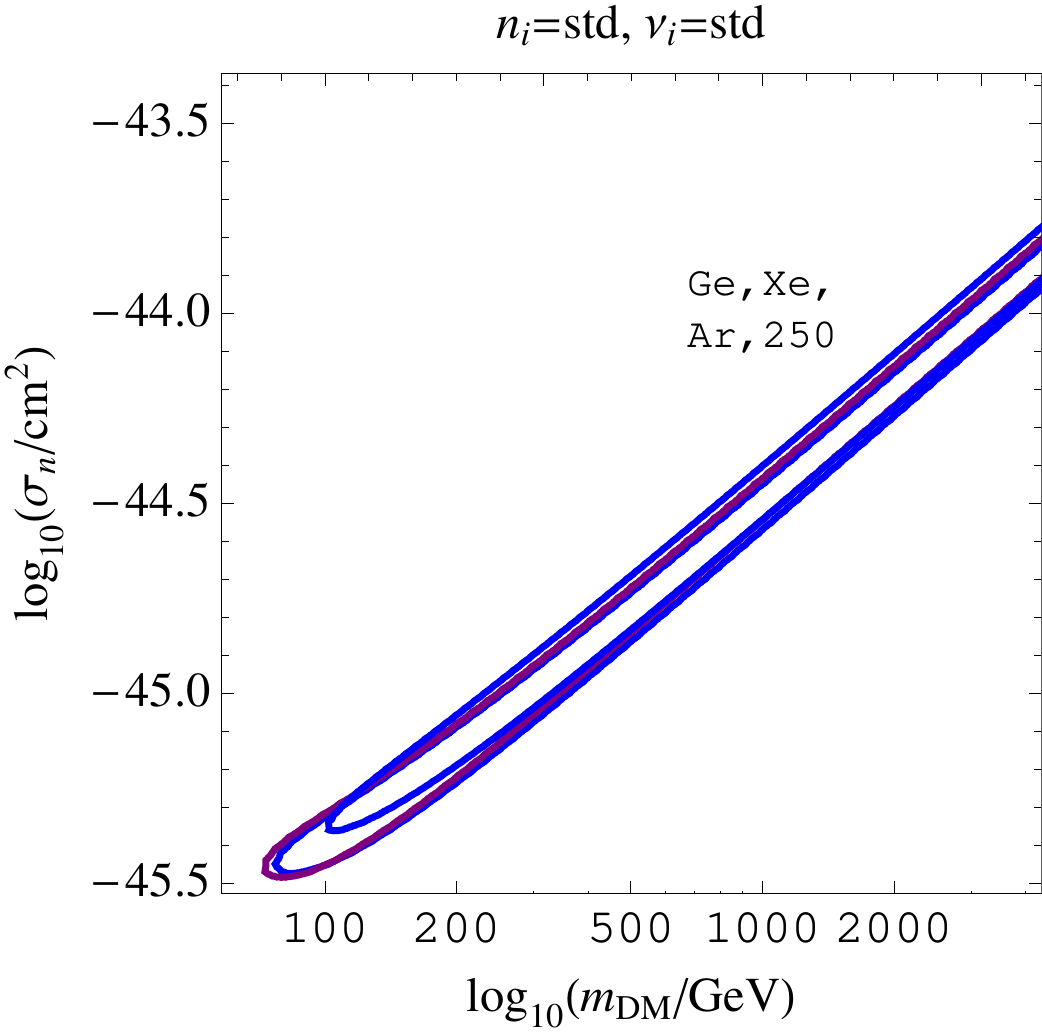}
\includegraphics[height=2.3in,width=2.5in]{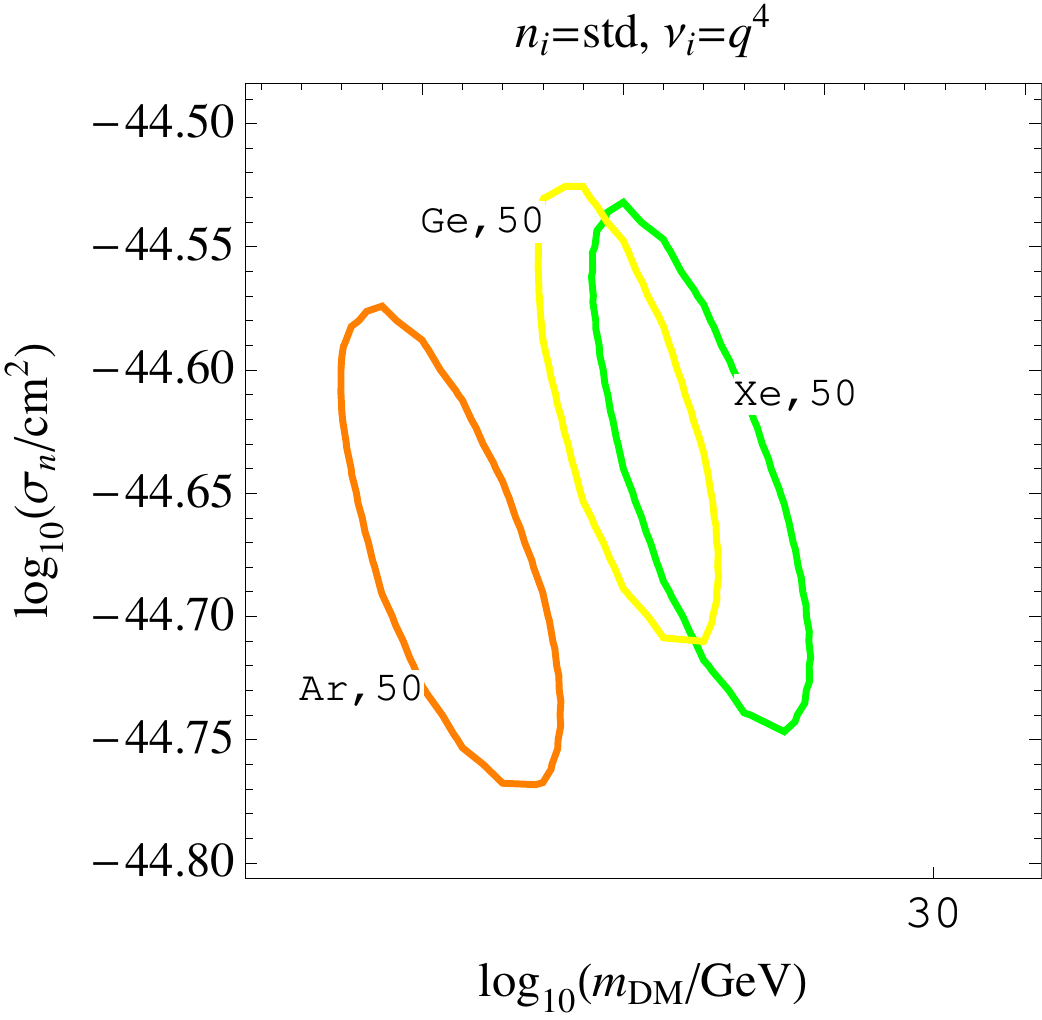}~~~~~~~~
\includegraphics[height=2.3in,width=2.5in]{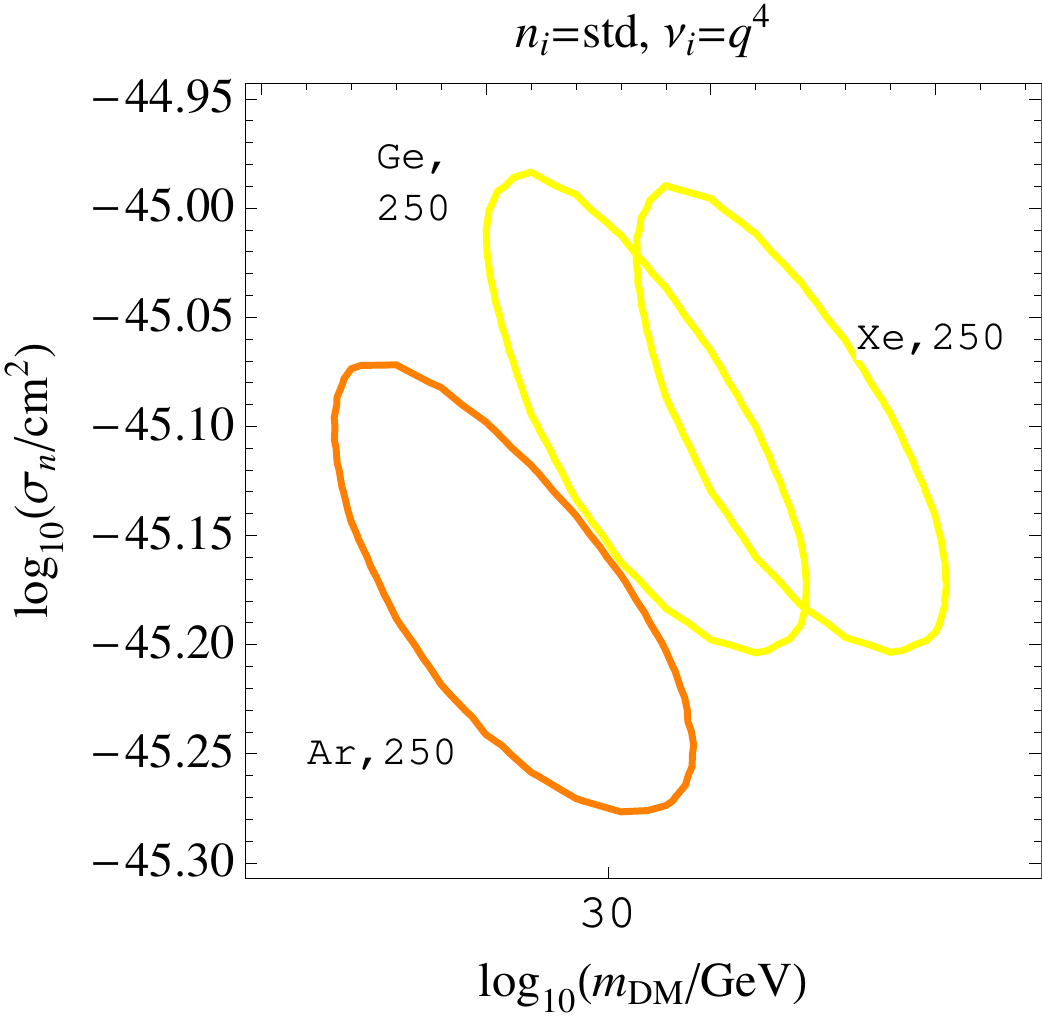}
\includegraphics[height=2.3in,width=2.5in]{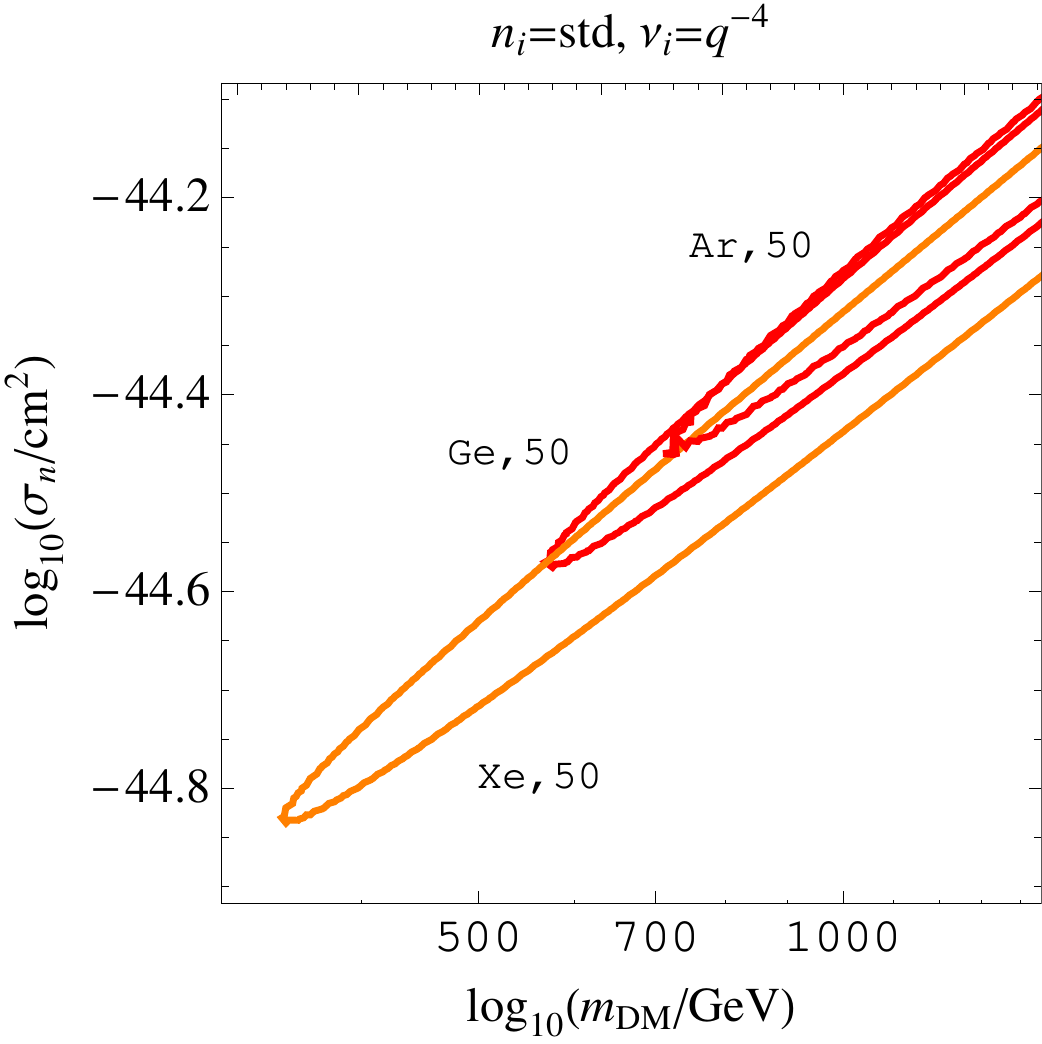}~~~~~~~~
\includegraphics[height=2.3in,width=2.5in]{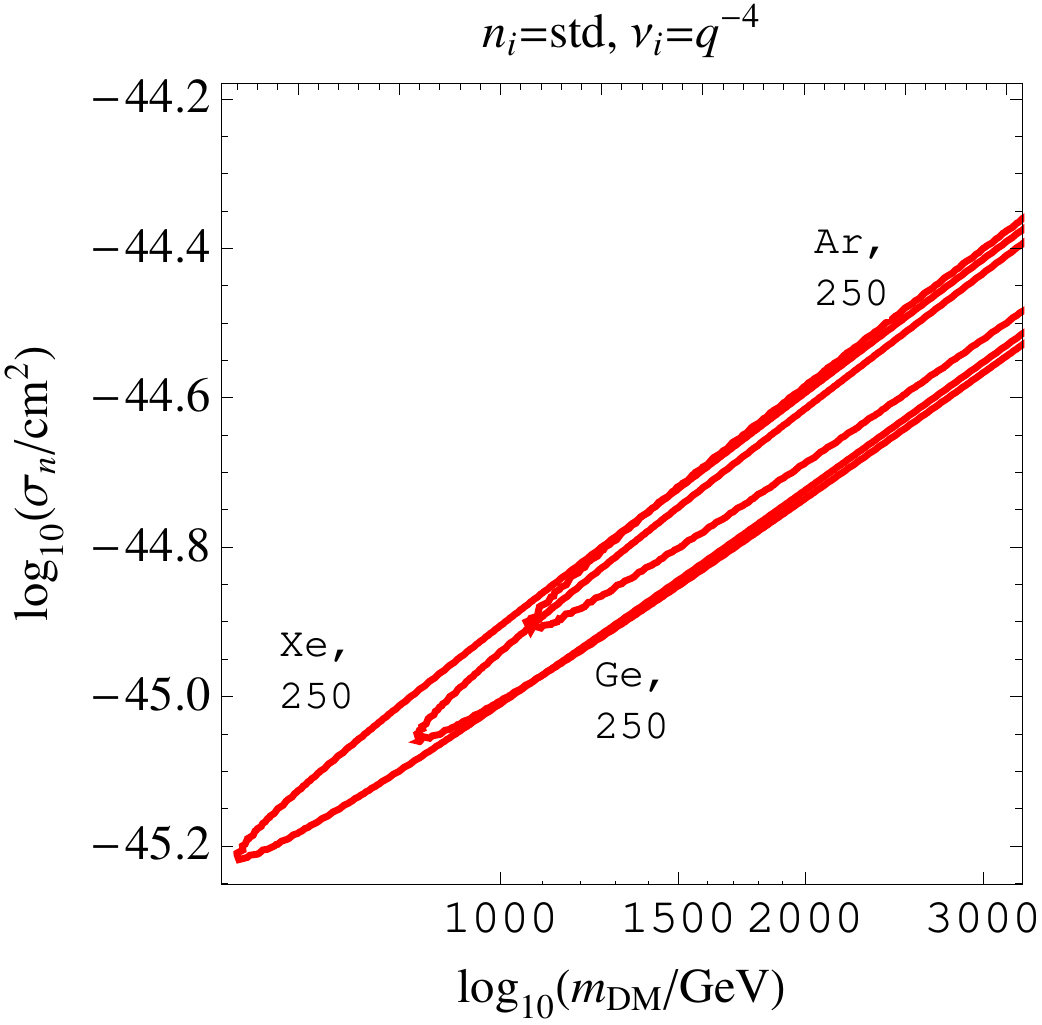}
\qquad \includegraphics[height=.5in,width=3.8in]{bar2.jpg}
\caption{95\% CLCs for 10, 50, and 250 GeV particles interacting through an $n_i=$ standard operator with 300 events on each target. Comparisons are made to $\nu_i=$ standard, $q^4$, and $q^{-4}$ operators. The colors represent the value of $\widetilde{L}_{\rm min}/$ d.o.f. The standard interaction can be distinguished from the $q^{-4}$ operators only via the values of $\widetilde{L}_{\rm min}/$d.o.f., unlike in Fig.~\eqref{SI} where CLC overlap also provides a test.}
\label{sizstd}
\end{center}
\end{figure}

\begin{figure}[h]
\begin{center}
\includegraphics[height=2.3in,width=2.5in]{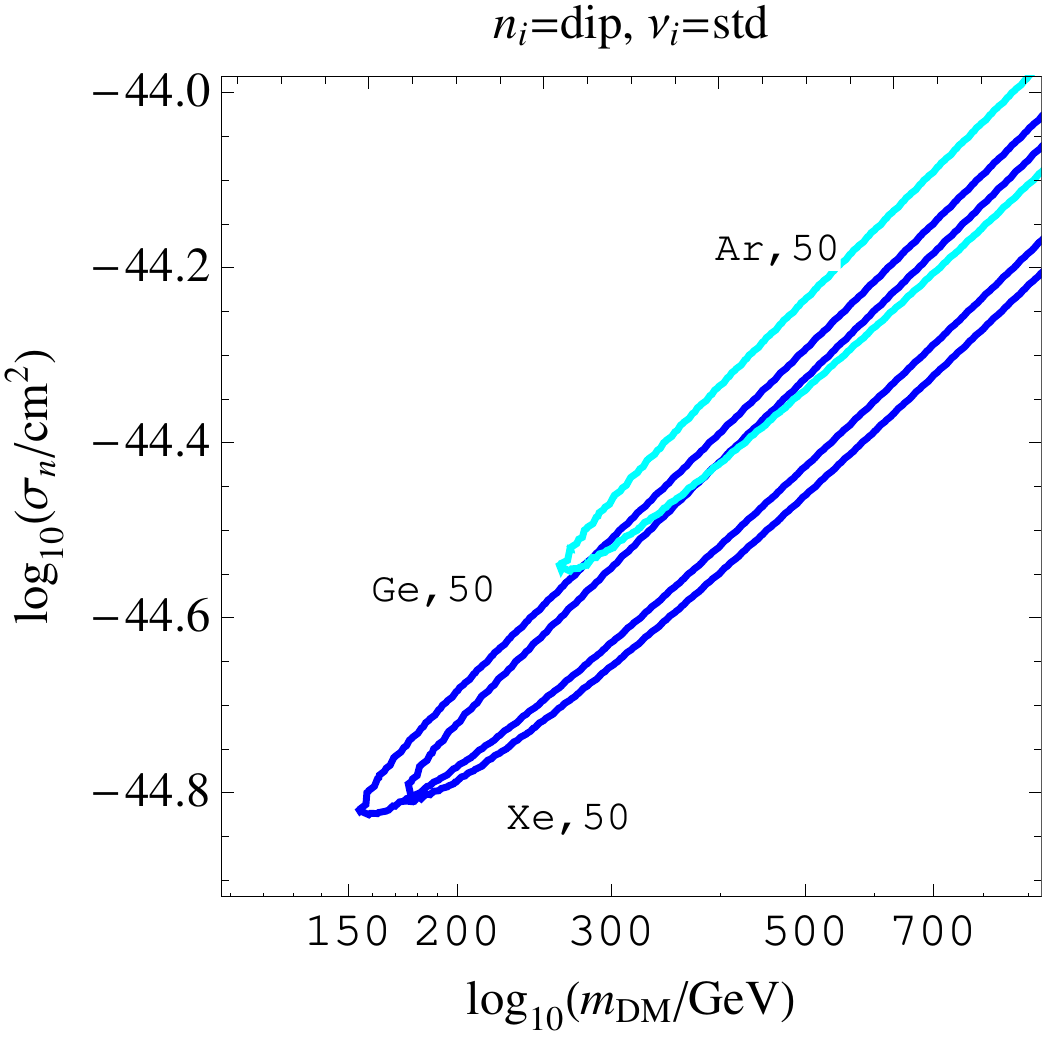}~~~~~~~~
\includegraphics[height=2.3in,width=2.5in]{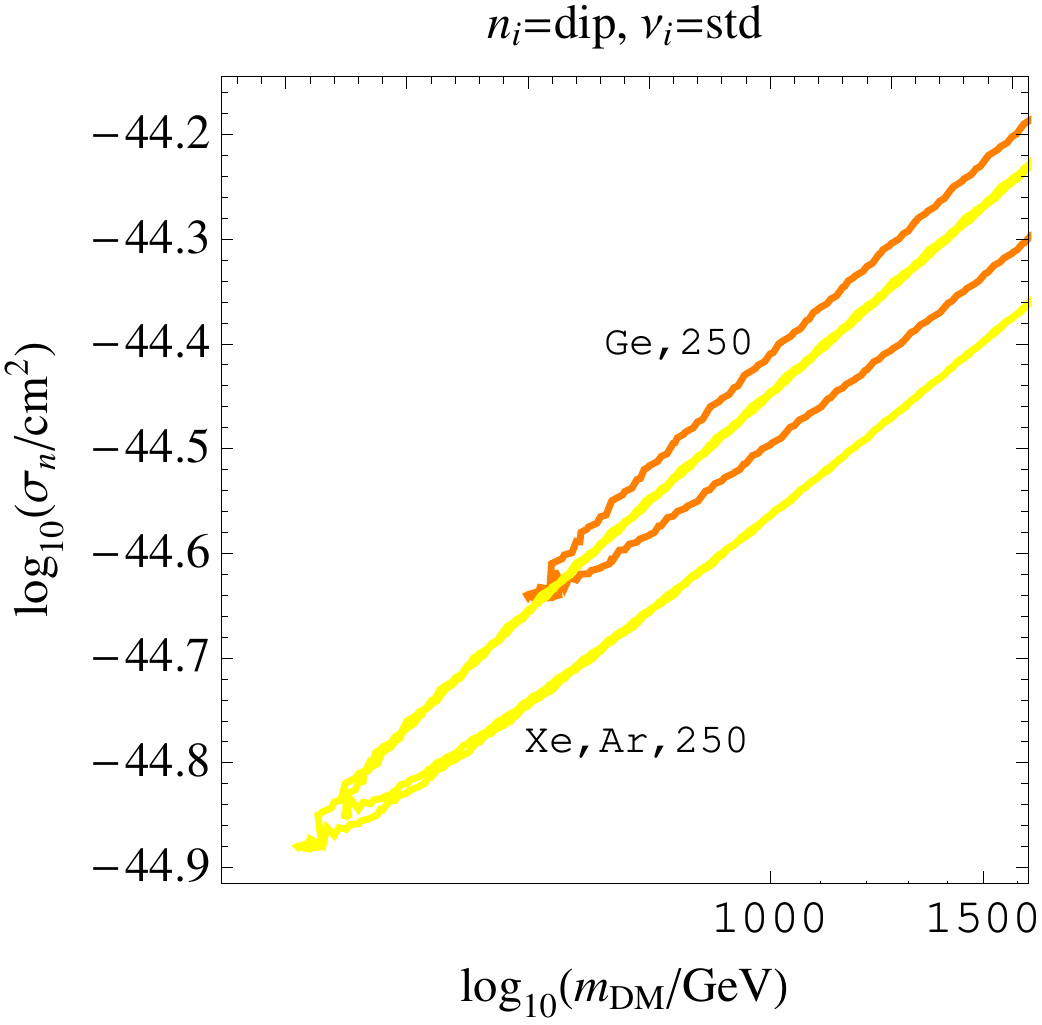}
\includegraphics[height=2.3in,width=2.5in]{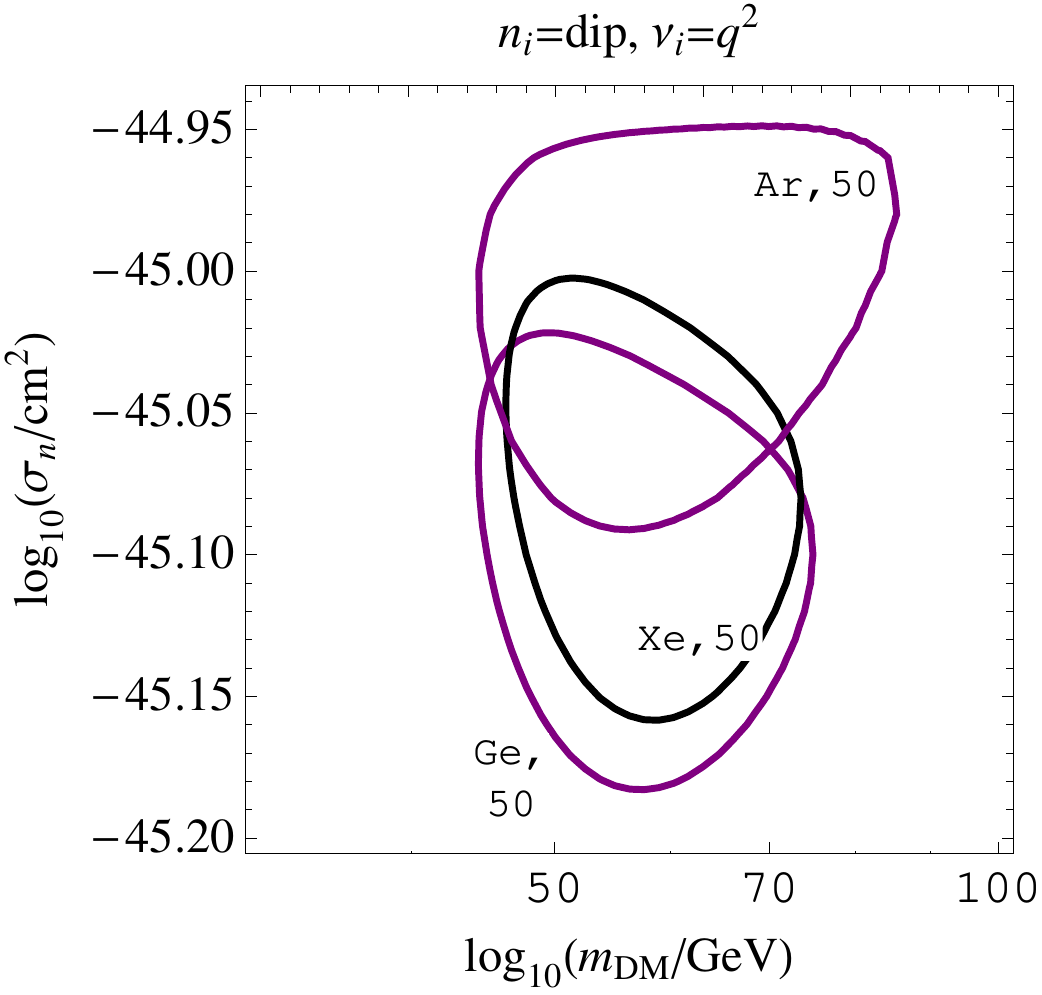}~~~~~~~~
\includegraphics[height=2.3in,width=2.5in]{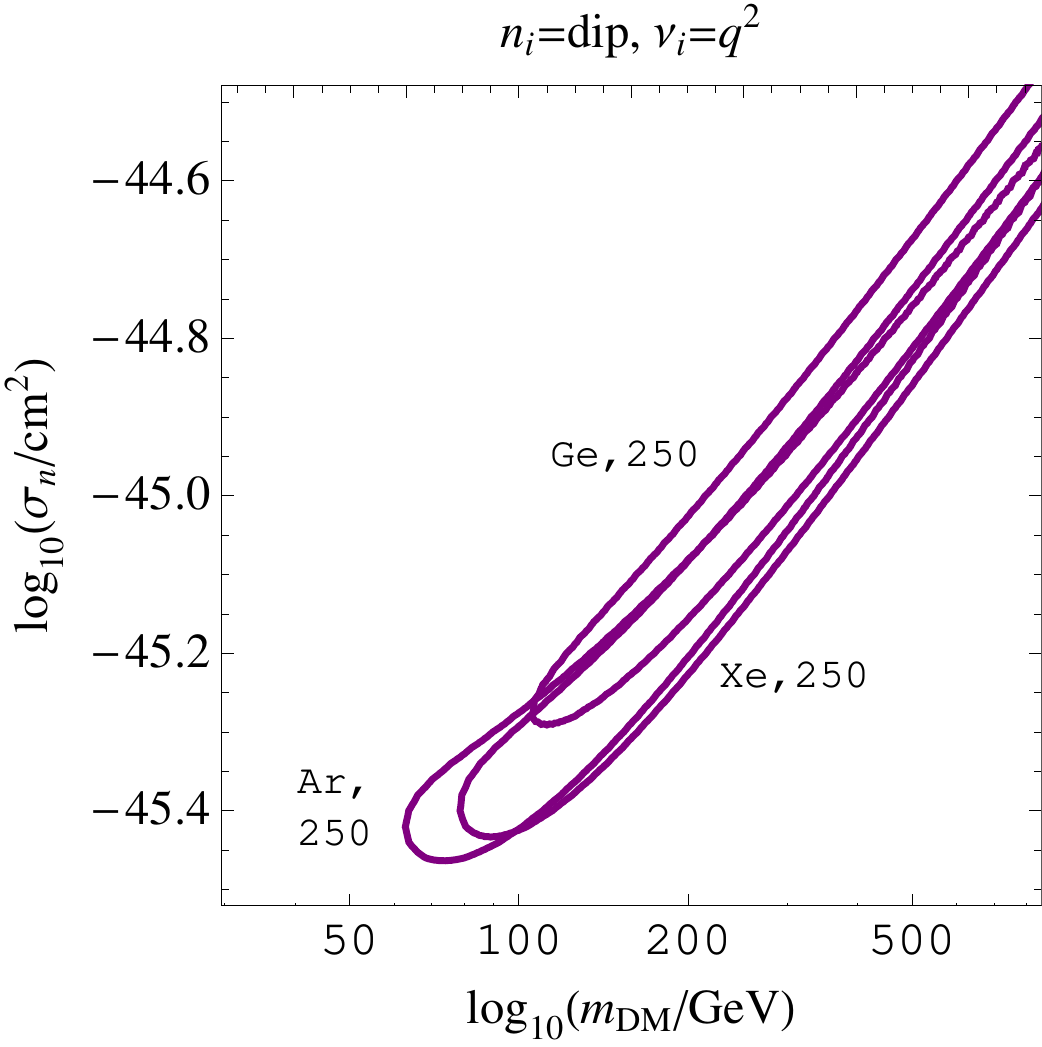}
\includegraphics[height=2.3in,width=2.5in]{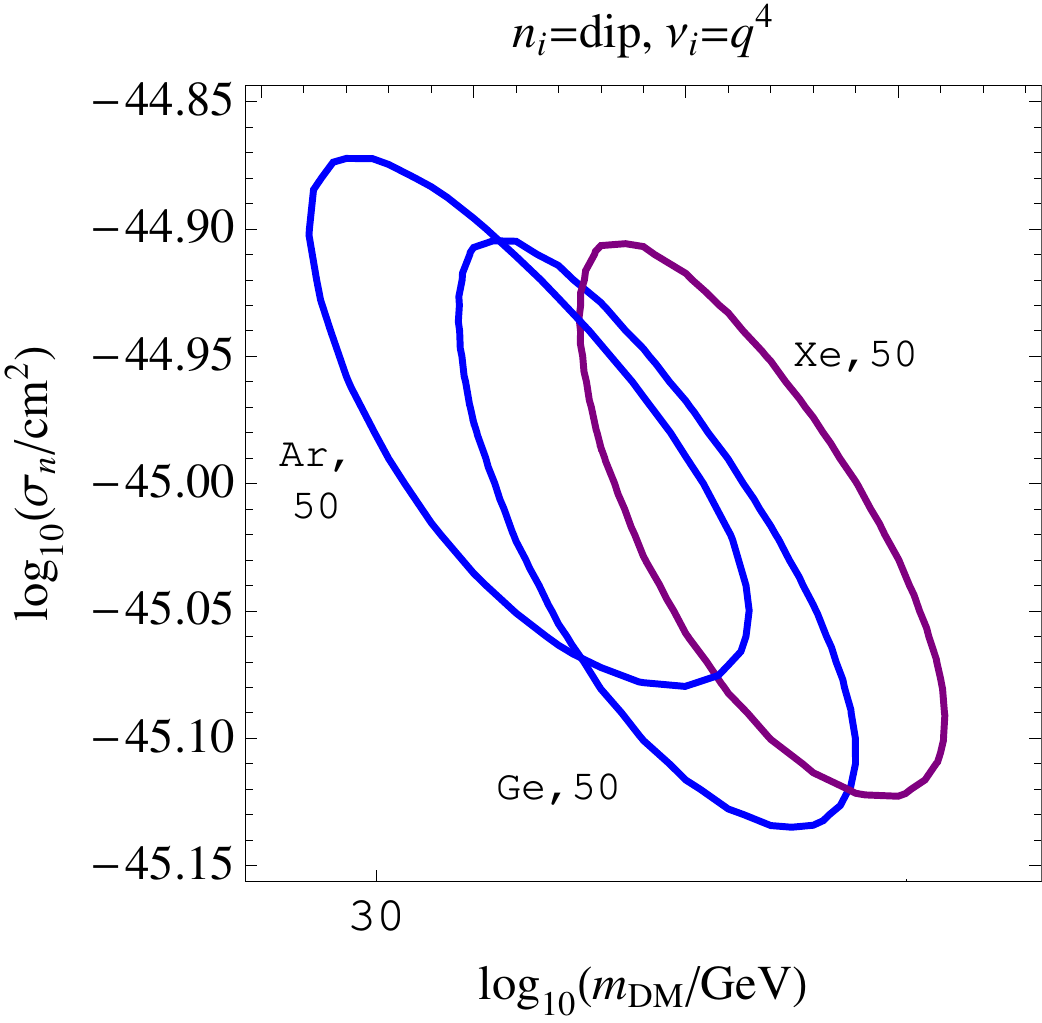}~~~~~~~~
\includegraphics[height=2.3in,width=2.5in]{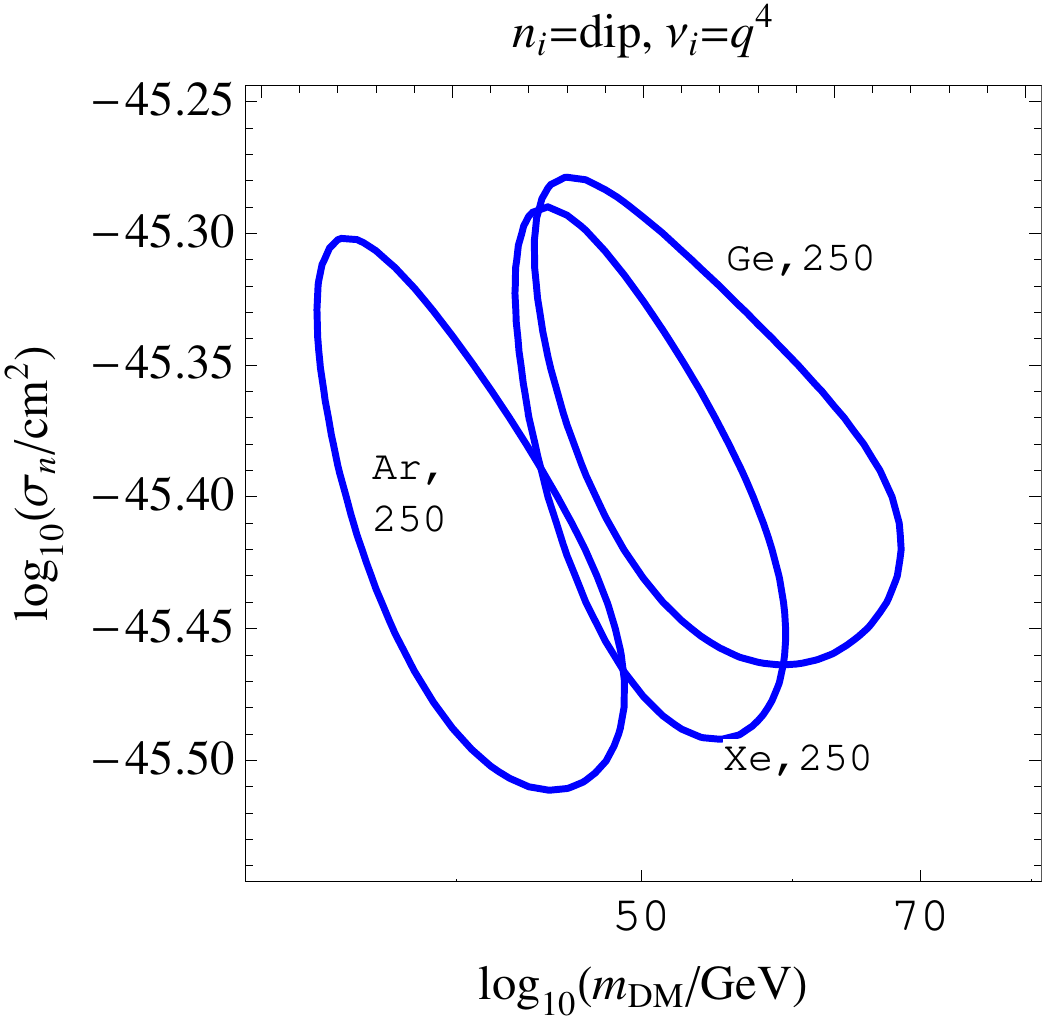}
\qquad \includegraphics[height=.5in,width=3.8in]{bar2.jpg}
\caption{95\% CLCs for a 10, 50, and 250 GeV particles interacting through an $n_i=$ dipole moment operator with 300 events on each target. Comparisons are made to $\nu_i=$ standard, $q^2$, and $q^{4}$ operators. The colors represent the value of $\widetilde{L}_{\rm min}/$d.o.f. Compared to the case shown in Fig.~\eqref{dipole} where exposures were fixed for all targets so that xenon was able to power the statistics, the ability to discriminate the dipole interaction from standard and $q^4$ operators using overlap or  the values of $\widetilde{L}_{\rm min}/$d.o.f.  is diminished.}
\label{sizdip}
\end{center}
\end{figure}

In Figs.~\eqref{sizstd} and \eqref{sizdip} we show the CLC plots for 50 GeV and 250 GeV candidates with 300 events on all targets to see if any discriminatory ability is lost. We omit the 10 GeV case because of the poor discrimination; we have checked that even by increasing the number of events in the 10 GeV candidate case by an order of magnitude, little improvement occurs.  To achieve these event numbers we simply reduce the exposure for xenon; for argon we increase the exposure in addition to lowering the energy threshold from 20 keV to 5 keV. This ``equal event number" normalization is less physically motivated than the ``equal exposure" normalization adopted above, but is necessary for understanding the robustness of our results.

Compared to the figures from the equal exposure scenario it is apparent that for high mass candidates the simple overlap test loses much of its capability to distinguish operators. There is no overlap when $n_i=$ standard and $\nu_i=q^4$, but there is overlap for $\nu_i=q^{-4}$. This problem is even more noticeable in the case with $n_i=$ dipole where with equal events there is mutual overlap in almost every instance. This loss of power is expected since the precision of the xenon CLCs has greatly diminished.

However, the $\widetilde{L}_{\rm min}$/d.o.f. test remains generally strong and even increases in relevance for some combinations of operators. Previously, only xenon and germanium were capable of differentiating operators on the basis of the  $\widetilde{L}_{\rm min}$/d.o.f. test.  By increasing the number of events on an argon detector we have a third viable test of the $\widetilde{L}_{\rm min}$/d.o.f. of each operator. This compensates for the weakening of the overlap test that is a consequence of decreasing the number of events on xenon. Yet, even though we are still able to extract the correct operator, it seems that our standards for discrimination have been changed by the statistics.

We would like to be able to use the same test on any combination of operators, elements, and exposures.  The fact that the overlap test is powered by the high $\widetilde{L}_{\rm min}$ values of xenon is an important clue. We will show that what seems to be a qualitative difference in how we draw conclusions in the equal exposure and equal event scenarios is just an illusion.

\subsection{Comparison of Models}

In this subsection, we present significance tests for different trial operators. We combine all three data sets and define $\widetilde{L}_{\rm min}^{\rm tot}$ to be the minimum log-likelihood value from this global fit.
\begin{figure}[b]
\begin{center}
\includegraphics[height=2.8in,width=3in]{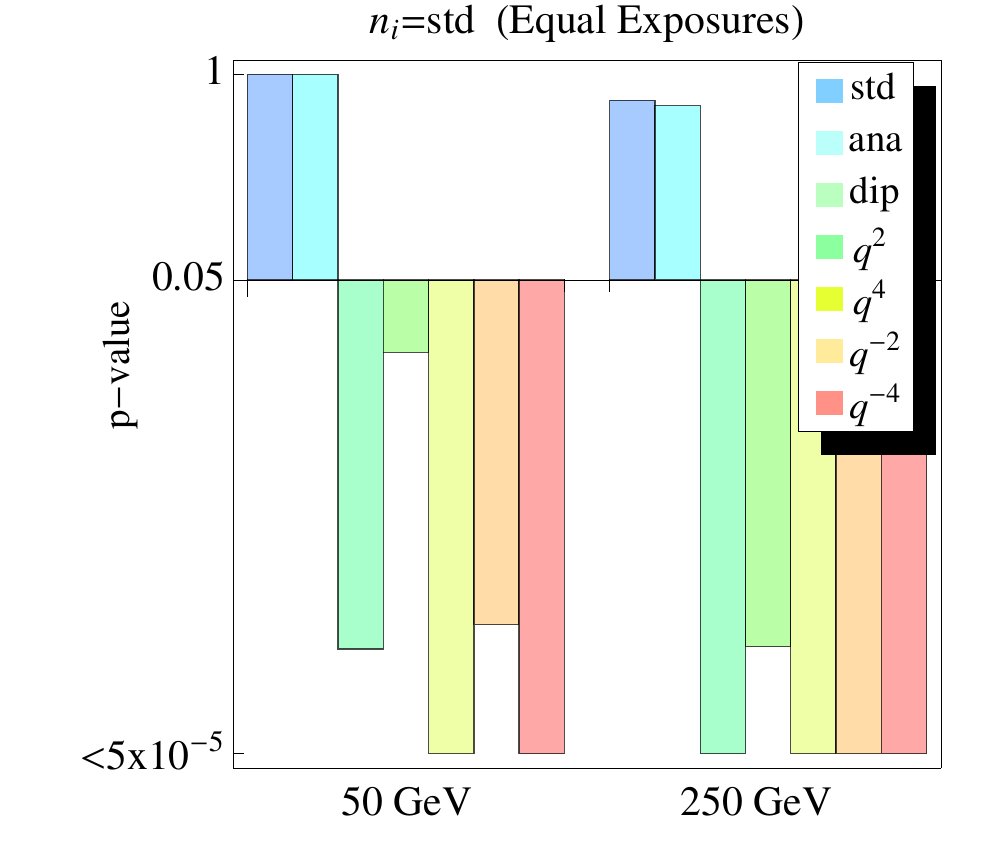}
\includegraphics[height=2.8in,width=3in]{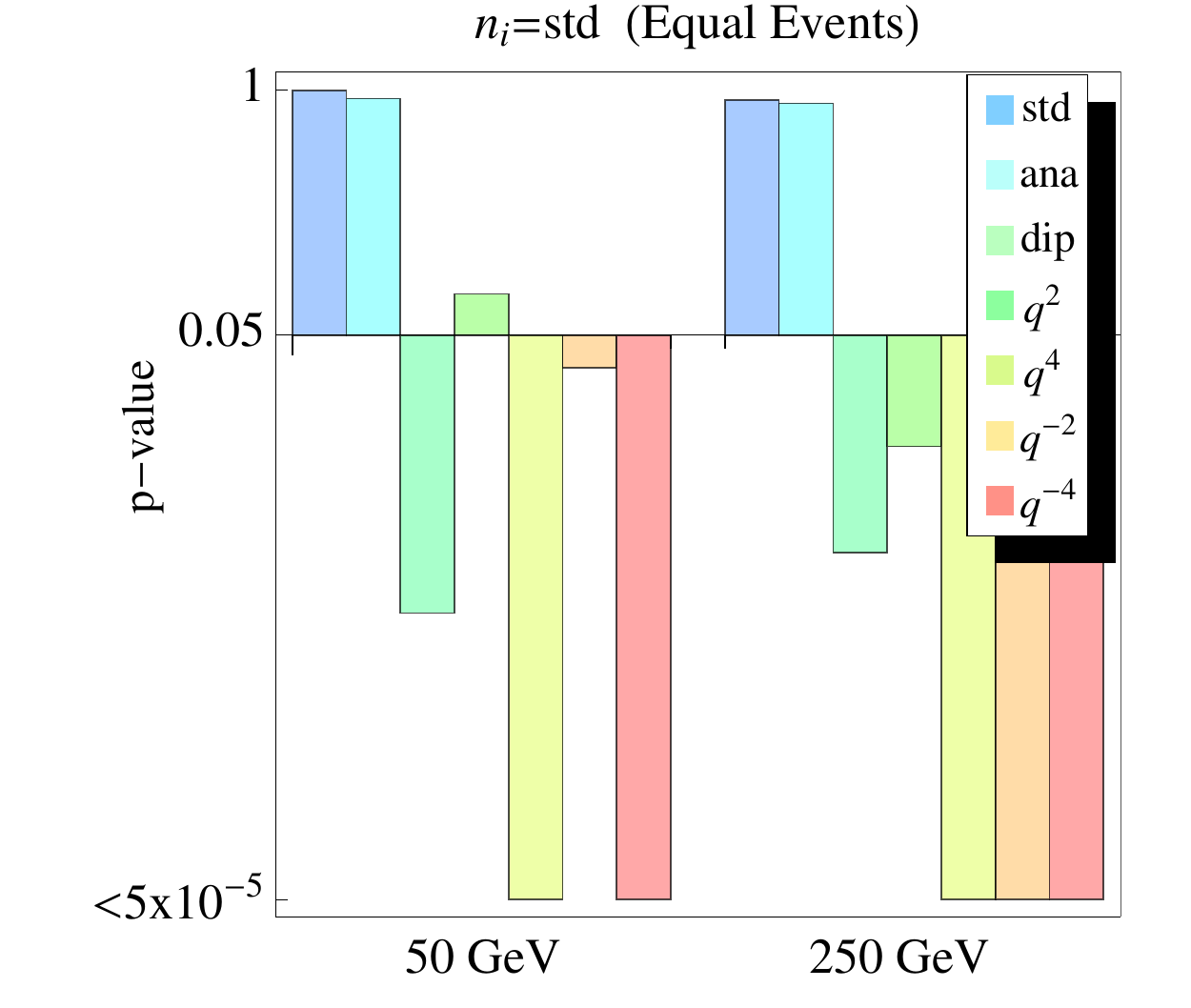}
\includegraphics[height=2.8in,width=3in]{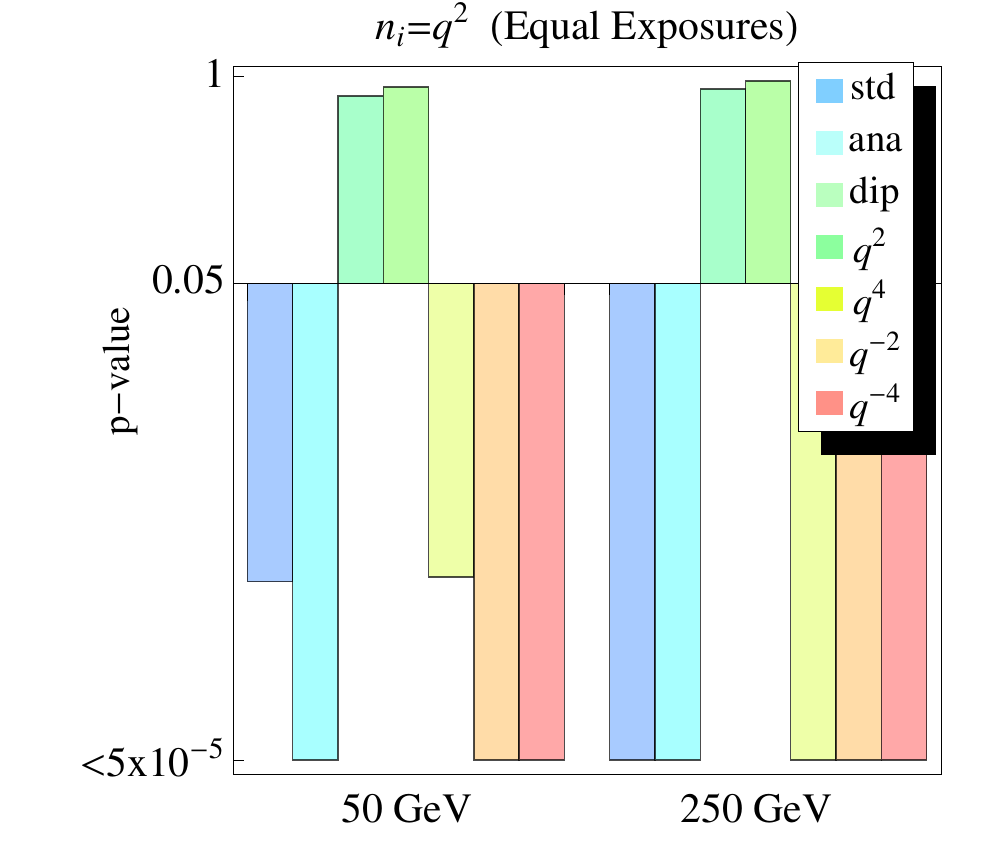}
\includegraphics[height=2.8in,width=3in]{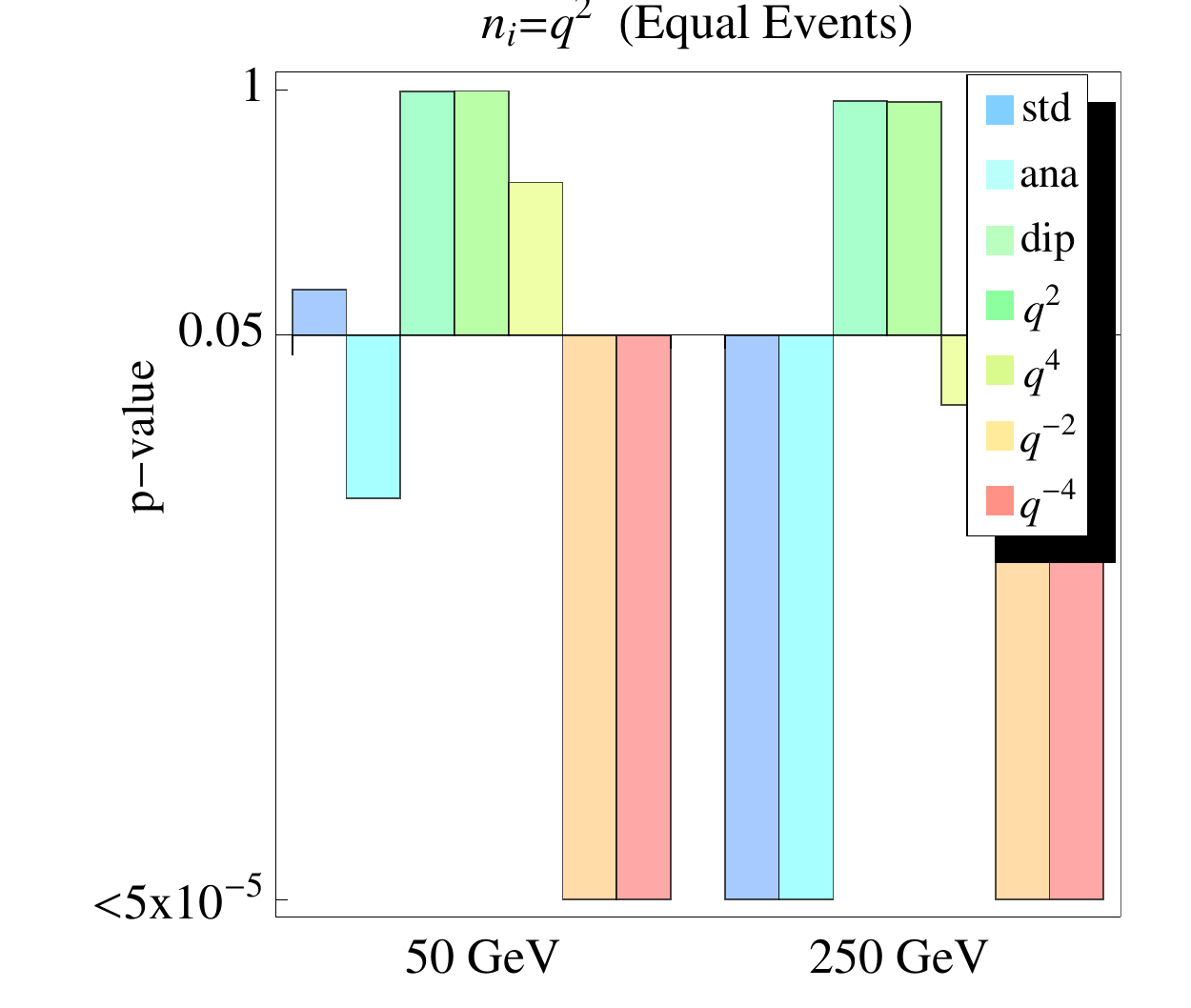}
\caption{The p-values of trial operators for $n_i=$ standard and $q^{2}$ operators for candidate masses of 50 and 250 GeV. We display the equal exposure and equal event bar charts side by side to underscore the robustness of the discrimination. For the visual purpose, the plot is normalized so that each bar starts at $5\%$ significance level.}
\label{smooth} 
\end{center}
\end{figure}
\begin{figure}[b]
\begin{center}
\includegraphics[height=2.8in,width=3in]{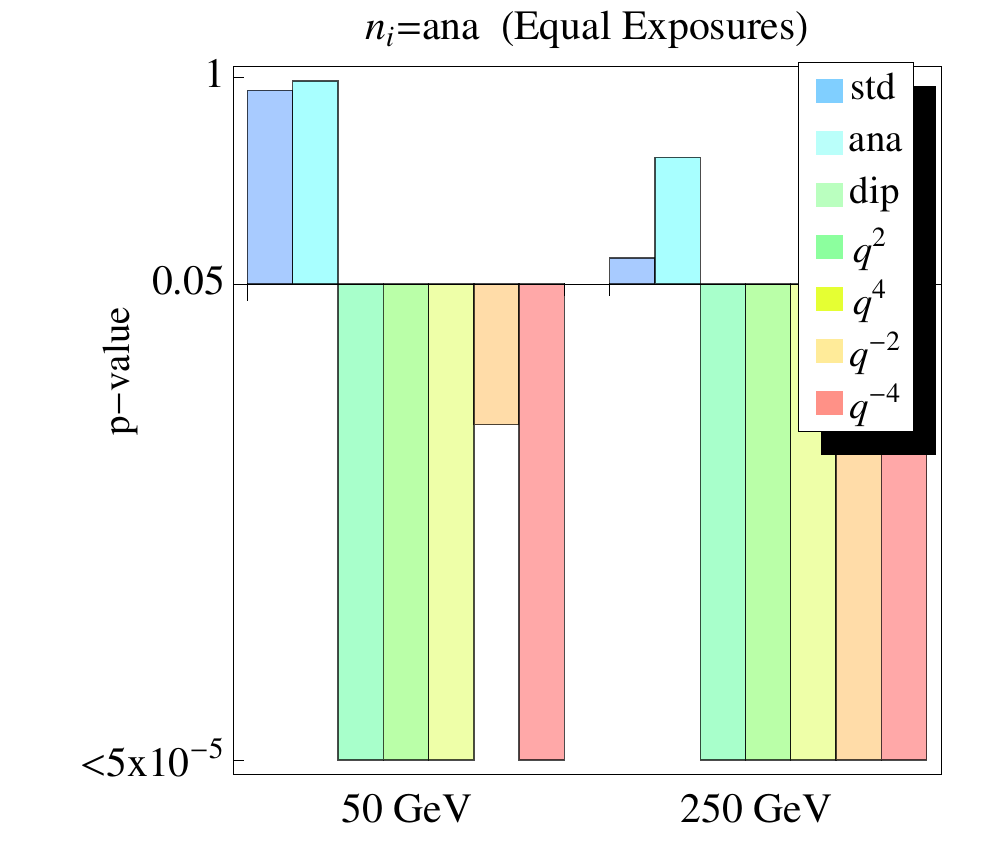}
\includegraphics[height=2.8in,width=3in]{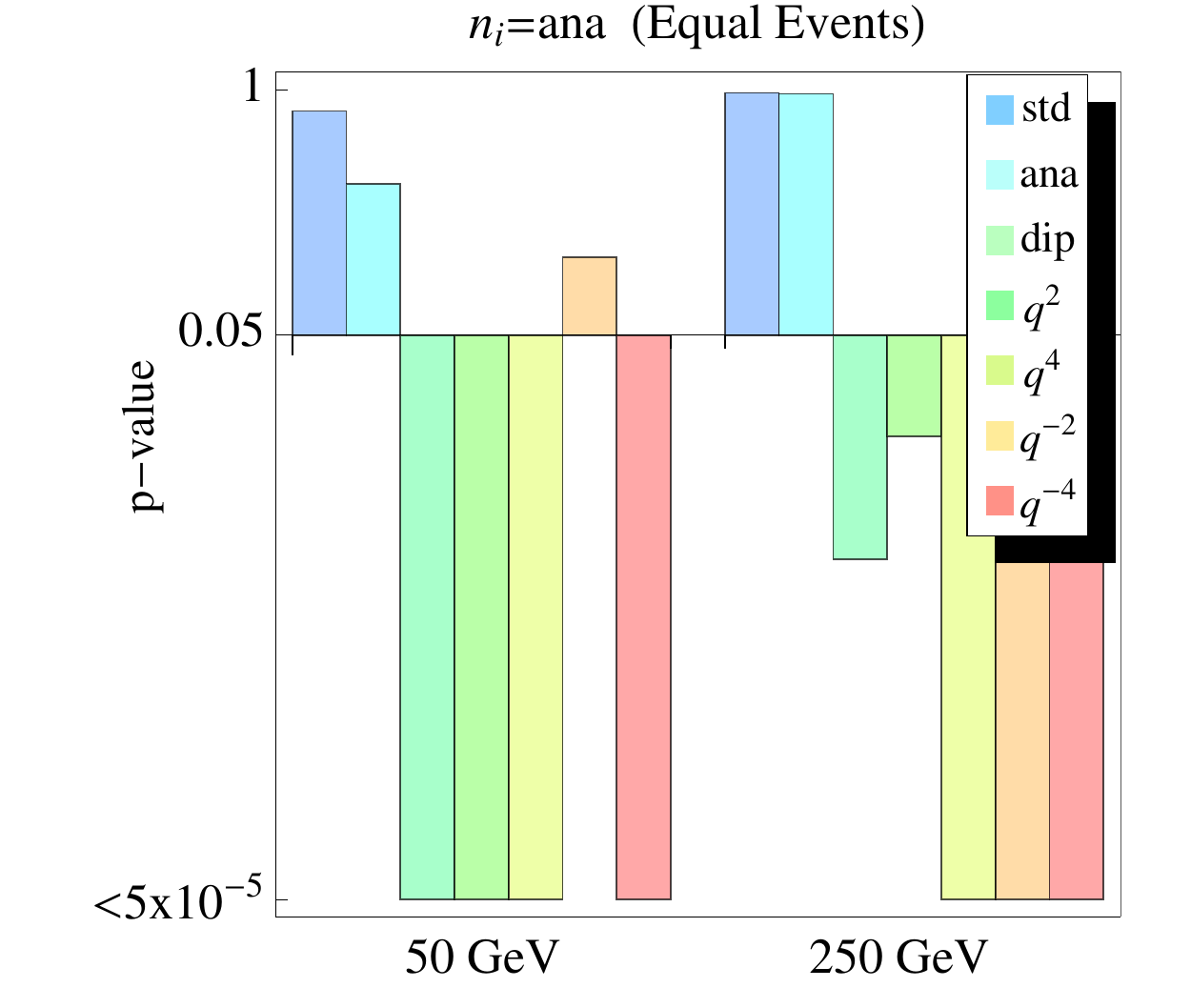}
\includegraphics[height=2.8in,width=3in]{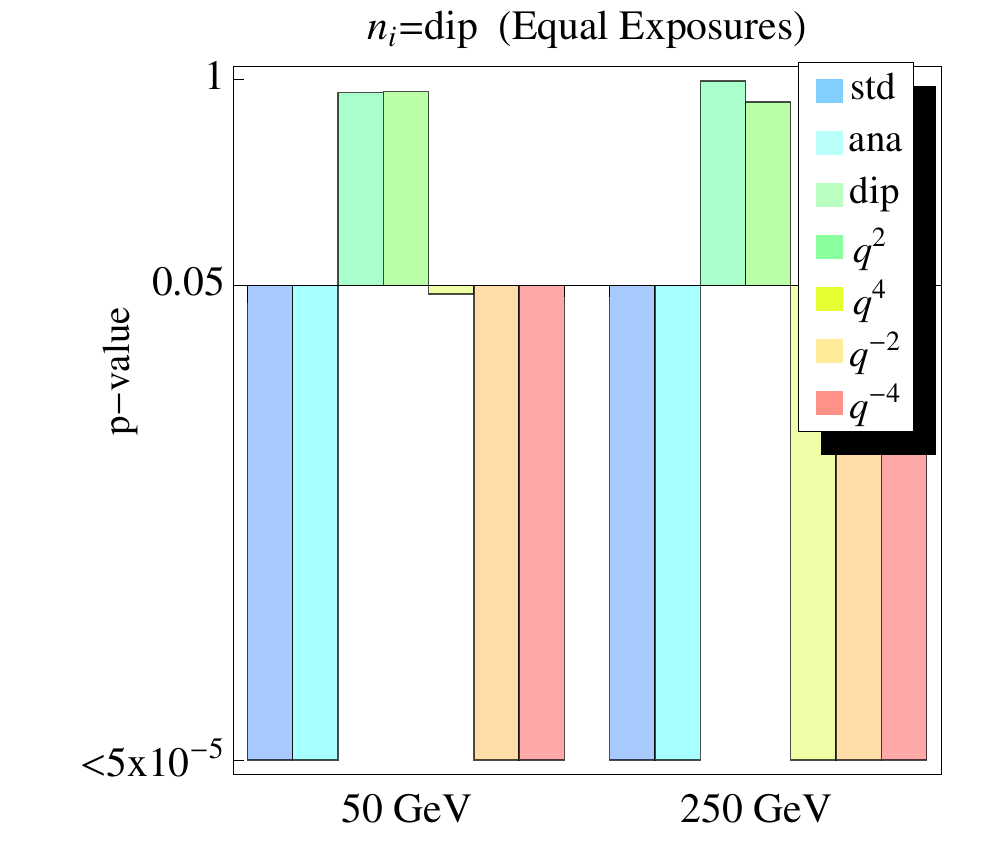}
\includegraphics[height=2.8in,width=3in]{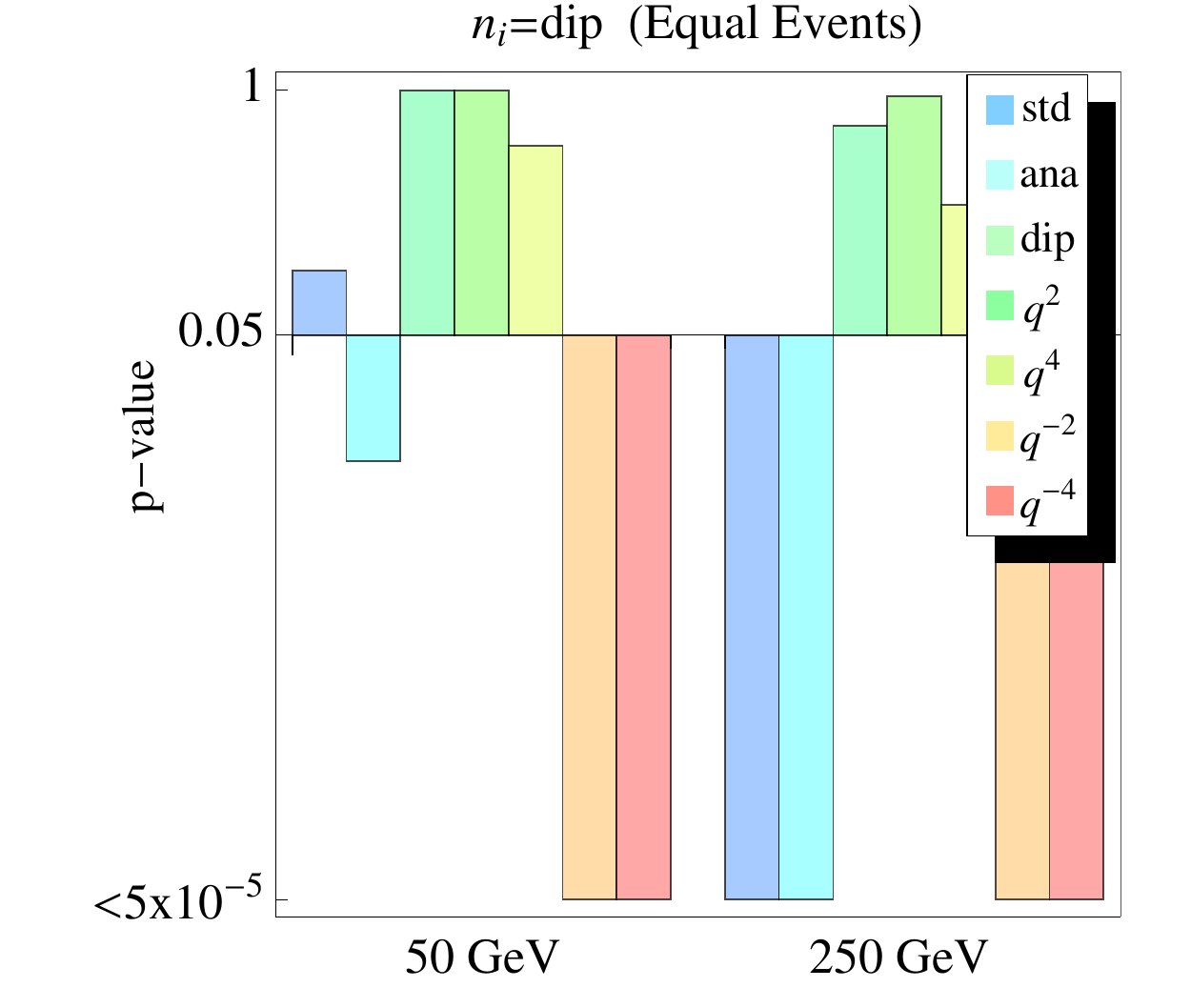}
\includegraphics[height=2.8in,width=3in]{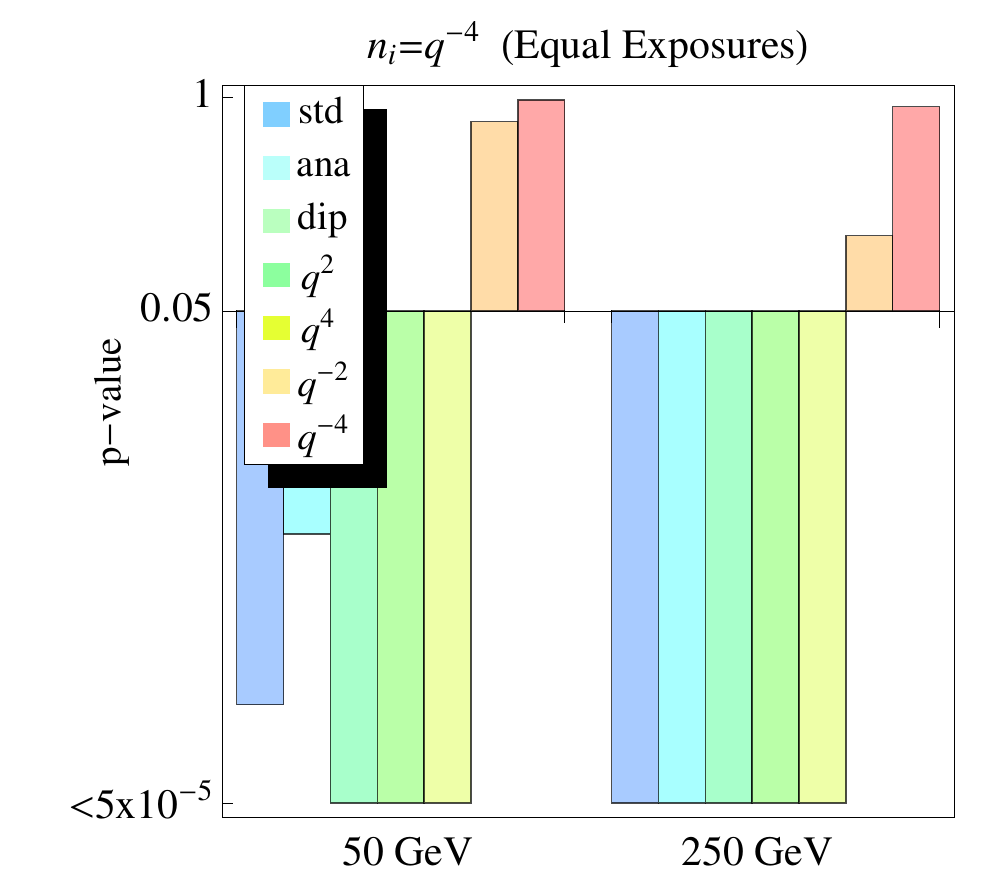}
\includegraphics[height=2.8in,width=3in]{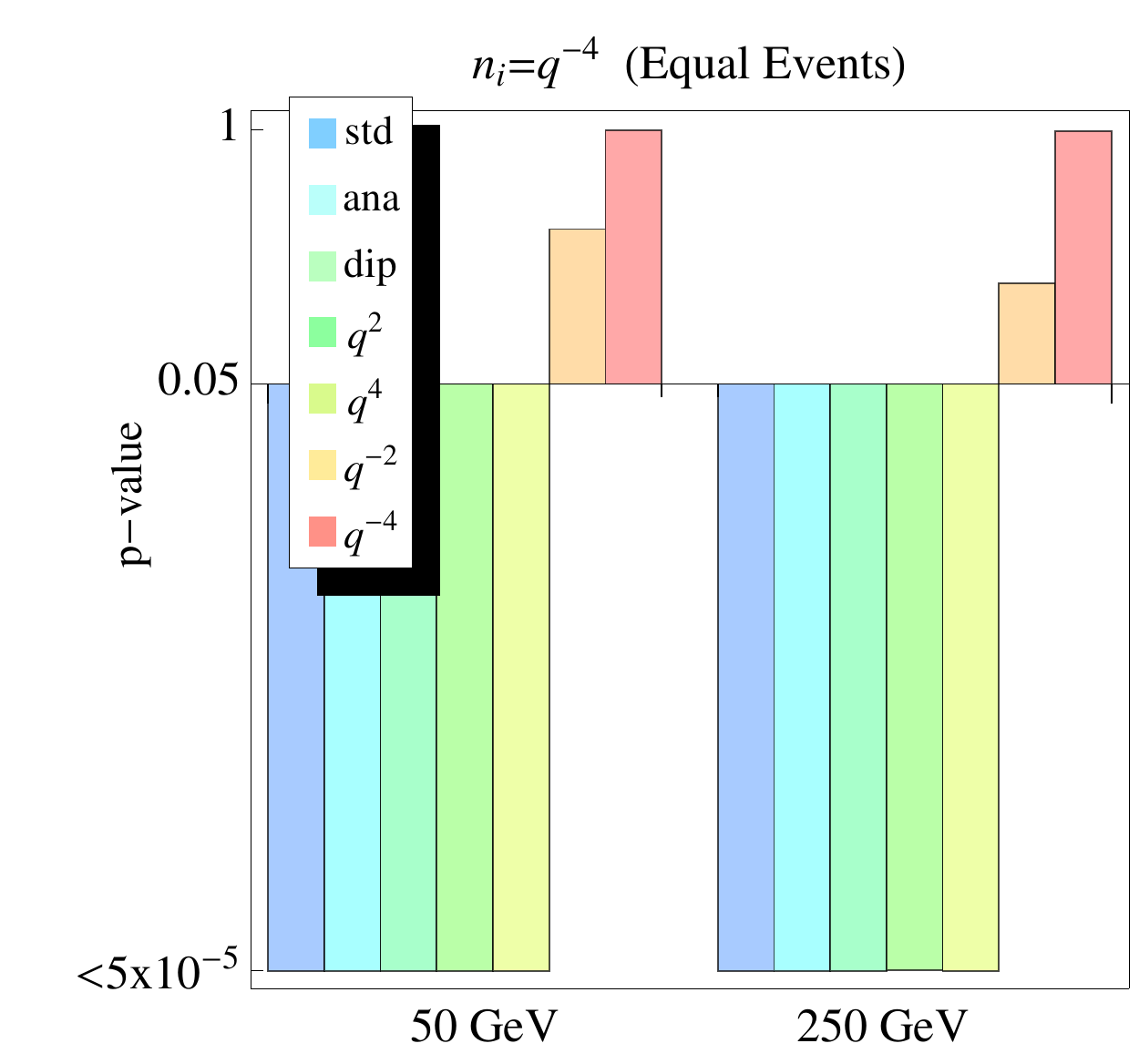}
\caption{The p-values of trial operators for $n_i=$ anapole, dipole, and $q^{-4}$ operators for candidate masses of 50 and 250 GeV. We display the equal exposure and equal event bar charts side by side to underscore the robustness of the discrimination. For the visual purpose, the plot is normalized so that each bar starts at $5\%$ significance level.}
\label{smooth2} 
\end{center}
\end{figure}
Here we make the approximation that $\widetilde{L}_{\rm min}^{\rm tot}$ values follow a $\chi^2$ distribution. This is valid when there are a large number of events, as in the cases studied here. Thus, we may derive a $p$-value for any given trial operator by defining \cite{PDG}
\beq
p(\widetilde{L}_{\rm min}^{\rm tot},n_{\rm d})=\int_{ \widetilde{L}_{\rm min}^{\rm tot}}^{\infty} \frac{x^{n_{\rm d}/2 -1}~ e^{-x/2} }{2^{ n_{\rm d}/2} \Gamma({ n_{\rm d}/2})} ~dx,
\eeq
where $n_{\rm d}$ is the number of degrees of freedom; this is the total number of nonzero bins for all three experiments minus the number of fit parameters. For a given trial operator, the $p$-value describes the probability of producing a fit that has a $\widetilde{L}^{\rm tot}$ larger than $\widetilde{L}_{\rm min}^{\rm tot}$, so that a high $p$-value for an operator indicates that the relevant model is able to fit the data well.

We display results for the standard, anapole, dipole, $q^2$, and $q^{-4}$ data sets in Figs.~\eqref{smooth} and \eqref{smooth2}. We omit the 10 GeV data from this comparison because the discrimination is obviously quite poor, and we are primarily interested in seeing if the higher mass candidates are sensitive to the statistical effect of changing the event numbers seen by the argon and xenon targets. 
One immediately notices that the scattering of 50 and 250 GeV candidates typically provides very good differentiation between correct and incorrect operators. In most of cases, we can reject the trial model at much better than $5\%$ significance level if operators do not match. Also as expected, we see that the standard and anapole operators are hard to distinguish, as are the $q^2$ and dipole operators. This is a consequence of the fact that the velocity-dependent contribution to the composite operators' spectra has weak momentum dependence, and thus the anapole and standard operators have approximately momentum independent spectra just as the dipole and $q^2$ operators have the same $q^2$ momentum dependence.

Modulo this degeneracy, we see that it is possible to extract the particle physics nature of the dark matter scattering events given that future ton-scale direct detection experiments observe ${\cal O}(100)$ dark matter events. Moreover, the $p$-value is capable of extracting the momentum dependence of the operator equally well for the two event normalizations (equal exposures or equal event numbers on all targets) studied here. Finally, we also have checked that the results obtained from the $p$-value test derived from the global-fit $\widetilde{L}_{\rm min}^{\rm tot}$ agree with the conclusions we would draw from calculating the $p$-values for each element individually, but the result from the global-fit has a better discrimination ability.

\section{Conclusions}

We have discussed the capability of direct detection experiments to extract the particle physics underlying DM scattering events.  We found that Poisson noise limits the ability of a single detector to determine the momentum dependence of the operator mediating the scattering.  When the exposures are equal for different elements and the data are from observations of high mass candidates, only xenon, due to its high atomic number, is capable of determining the momentum dependence of the interaction on its own.  Under these conditions, examining the preferred regions from data generated by more than one element allows one to extract the correct momentum dependence; this can be seen in Figs.~\eqref{SI} and \eqref{dipole}. When two operators give a similarly good fit to the data, it is because they have a very similar momentum dependence. We found that much of the power of this ``overlap test" is derived from xenon's capability to discriminate operators.  When all event numbers are held constant, so that the xenon exposure is decreased and the argon exposure is scaled up, we found that the overlap test is less powerful, but the combination of data from all three targets still provides good operator discrimination.  We found that the minimum log-likelihood value for the global fit summarizes all of this information succinctly and robustly.  The minimum log-likelihood value for the global fit is sensitive both to overlap and to the individual goodness of fit information, and can be put in one-to-one correspondence with the $p$-value, allowing us to make quantitative statements about the ability of each operator to fit the data.  For a light DM candidate on the other hand, the effects of the energy threshold, the energy resolution, and the scattering kinematics combine in such a way that the operator mediating the scattering cannot be extracted even for very high event numbers.

We thank Tim Cohen for discussions.  The work of SDM and KMZ was supported by NSF CAREER award PHY 1049896, and the work of HBY and KMZ was supported by NASA Astrophysics Theory Program grant NNX11AI17G.

\appendix

\newpage

\section{Plots}

Here we compile plots for analyses of all five data sets ($n_i=$ standard, anapole, dipole, $q^{2}{\rm~and~}q^{-4}$) with equal exposure. Due to space constraints, we show fits to six of the seven operators ($\nu_i=$ standard, anapole, dipole, $q^{\pm2}{\rm~and~}q^{\pm4}$) on each page. More comprehensive plots, including plots from data sets with equal event numbers, are shown online \cite{dminv}.

The plots illustrate the qualitative and quantitative features described in Section IV above.
Due primarily to the power of the discrimination of the xenon target, analysis that results in three overlapping CLCs will indicate an operator that fits the data well.
For low mass DM all operators are effectively indistinguishable in the sense that all analyses result in overlapping CLCs.
When the DM candidate is more massive, the distinguishing features of its interaction become more pronounced and harder to mimic, so a particular type of operator is selected.

Within each plot we display CLCs for all three detector elements and all three DM candidate masses, with the exception of the argon contour with the 10 GeV candidate.
None of the 10 GeV argon CLCs close because argon sees so few events. The argon ``exclusion curve" fits comfortably around the xenon and germanium CLCs in all cases.
To reduce clutter, these argon curves are omitted from the plots.

\begin{figure}[b]
\begin{center}
\includegraphics[height=2.3in,width=2.5in]{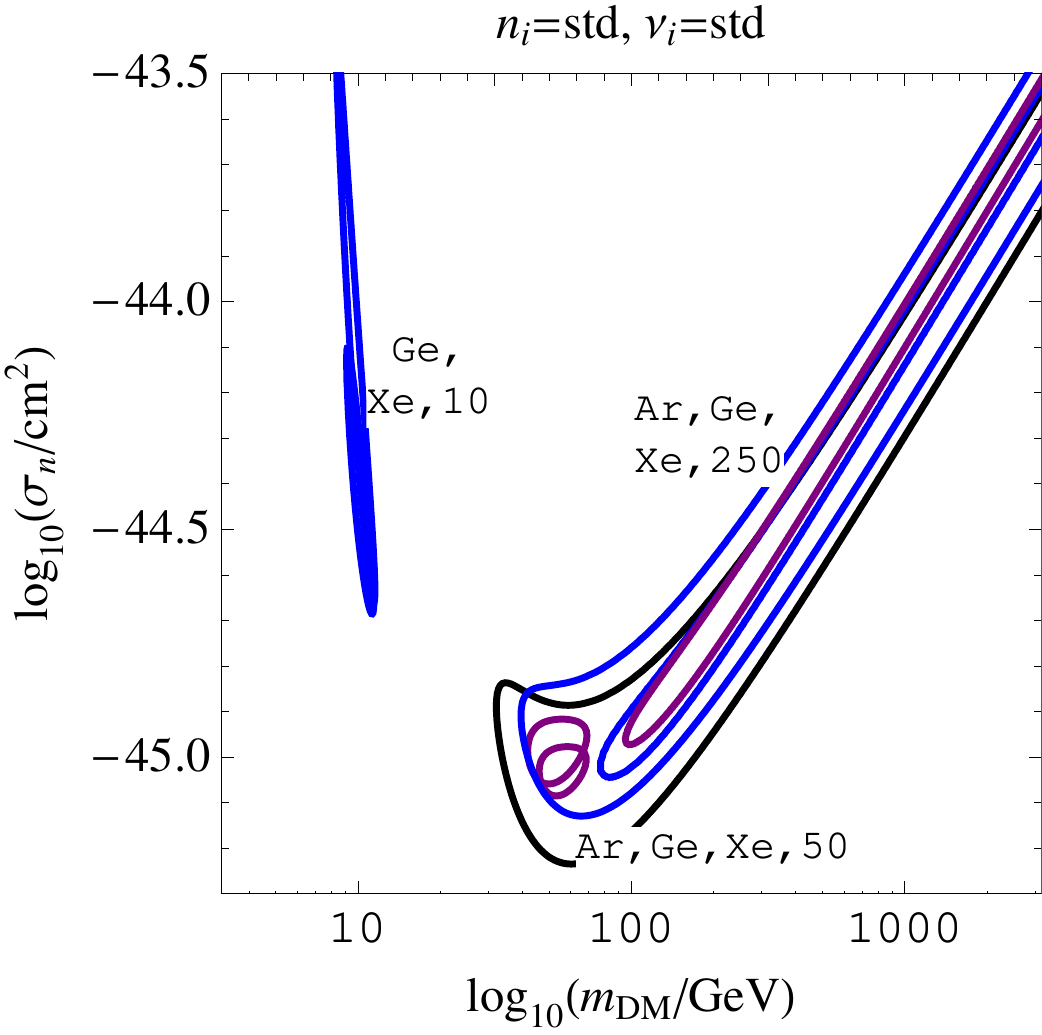}~~~~~~~~
\includegraphics[height=2.3in,width=2.5in]{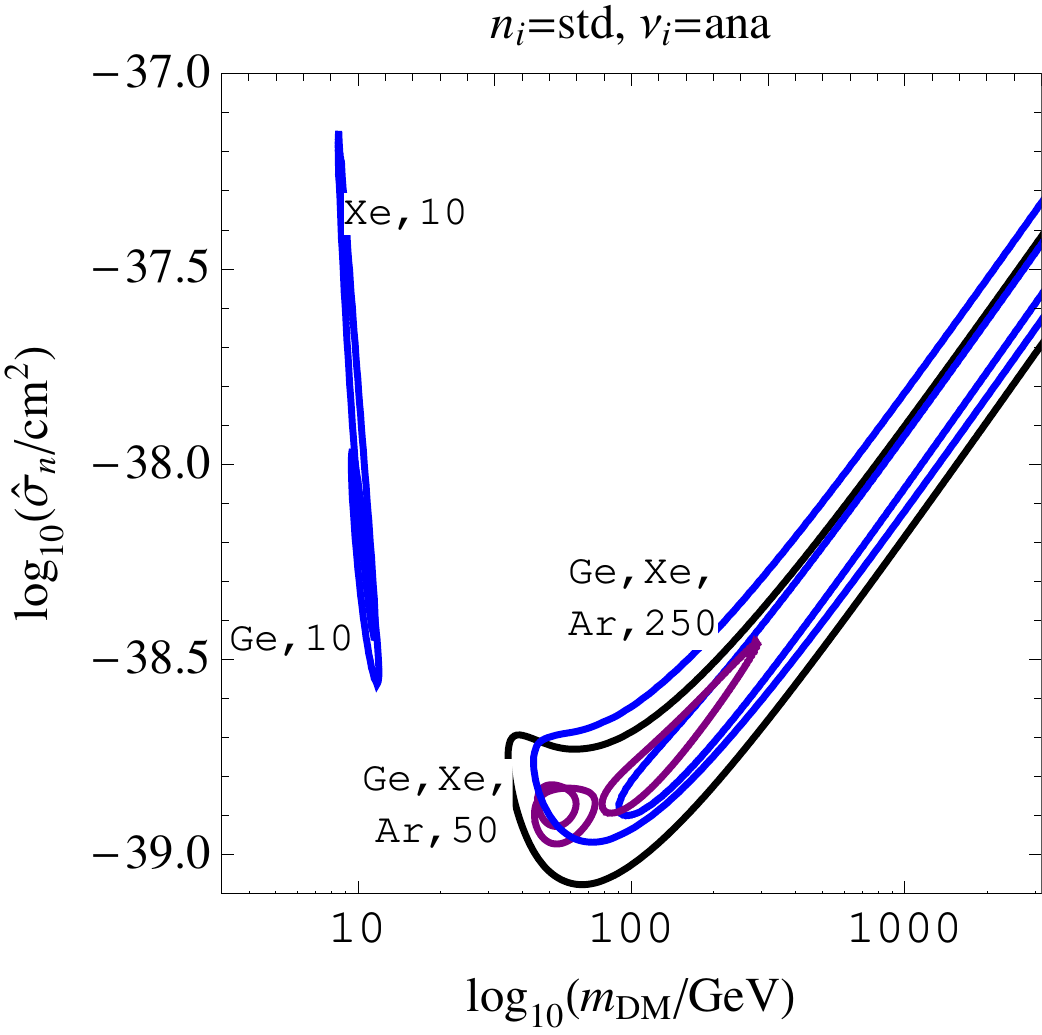}
\includegraphics[height=2.3in,width=2.5in]{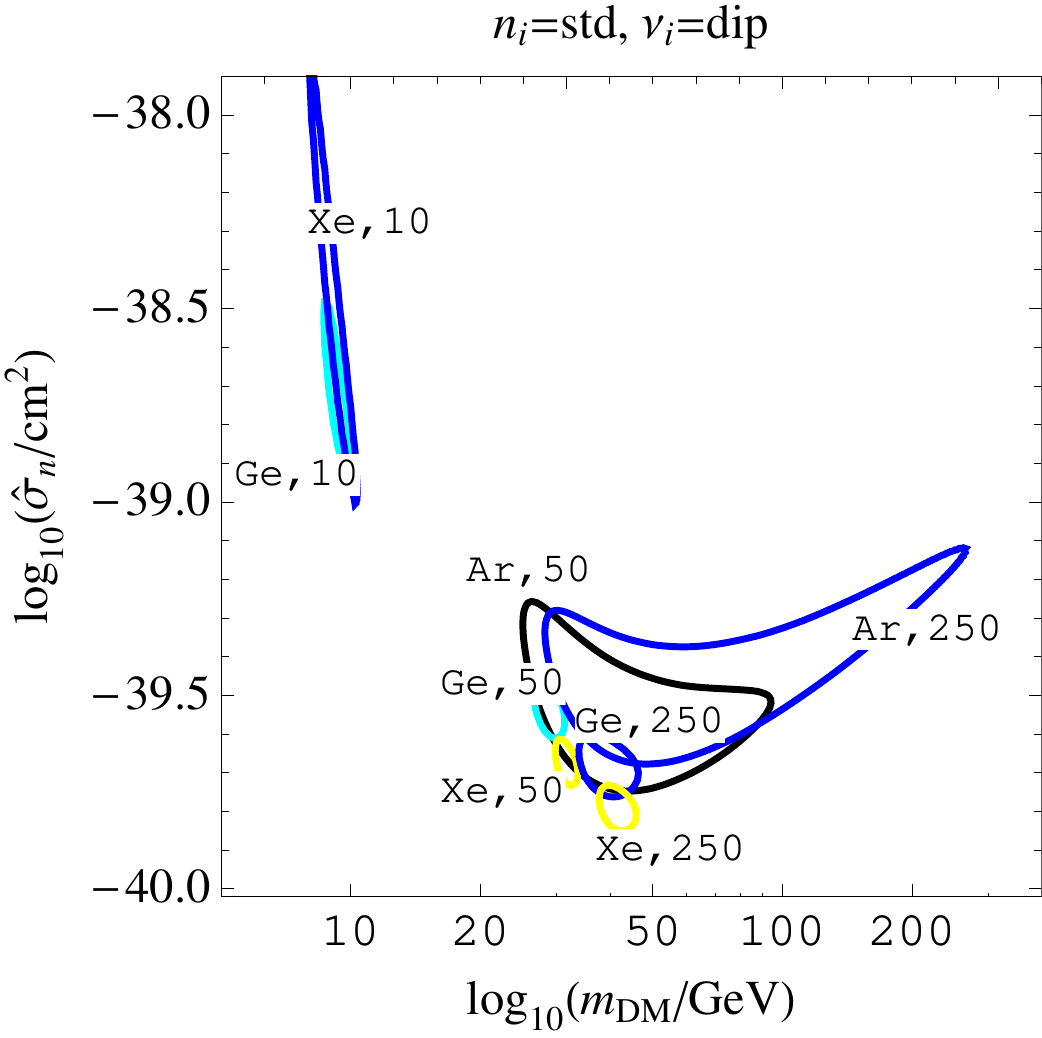}~~~~~~~~
\includegraphics[height=2.3in,width=2.5in]{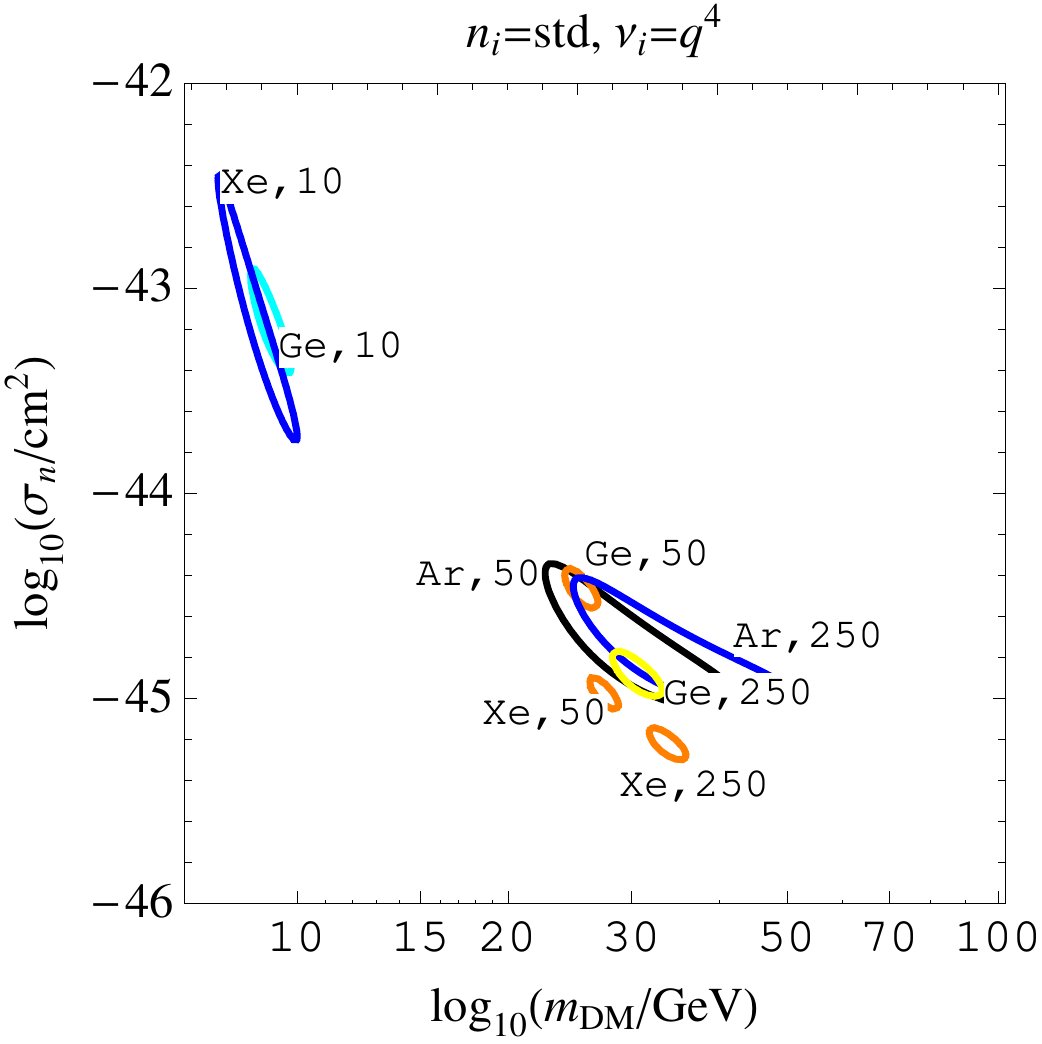}
\includegraphics[height=2.3in,width=2.5in]{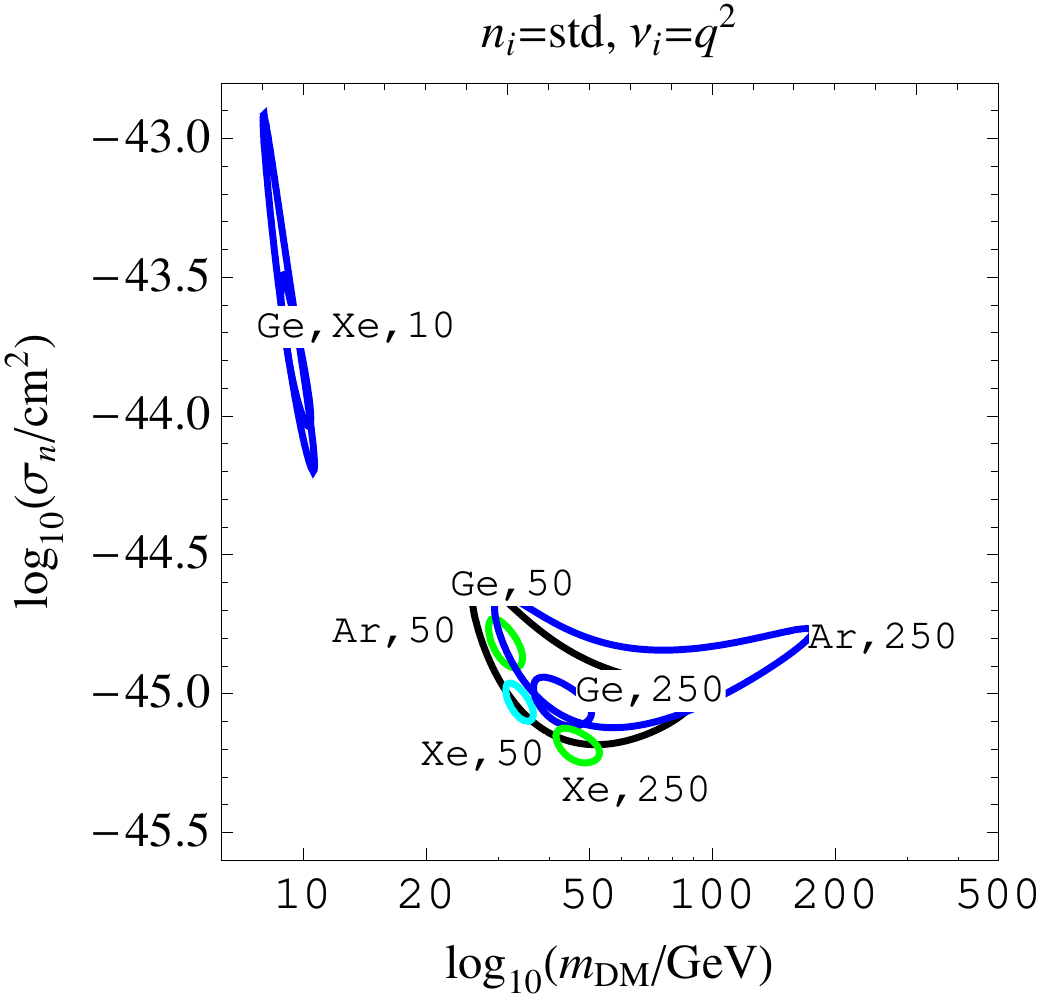}~~~~~~~~
\includegraphics[height=2.3in,width=2.5in]{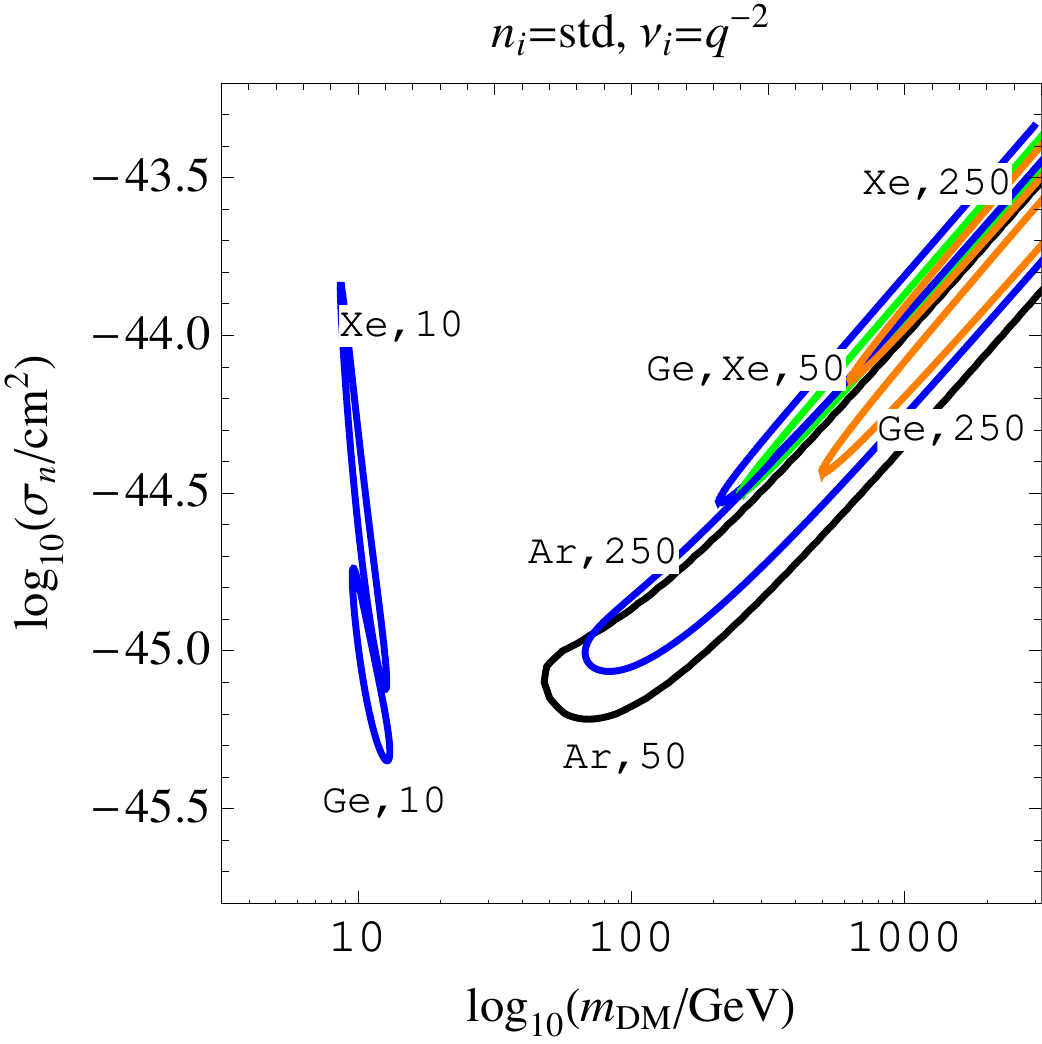}
\includegraphics[height=.6in,width=3.9in]{bar2.jpg}
\caption{95\% CLCs for a 10, 50, and 250 GeV particle interacting through an $n_i=$ standard operator. Comparisons are made to $\nu_i=$ standard, anapole, dipole, $q^4$, $q^2$, and $q^{-2}$ operators. The colors represent the value of $\widetilde{L}_{\rm min}/$ d.o.f. As described above, cyan and lighter colors correspond to 95\% or worse disagreement with the data.}
\end{center}
\end{figure}

\begin{figure}[b]
\begin{center}
\includegraphics[height=2.3in,width=2.5in]{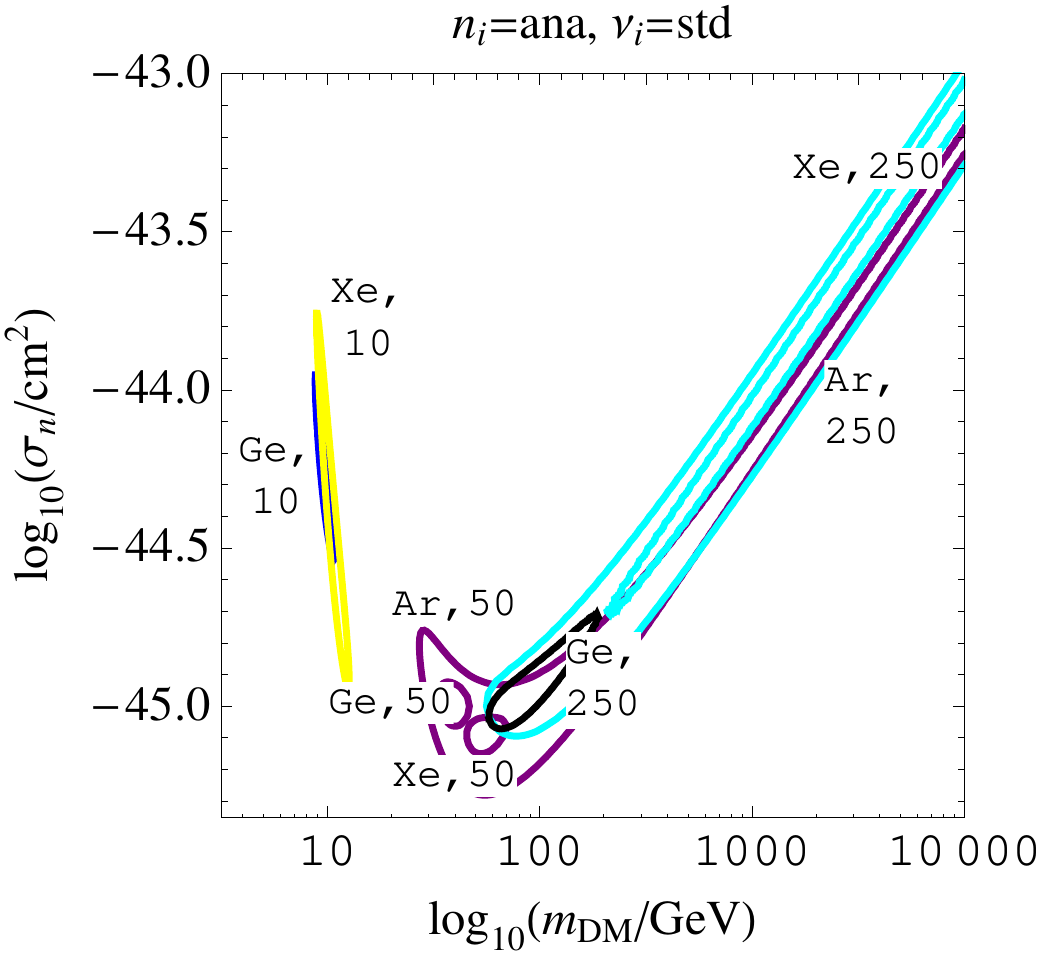}~~~~~~~~
\includegraphics[height=2.3in,width=2.5in]{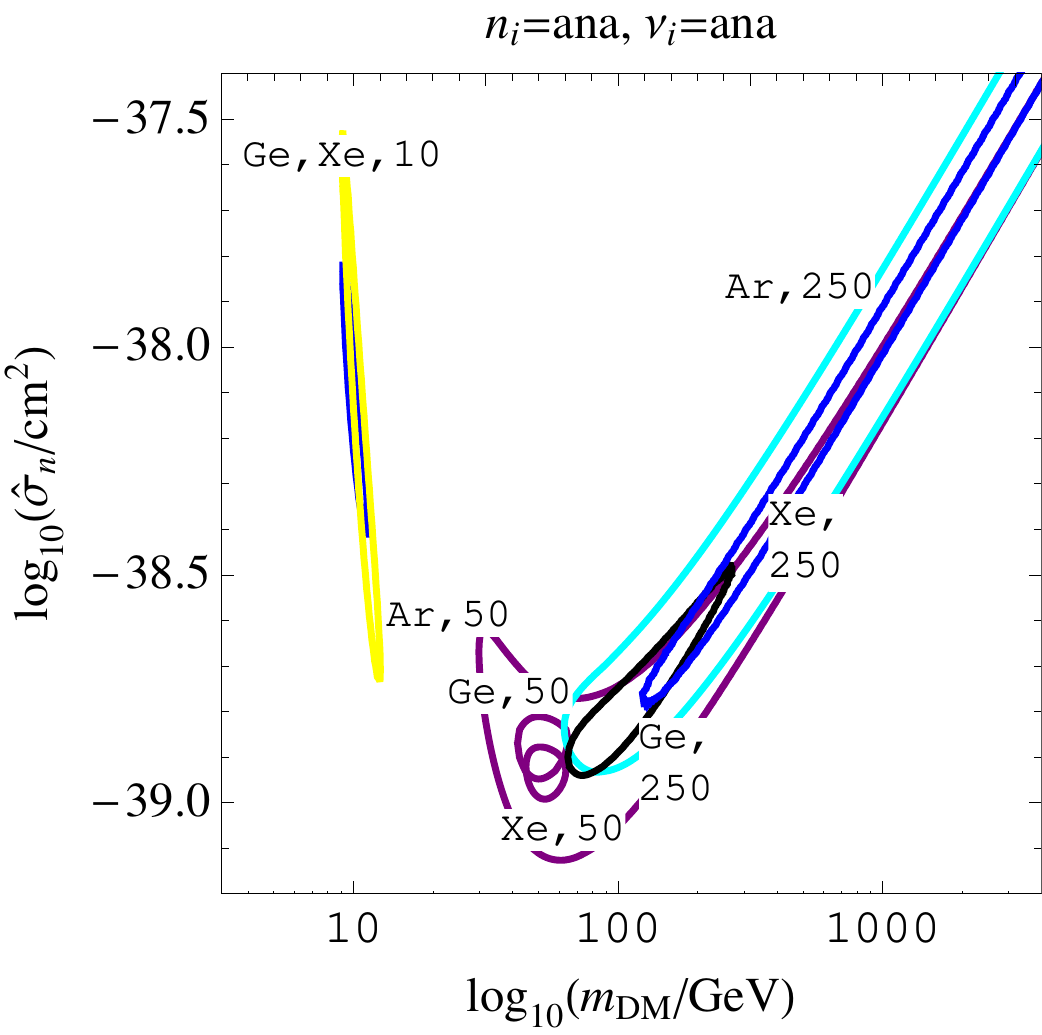}
\includegraphics[height=2.3in,width=2.5in]{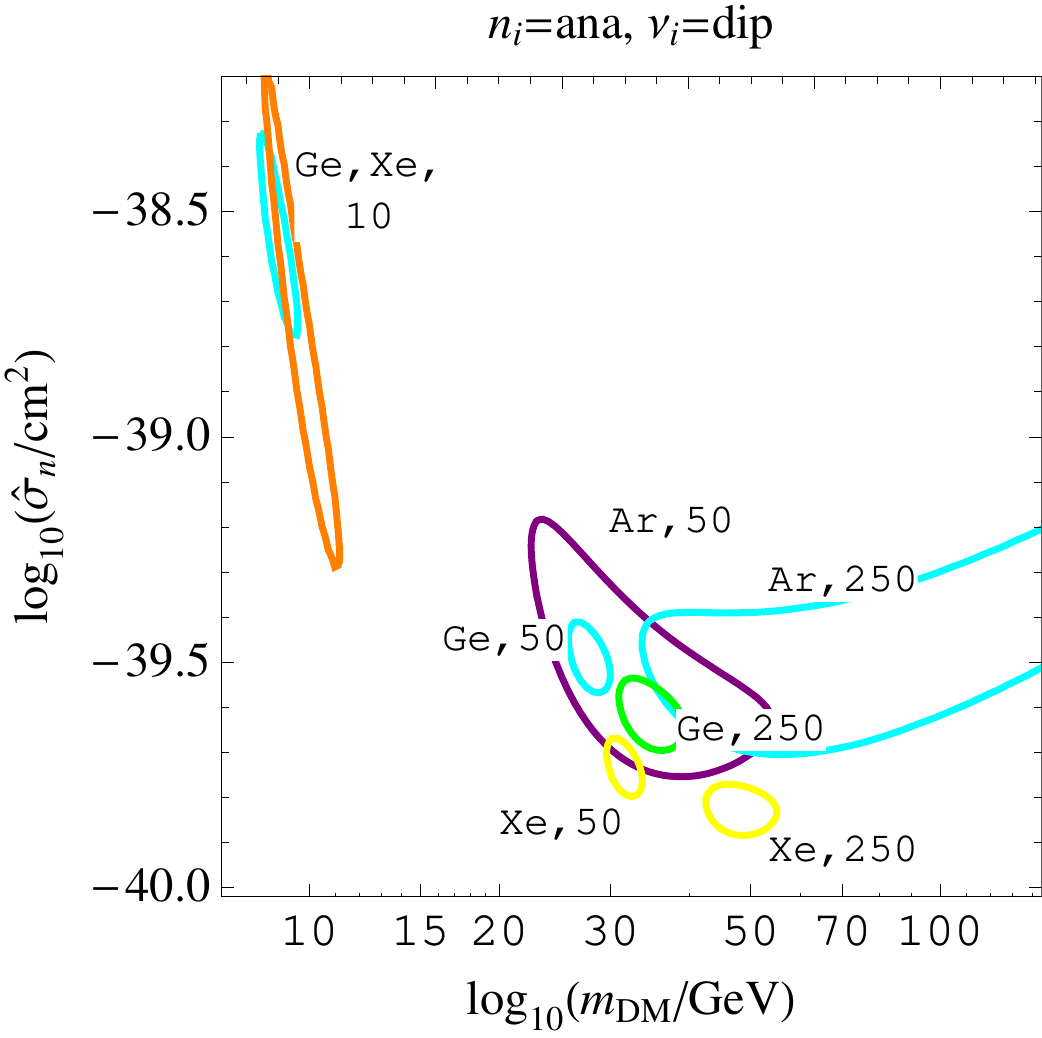}~~~~~~~~
\includegraphics[height=2.3in,width=2.5in]{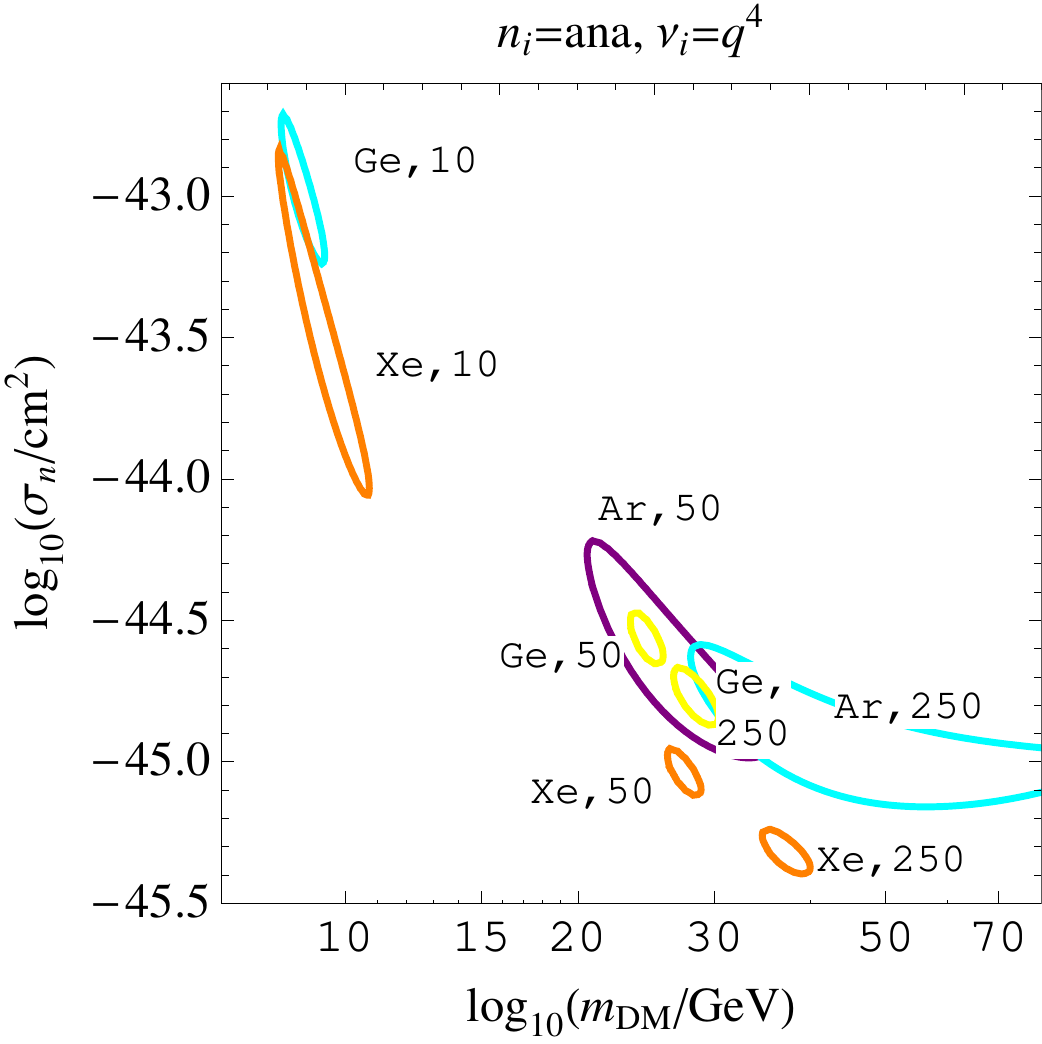}
\includegraphics[height=2.3in,width=2.5in]{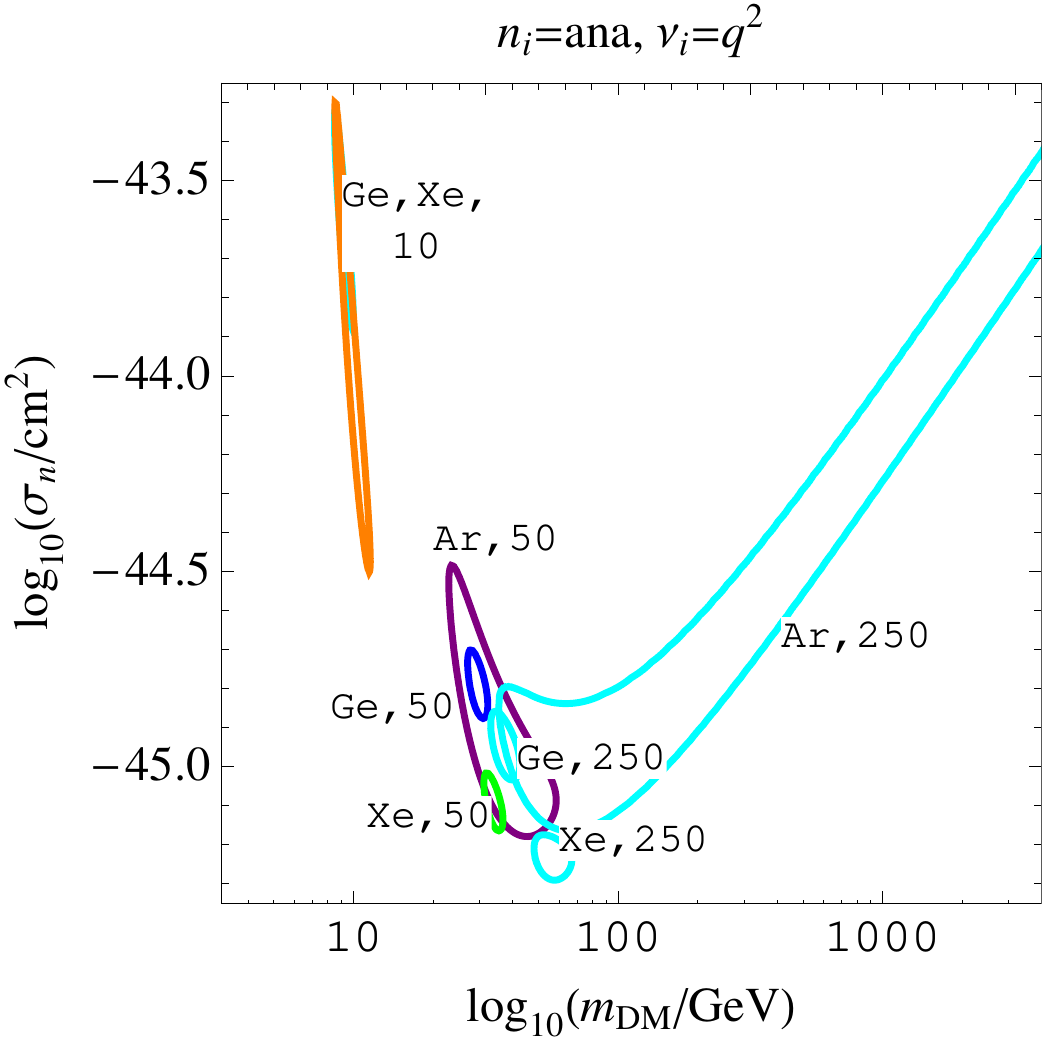}~~~~~~~~
\includegraphics[height=2.3in,width=2.5in]{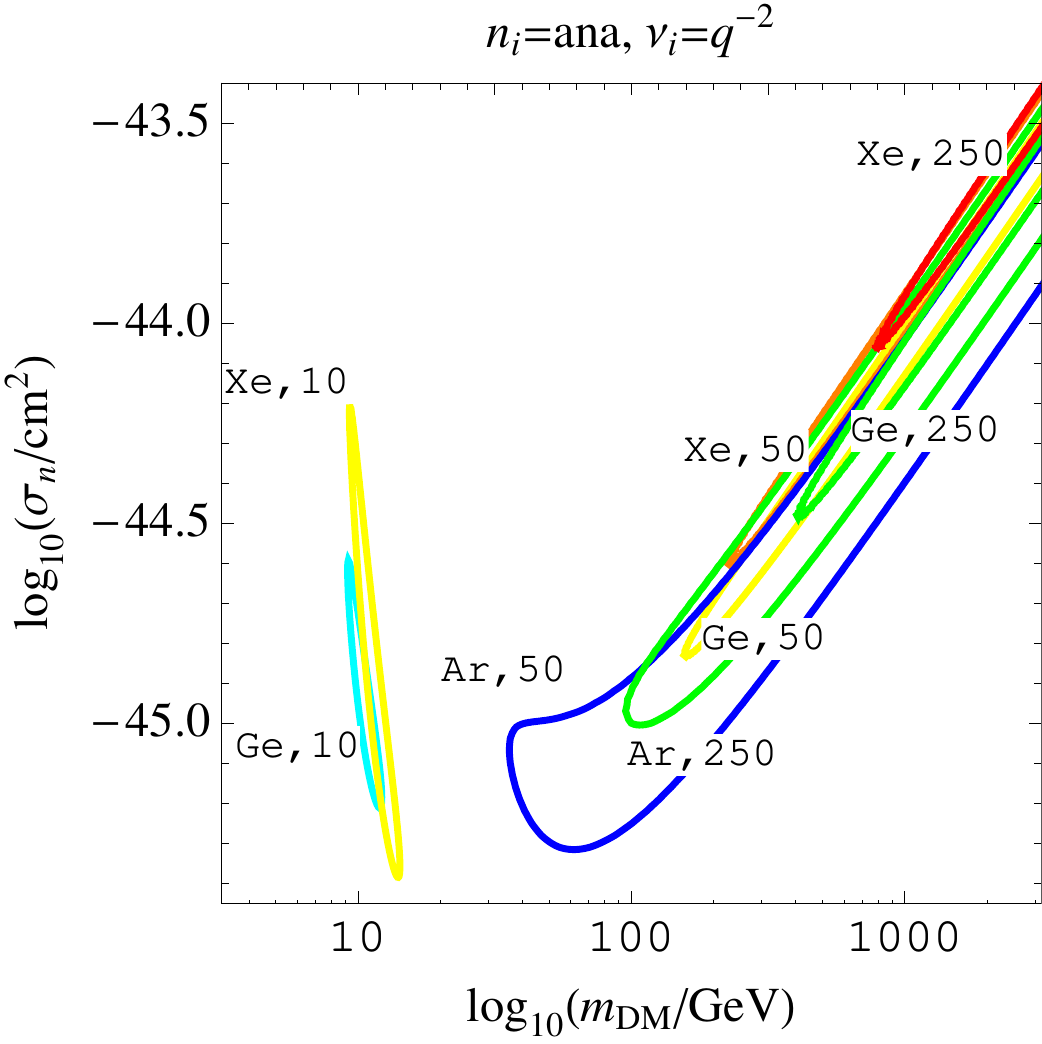}
\includegraphics[height=.6in,width=3.9in]{bar2.jpg}
\caption{95\% CLCs for a 10, 50, and 250 GeV particle interacting through an $n_i=$ anapole moment operator. Comparisons are made to $\nu_i=$ standard, anapole, dipole, $q^4$, $q^2$, and $q^{-2}$ operators. The colors represent the value of $\widetilde{L}_{\rm min}/$ d.o.f. As described above, cyan and lighter colors correspond to 95\% or worse disagreement with the data.}
\end{center}
\end{figure}

\begin{figure}[b]
\begin{center}
\includegraphics[height=2.3in,width=2.5in]{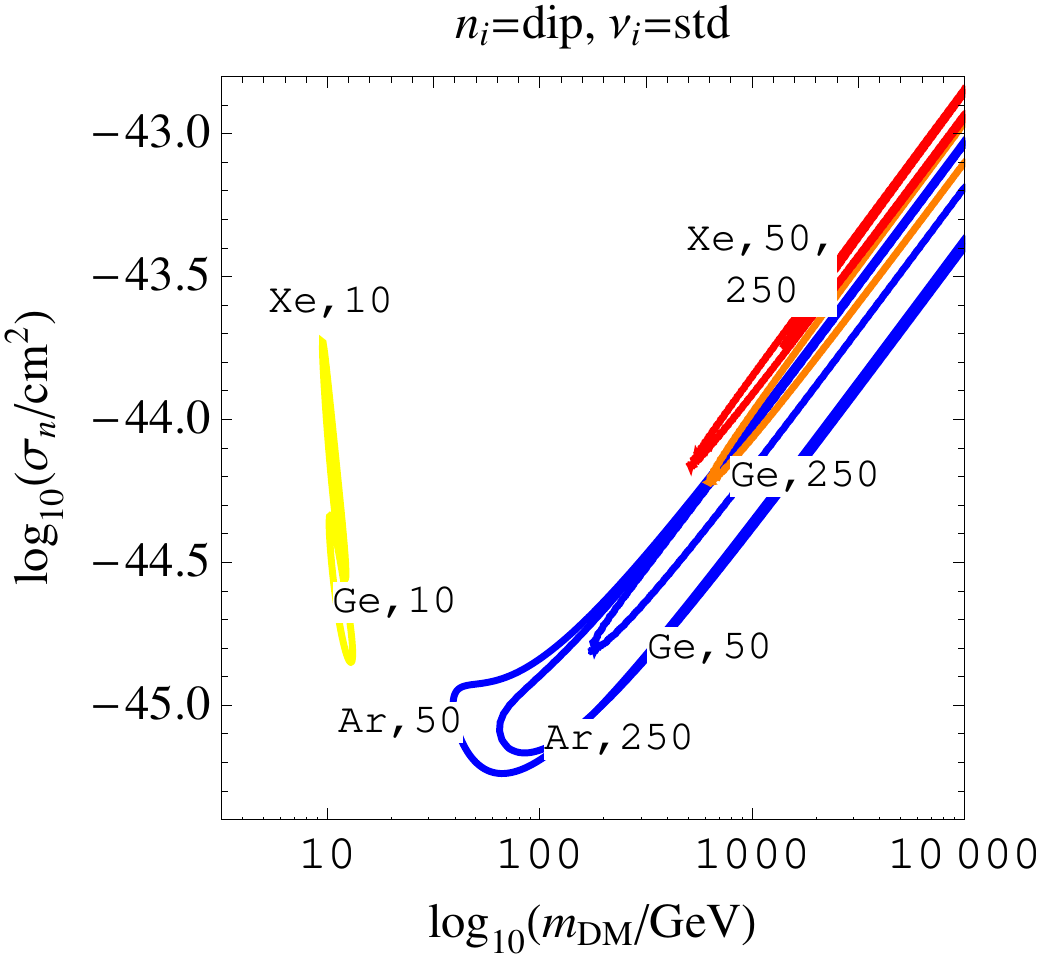}~~~~~~~~
\includegraphics[height=2.3in,width=2.5in]{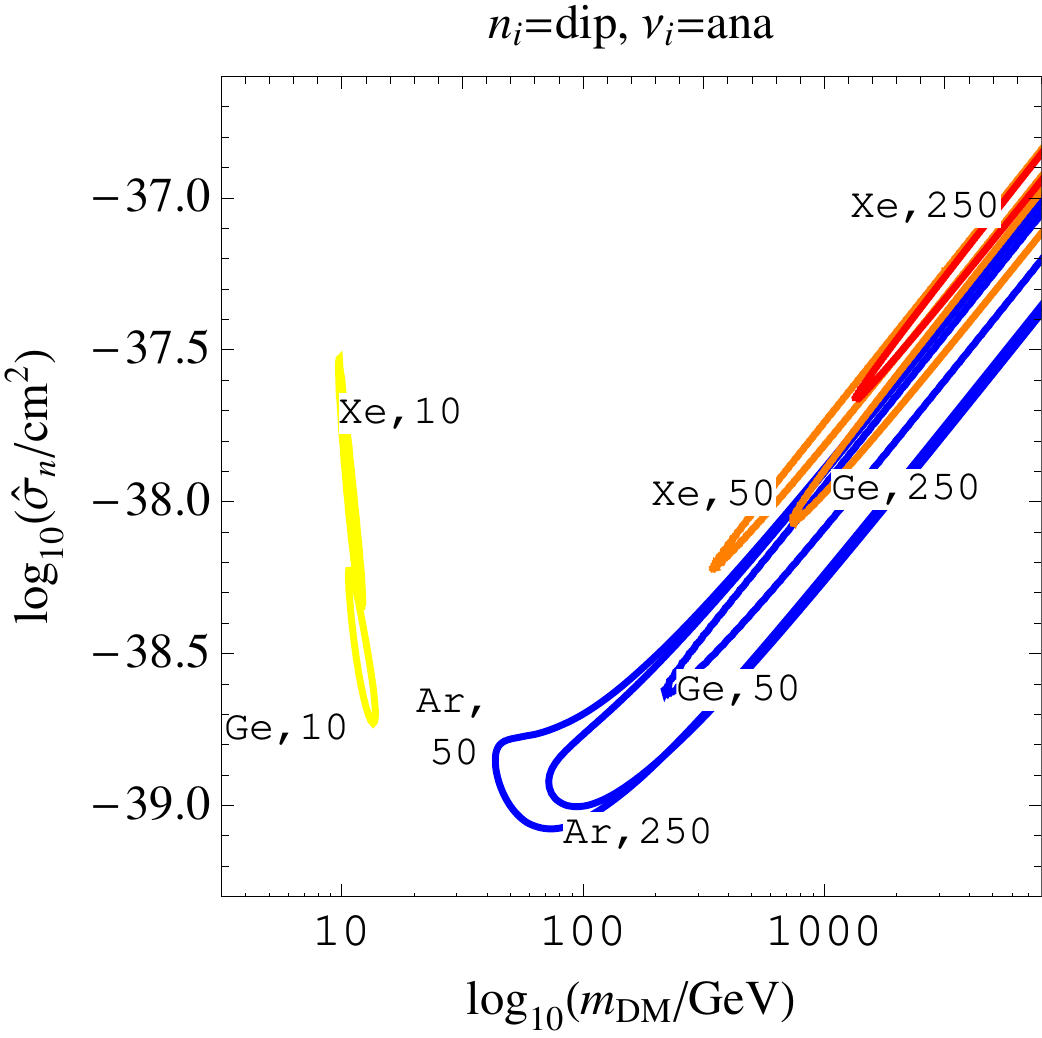}
\includegraphics[height=2.3in,width=2.5in]{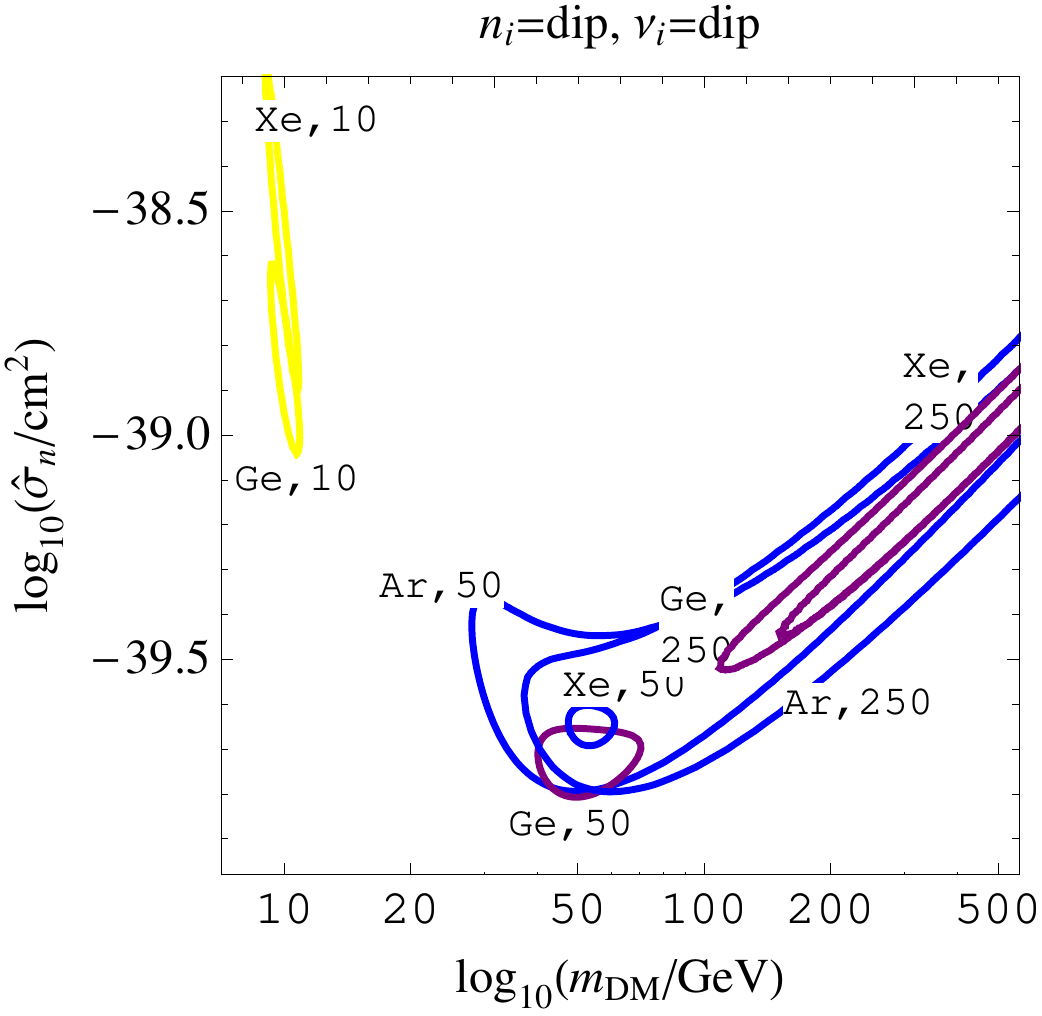}~~~~~~~~
\includegraphics[height=2.3in,width=2.5in]{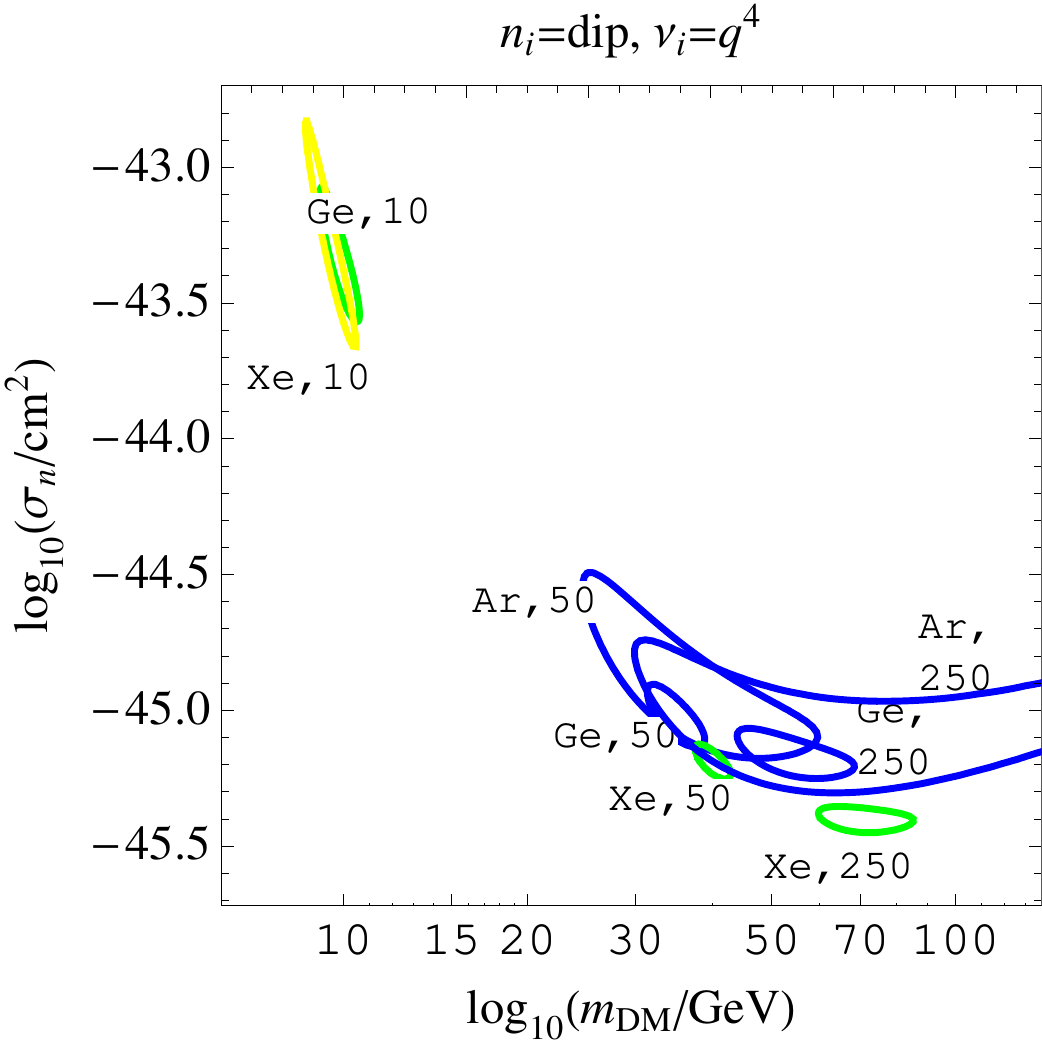}
\includegraphics[height=2.3in,width=2.5in]{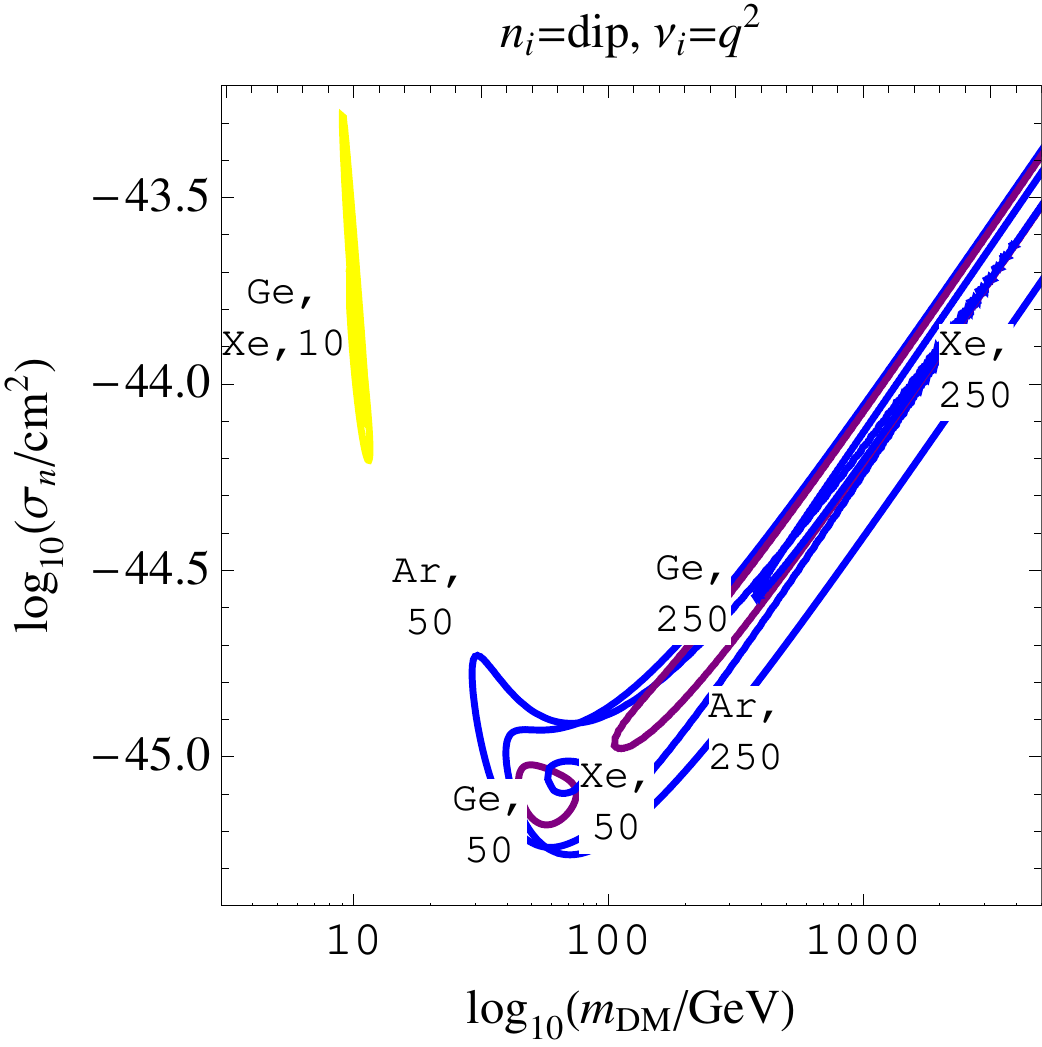}~~~~~~~~
\includegraphics[height=2.3in,width=2.5in]{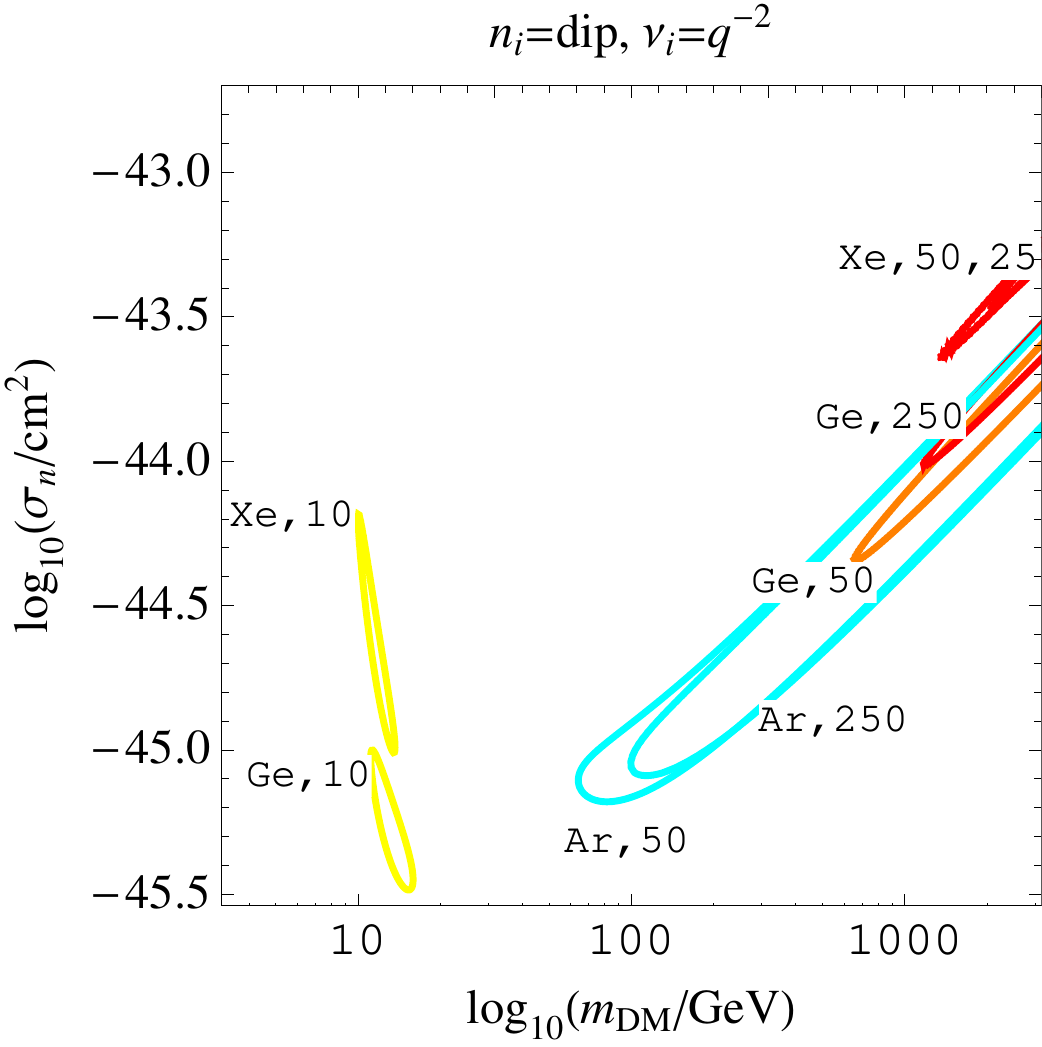}
\includegraphics[height=.6in,width=3.9in]{bar2.jpg}
\caption{95\% CLCs for a 10, 50, and 250 GeV particle interacting through an $n_i=$ dipole moment operator. Comparisons are made to $\nu_i=$ standard, anapole, dipole, $q^4$, $q^2$, and $q^{-2}$ operators. The colors represent the value of $\widetilde{L}_{\rm min}/$ d.o.f. As described above, cyan and lighter colors correspond to 95\% or worse disagreement with the data.}
\end{center}
\end{figure}

\begin{figure}[b]
\begin{center}
\includegraphics[height=2.3in,width=2.5in]{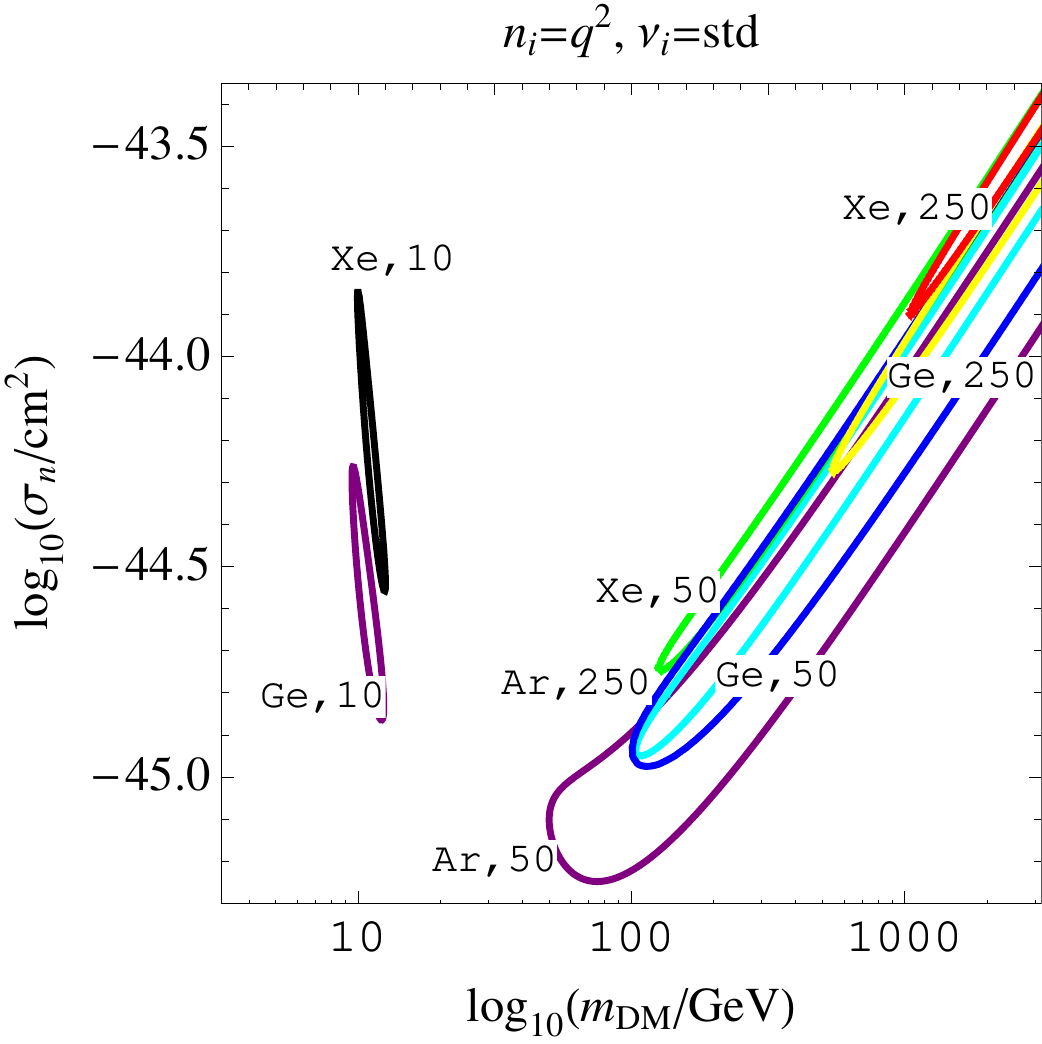}~~~~~~~~
\includegraphics[height=2.3in,width=2.5in]{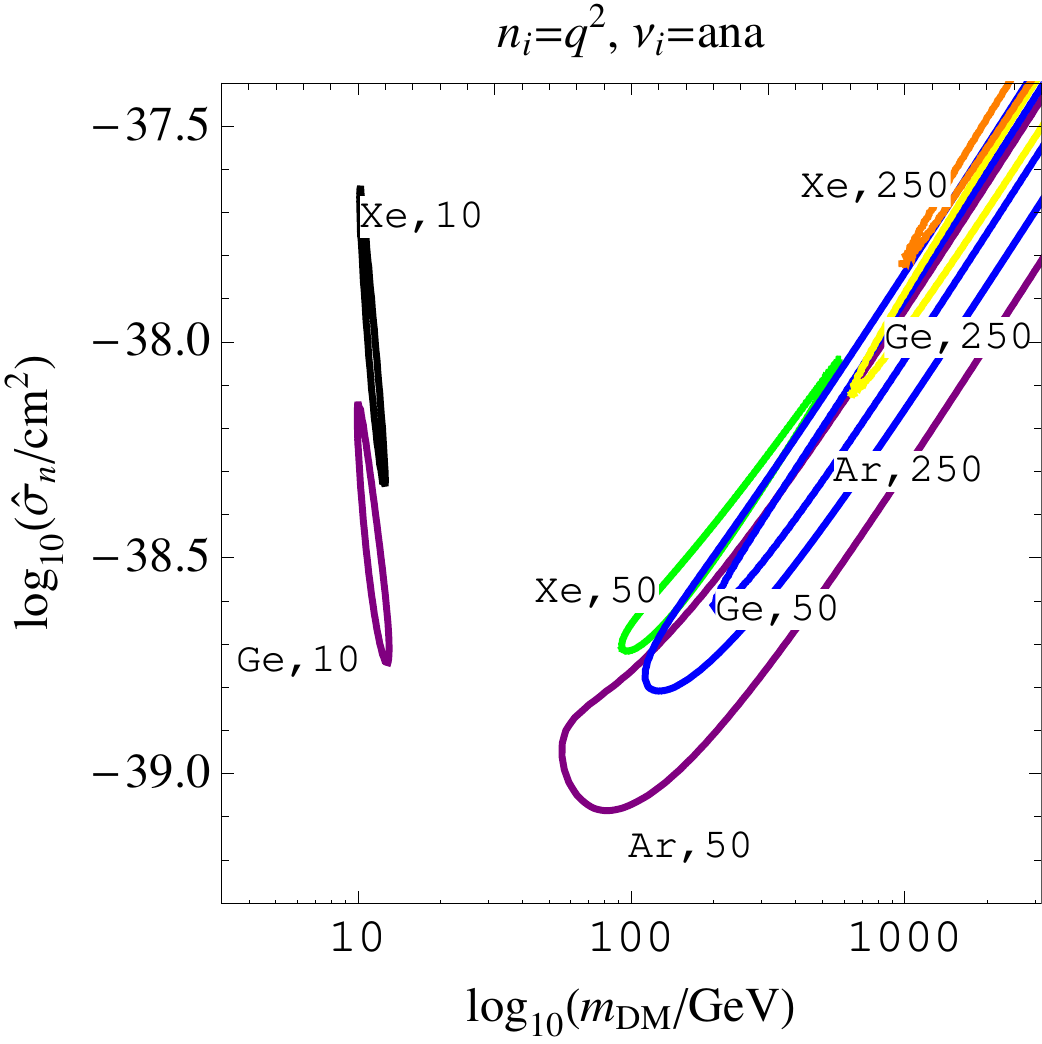}
\includegraphics[height=2.3in,width=2.5in]{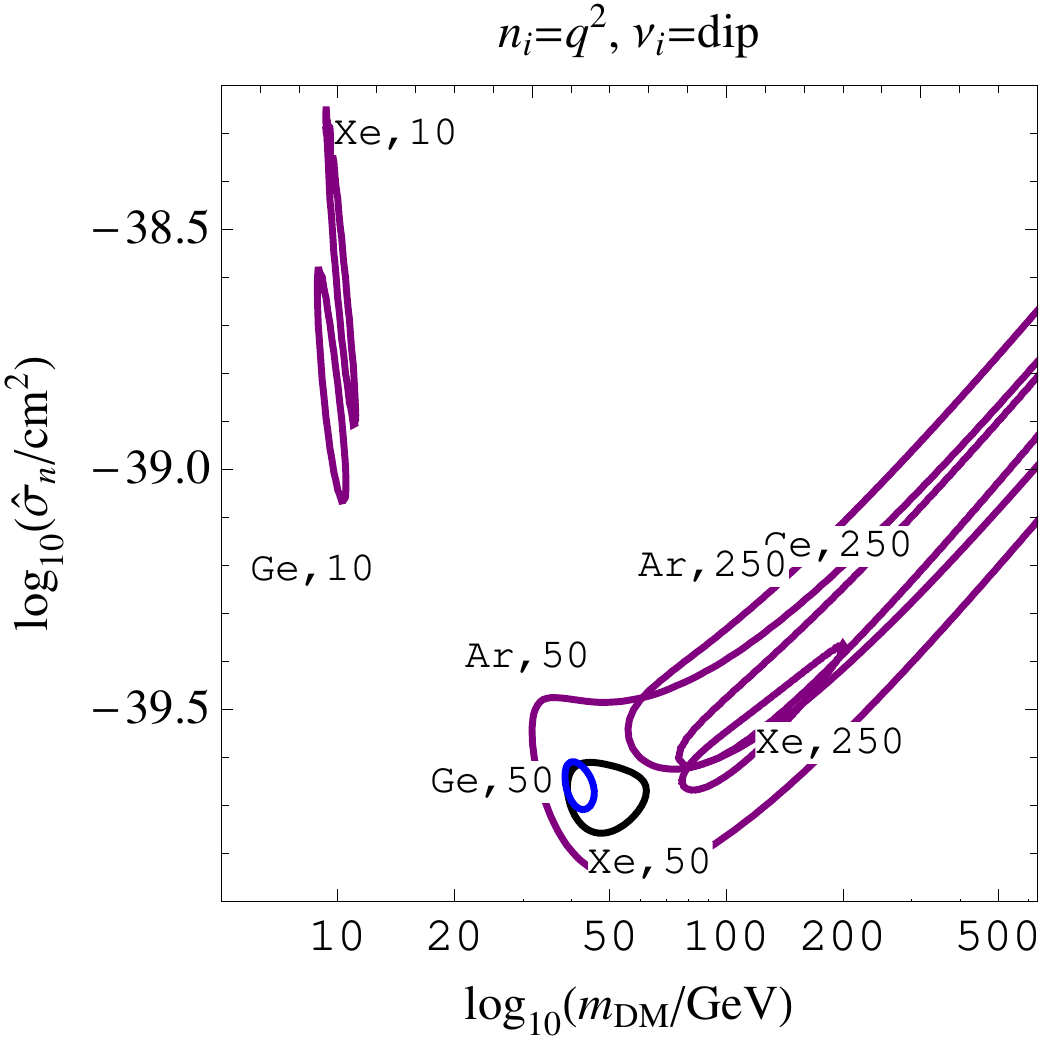}~~~~~~~~
\includegraphics[height=2.3in,width=2.5in]{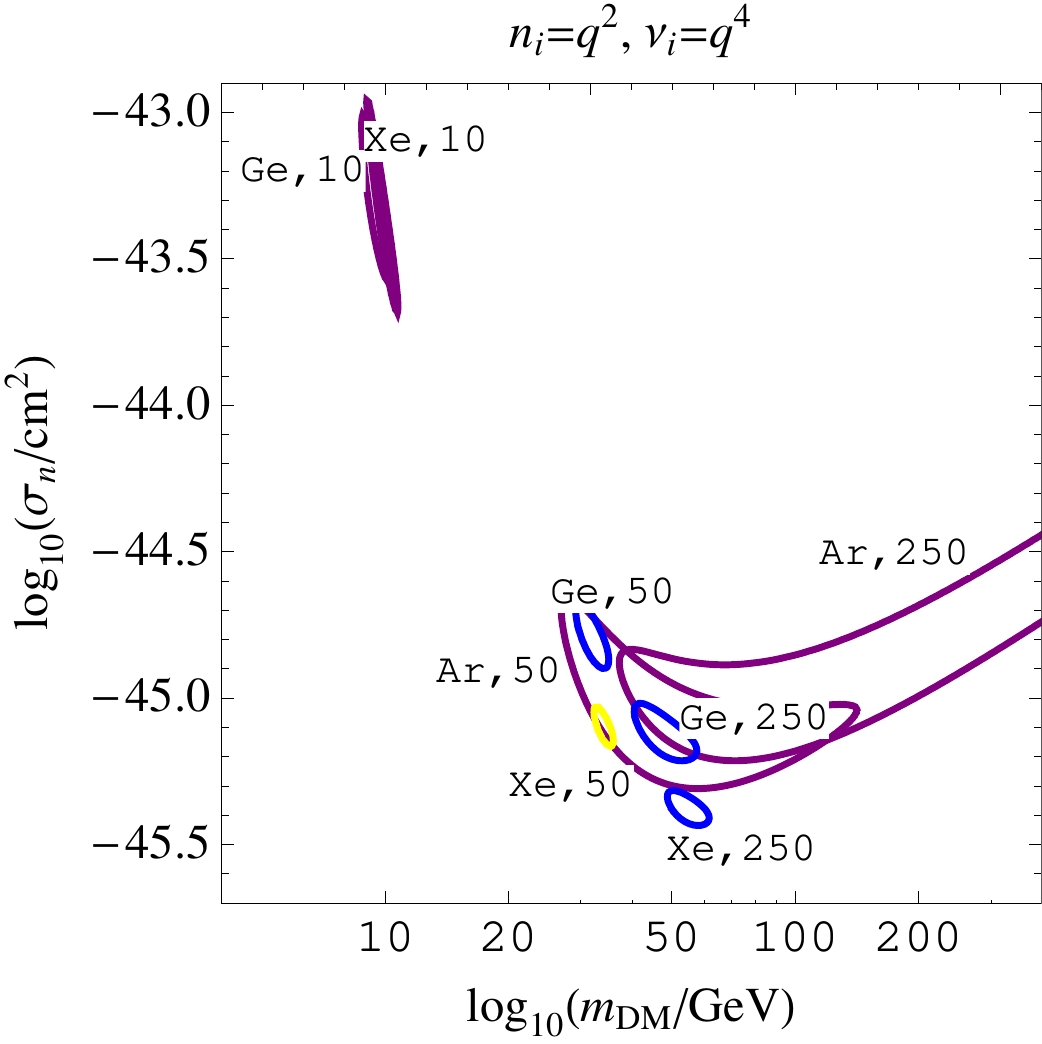}
\includegraphics[height=2.3in,width=2.5in]{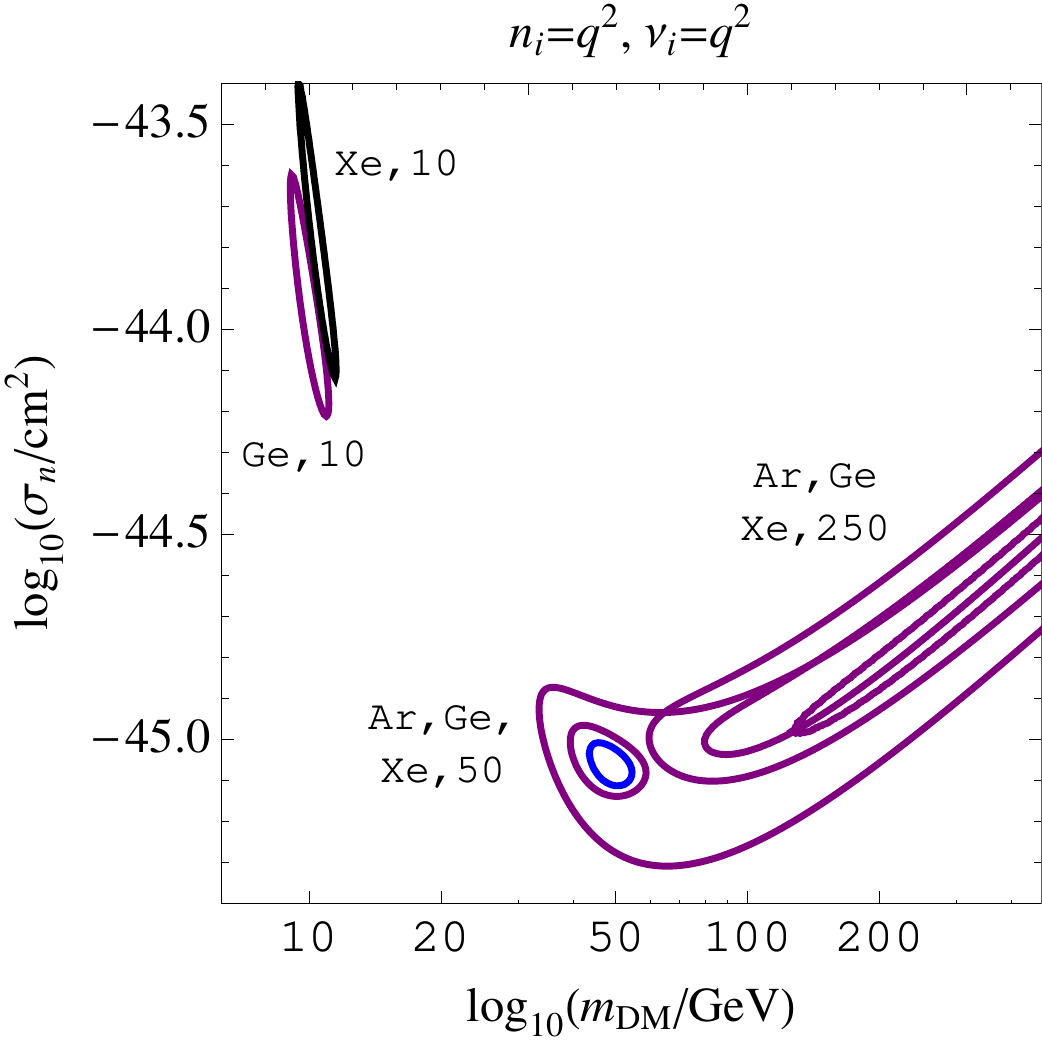}~~~~~~~~
\includegraphics[height=2.3in,width=2.5in]{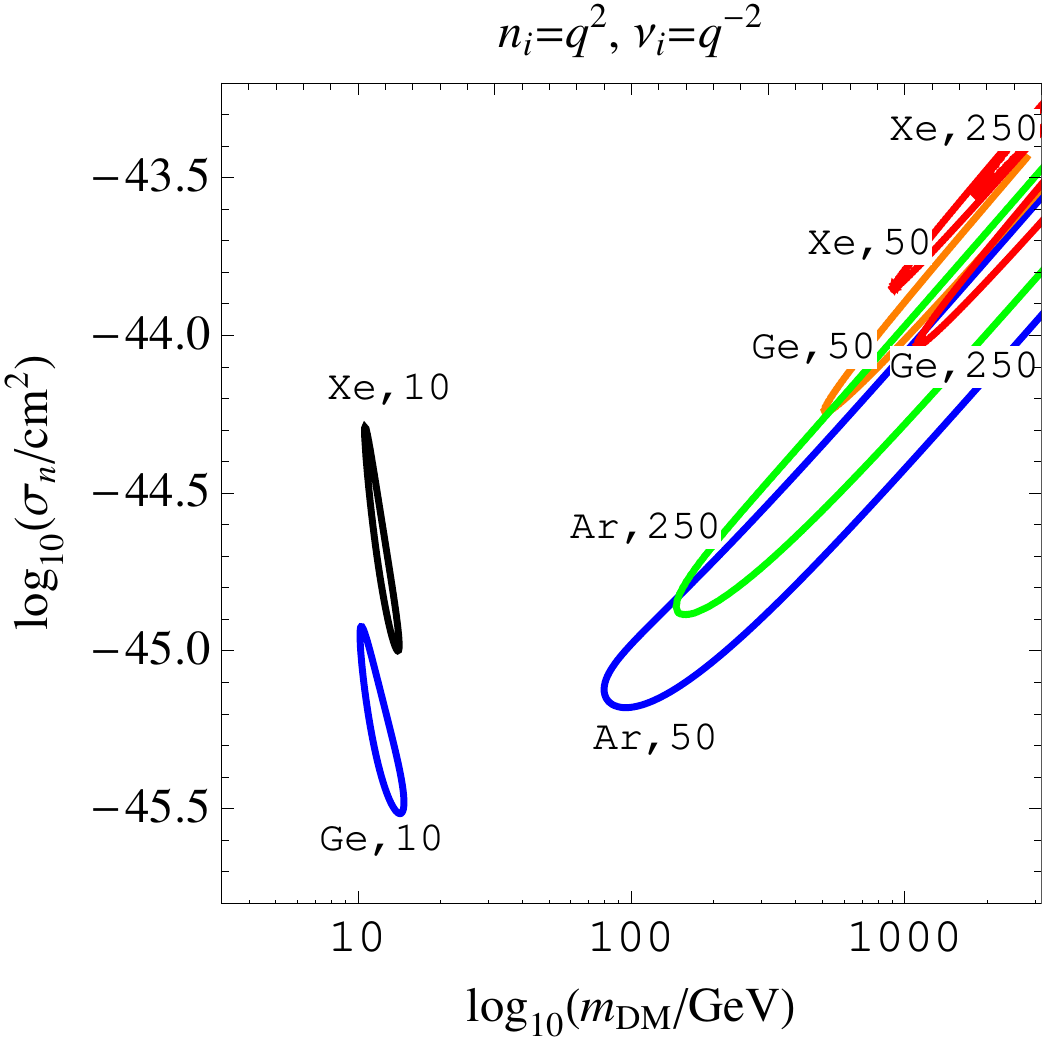}
\includegraphics[height=.6in,width=3.9in]{bar2.jpg}
\caption{95\% CLCs for a 10, 50, and 250 GeV particle interacting through an $n_i=q^2$ operator. Comparisons are made to $\nu_i=$ standard, anapole, dipole, $q^4$, $q^2$, and $q^{-2}$ operators. The colors represent the value of $\widetilde{L}_{\rm min}/$ d.o.f. As described above, cyan and lighter colors correspond to 95\% or worse disagreement with the data.}
\end{center}
\end{figure}

\begin{figure}[b]
\begin{center}
\includegraphics[height=2.3in,width=2.5in]{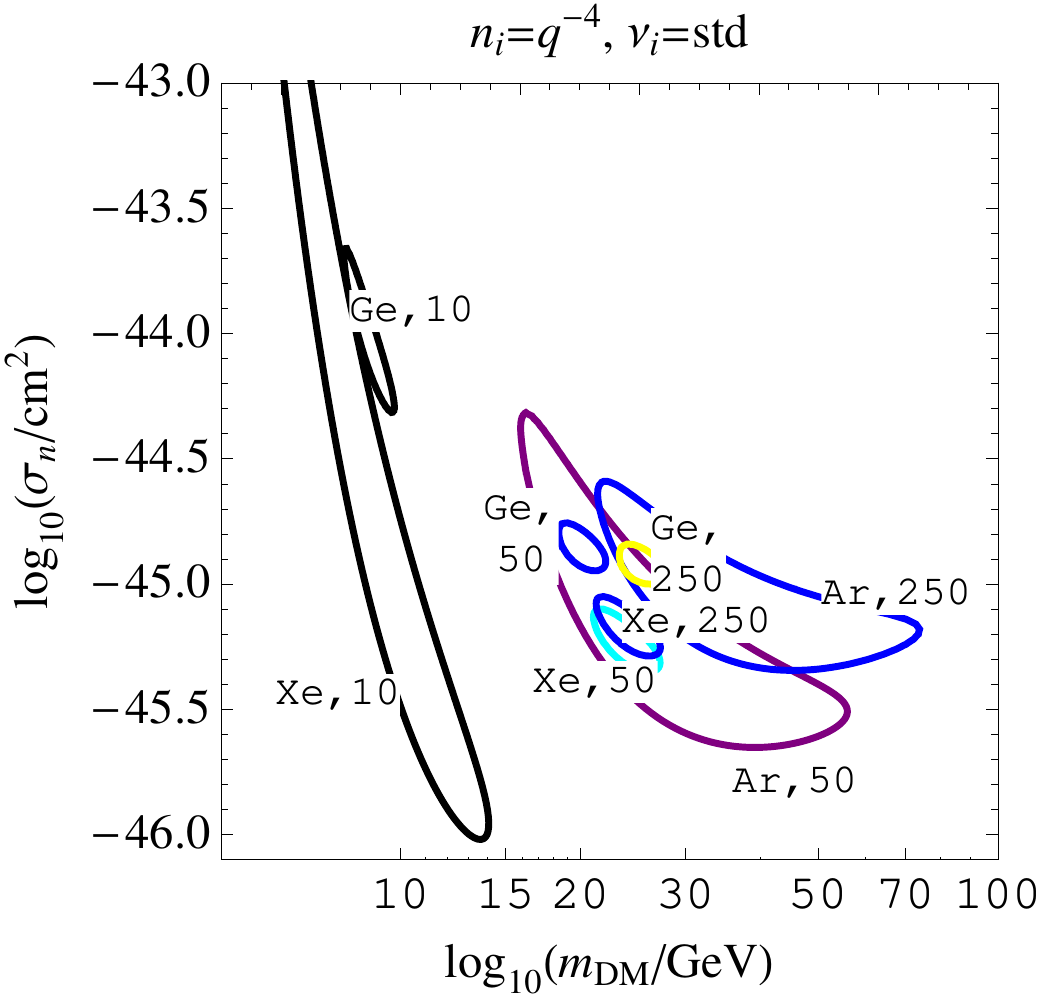}~~~~~~~~
\includegraphics[height=2.3in,width=2.5in]{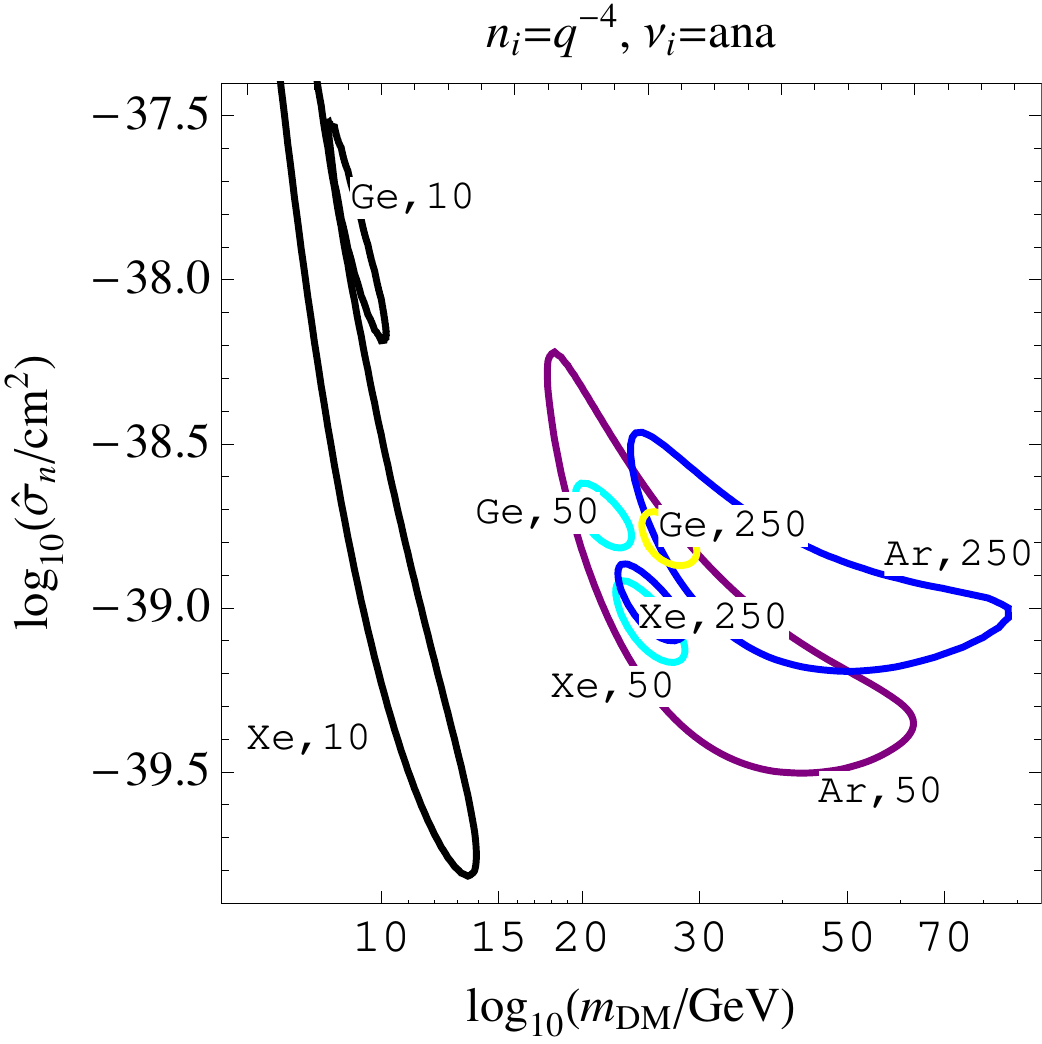}
\includegraphics[height=2.3in,width=2.5in]{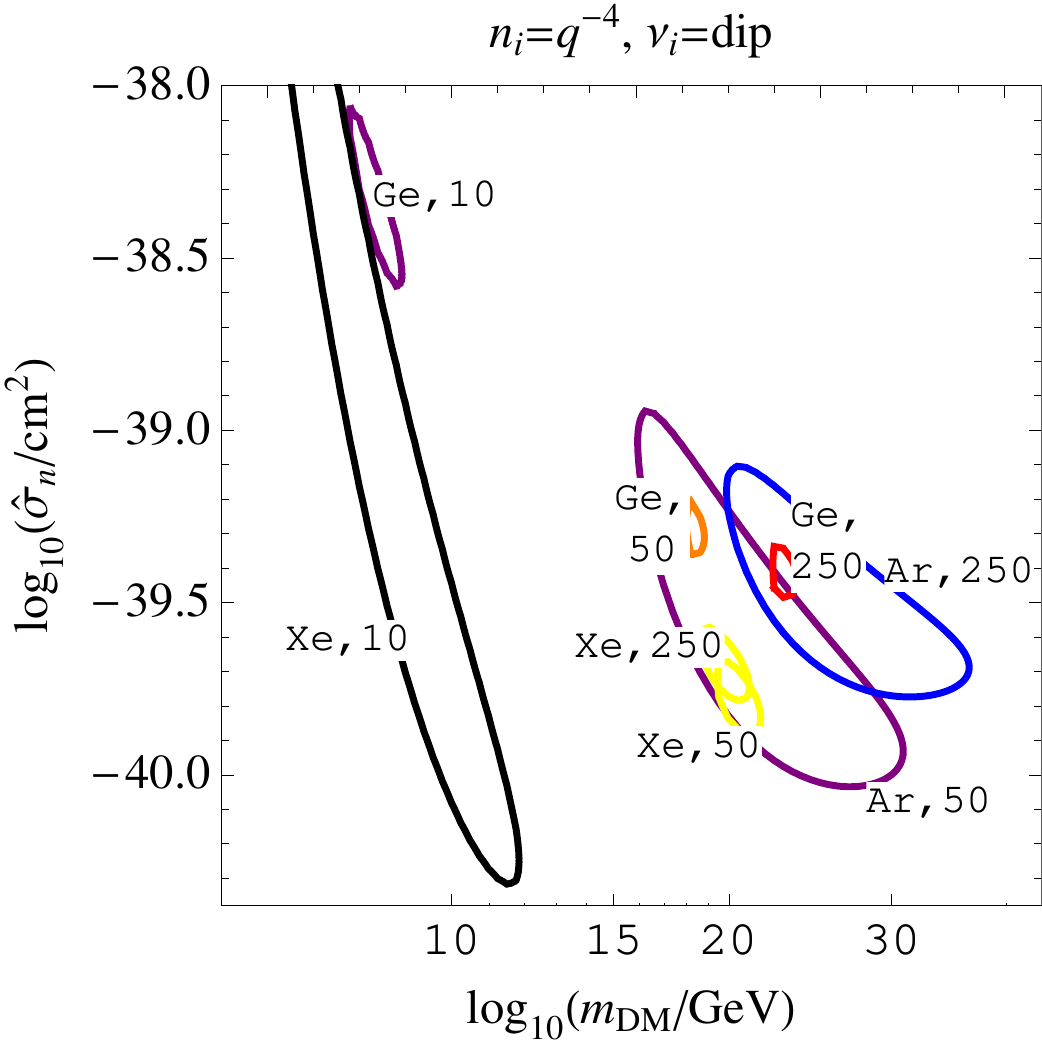}~~~~~~~~
\includegraphics[height=2.3in,width=2.5in]{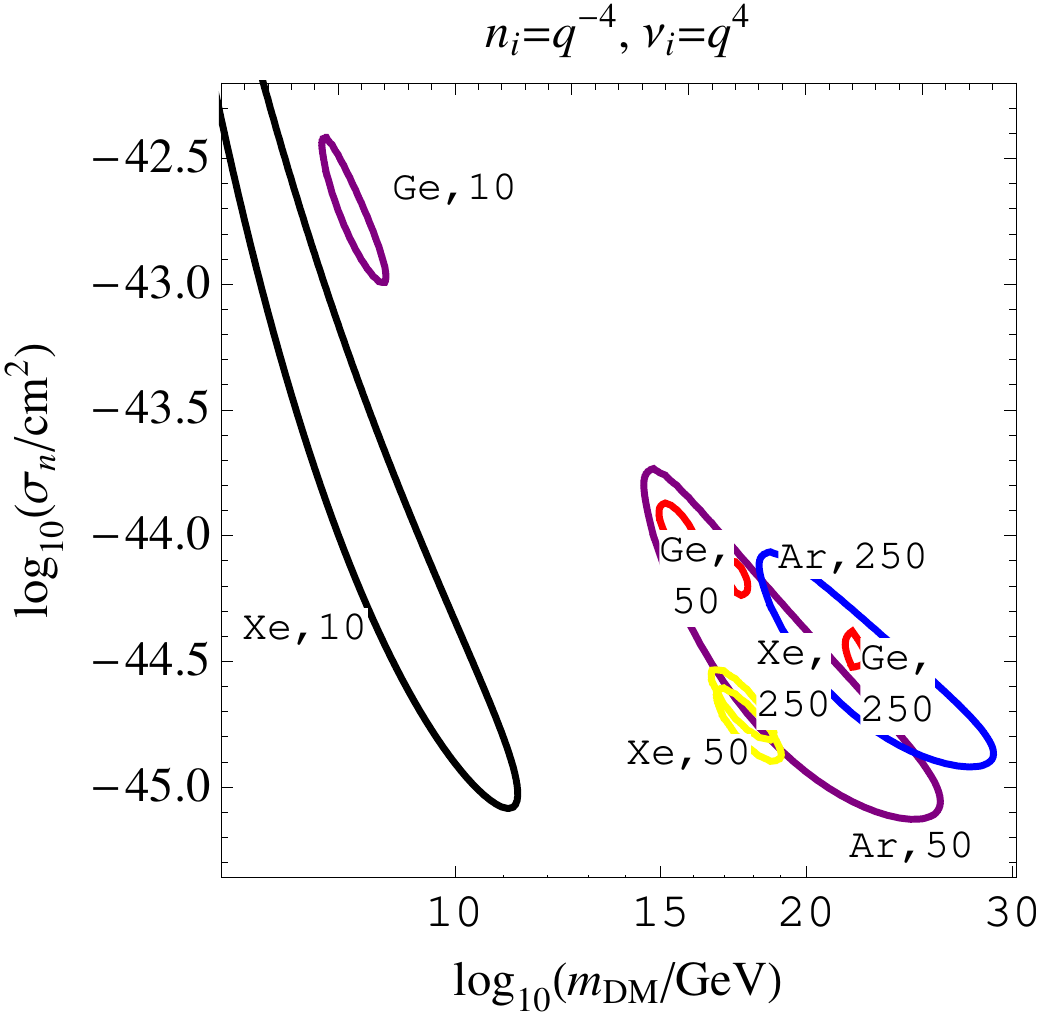}
\includegraphics[height=2.3in,width=2.5in]{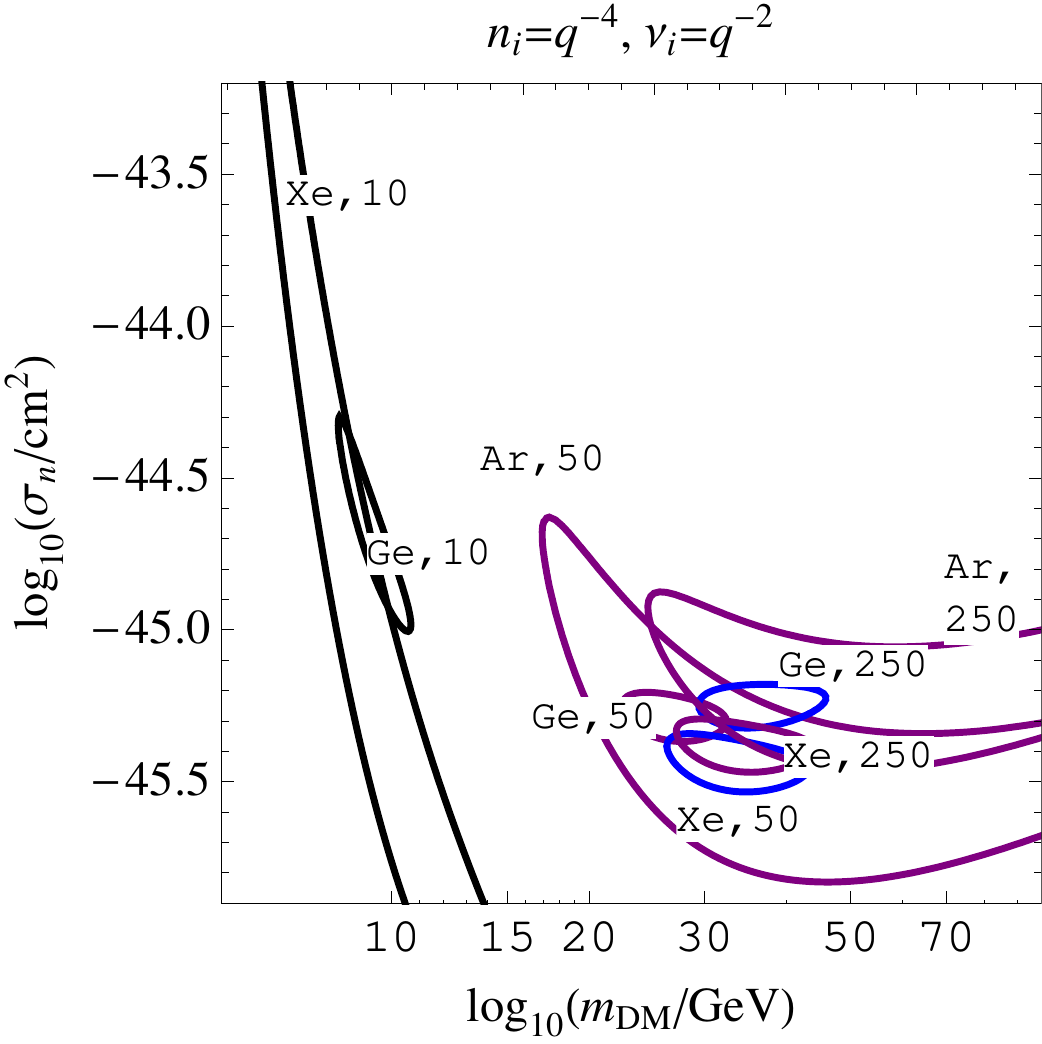}~~~~~~~~
\includegraphics[height=2.3in,width=2.5in]{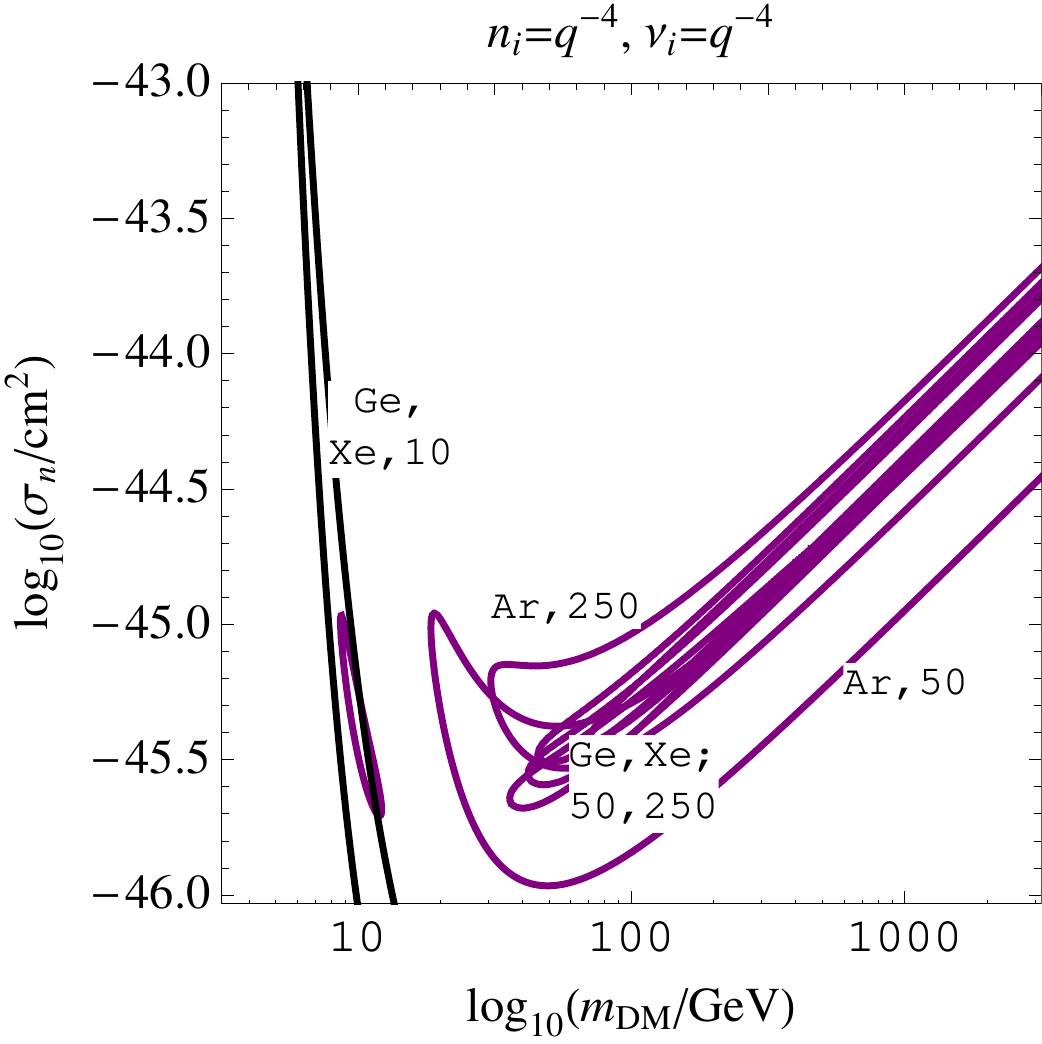}
\includegraphics[height=.6in,width=3.9in]{bar2.jpg}
\end{center}
\caption{95\% CLCs for a 10, 50, and 250 GeV particle interacting through an $n_i=q^{-4}$ operator. Comparisons are made to $\nu_i=$ standard, anapole, dipole, $q^4$, $q^{-2}$, and $q^{-4}$ operators. The colors represent the value of $\widetilde{L}_{\rm min}/$ d.o.f. As described above, cyan and lighter colors correspond to 95\% or worse disagreement with the data.}
\end{figure}

\end{document}